\documentclass[a4paper,11pt]{article}
\pdfoutput=1 % if your are submitting a pdflatex (i.e. if you have
\usepackage{jheppub} % for details on the use of the package, please
\usepackage[T1]{fontenc} % if needed

\usepackage{amsmath,amssymb,amsfonts,slashed,comment}
\usepackage{cancel}
\usepackage{mathtools}
\usepackage[normalem]{ulem}

\title{\boldmath  Running coupling effects in the anti-collinear resummation in high energy evolution}

\author[a]{Alex Kovner,}
\author[b]{Michael Lublinsky,}
\author[b,1]{Maxim Nefedov\note{Corresponding author.}}
\author[c]{and Vladimir Skokov}

\affiliation[a]{Physics Department, University of Connecticut, \\
196A Auditorium road, Storrs, CT 06269-3046, U.S.A.}
\affiliation[b]{Physics Department, Ben-Gurion University of the Negev, \\
Beer Sheva 84105, Israel}
\affiliation[c]{Department of Physics, North Carolina State University, \\
Raleigh, NC 27695, U.S.A.}

\emailAdd{alexander.kovner@uconn.edu}
\emailAdd{lublinm@bgu.ac.il}
\emailAdd{nefedov@post.bgu.ac.il}
\emailAdd{vskokov@ncsu.edu}

\abstract{We study the effects of the running of the QCD coupling  on the  anti-collinear resummation in  JIMWLK evolution in the linear (BFKL) regime.  We determine the appropriate  scale {choice} for the coupling {entering} the JIMWLK kernel, and derive the anti-collinearly resummed BFKL kernel, {which} includes running-coupling effects {both} in the resummation equation  ({i.e.} DGLAP) {and} in the JIMWLK kernel proper. We find that {the} running {of the coupling}  generally  {further slows down} BFKL evolution, as expected. Surprisingly however the value of the generalized characteristic function at $\gamma=1$ is unaffected by the running {coupling} {owing to} subtle cancellations. We {develop} an approximation that allows us to use the generalized characteristic function to study the BFKL Green's function. Within this approximation we find that the Pomeron intercept in the anti-collinear regime is significantly reduced {by both}, the resummation and the running of the coupling.}

\newcommand{\T}[1]{\boldsymbol{#1}}  % Transverse vector
\newcommand{\Tb}[1]{\bar{\boldsymbol{#1}}} % Transverse vector with a bar

\newcommand{\tr}{\mathop{\mbox{tr}}\nolimits}

\newcommand{\ket}[1]{\left\vert #1 \right\rangle}
\newcommand{\bra}[1]{\left\langle #1 \right\vert}

\newcommand{\Gtil}{g}

\begin{document} 
\maketitle
\flushbottom

%%%%%%%%%%%%%%%%%%%%%%%%%%%%%%%%%%%%%%%%%%%%%%%%%%%%%%%%%%%%%%%%%%%%%%%%%%%%%%%%%%%%%%%%%%%

\section{Introduction}\label{sec:intro}
{In Ref.~\cite{Kovner:2023vsy}, an approach to resumming large anti-collinear logarithms within JIMWLK evolution \cite{Jalilian-Marian:1997qno,Jalilian-Marian:1997ubg,Jalilian-Marian:1997jhx,Kovner:2000pt,Kovner:1999bj,Iancu:2000hn,Iancu:2001ad,Ferreiro:2001qy,Weigert:2000gi} at high energy was formulated.}
Such a resummation is {necessary} since JIMWLK evolution,  
{like} its BFKL limit at NLO~\cite{Fadin:1998py,NLOCiafaloni1,NLOCiafaloni2,Salam98,Ciafaloni:2003rd}, {suffers} from  {the appearance} of large transverse logarithms, which render the  {high-energy} evolution  {unstable for some classes of initial conditions}, leading to unphysical results for the cross section. Resummation of these transverse logarithms is necessary to {control} the energy evolution {when} the projectile and target are {characterized} by vastly different transverse scales, {$Q_P$ and $Q_T$ respectively.} 

This resummation was then studied in detail in Ref.~\cite{Kovner:2026qfm} in the weak-field,  {i.e.} BFKL, limit. {The {\it anti-collinear} logarithms are large when $Q_T\gg Q_P$ and the resummation parameter in this limit is $\alpha_s \ln (Q_T/Q_P)$. Keeping only terms $\sim \alpha_s^n  \ln^n (Q_T/Q_P)$ is referred to as the Leading Logarithmic Approximation (LLA). At the level of the BFKL characteristic function $\chi(\gamma)$, those correspond to the terms $\sim \alpha_s^n/(1-\gamma)^{n+1}$ in the limit $\gamma\rightarrow 1$.}
We {showed} in Ref.~\cite{Kovner:2026qfm} that the resummation of Ref.~\cite{Kovner:2023vsy}, when applied to the BFKL equation, reproduces many of the results of  previous approaches aimed at {improving the (anti)collinear behavior of} the BFKL evolution. We found that resummation of {anti-collinear} logarithms leads to {a slowdown} of the BFKL evolution. In particular, the {anti-collinear} pole in the characteristic function $\chi(\gamma)$ of the BFKL equation at $\gamma=1$ disappears, and the value of $\chi(\gamma)$ at $\gamma=1$ is finite and inversely proportional to $\alpha_s$.

In the calculations of Ref.~\cite{Kovner:2026qfm} we  considered the coupling constant $\alpha_s$ to be fixed. The aim of the present paper is to extend the approach of Ref.~\cite{Kovner:2026qfm} by including the running of the coupling. {Note that within the anti collinearly resummed JIMWLK evolution, the coupling constant enters in two distinct ways. First, it appears as the prefactor of the (resummed) JIMWLK Hamiltonian, or equivalently of the resummed BFKL kernel. Second, the anti-collinear logarithms are resummed through a DGLAP-like evolution equation, which involves its own coupling constant. While the running of the coupling in the BFKL kernel first contributes at NLO accuracy, the running of the coupling associated with the DGLAP resummation affects the evolution starting from NNLO. The two couplings run at different characteristic scales, and their impact on the final resummed kernel is qualitatively different. In the following, we distinguish between these two effects and discuss them separately.}

The paper has the following structure. We start in Sec.~\ref{sec:AC-res} by recapping the approach of Ref.~\cite{Kovner:2023vsy} to the resummation (Sec.~\ref{sec:resumm-general}). We also point out in Sec.~\ref{sec:muR-scale-setting} that {the scale choice} in the running coupling suggested in Ref.~\cite{Kovner:2023vsy}, although adequate for leading terms that contribute to the total cross section (as argued in \cite{Kovner:2023vsy}), is not the right one for all possible observables described by JIMWLK evolution. We explain what {the correct scale choice is}, and note an interesting interpretation of this scale in terms of the CSS evolution~\cite{CollinsQCD} in transverse resolution scale (Sec.~\ref{sec:scale-CSS}). Thus the choice of scale is intimately related {to} the  DGLAP cascade, which is the basis of our anti-collinear resummation.
In Sec.~\ref{sec:res-BFKL} we discuss the resummed BFKL kernel which includes the correct scale in the running coupling in the coordinate space.

In Sec.~\ref{sec:mom-BFKL} we derive the kernel in momentum space (Sec.~\ref{sec:K-BFKL-Fourier}), and solve the resummation equations for the gluon density and density of gluon pairs in the presence of a  running coupling in the resummation equations (Sec.~\ref{sec:evol-R2}). This is analogous to solving the DGLAP equation with a running coupling.

In Sec.~\ref{sec:chi-RC-main} we calculate the generalized characteristic function of the BFKL equation including the resummation and running coupling effects. We observe that the running coupling {generally} produces a stronger suppression of the characteristic function. However, as we show in Sec.~\ref{sec:RC-resumm}, for the value of the characteristic function at $\gamma=1$ there is a complete cancellation between the effect of {the} running  coupling  in the JIMWLK kernel and the effect of the running coupling in the DGLAP resummation. Thus the value of $\chi(\gamma=1)$ is unaffected by the running and we obtain the same value of this quantity as in Ref.~\cite{Kovner:2023vsy}. 

In Sec.~\ref{BFKLGF} we address the question of how the generalized characteristic function can be used to obtain the BFKL Green's function. In Sec.~\ref{sec:anti-coll-LLA-G} we suggest an approximate expression for the Green's function, which incorporates all the running-coupling and anti-collinear resummation effects at LLA, and analyze it using the saddle-point approximation in Sec.~\ref{sec:saddle}. We find that the effect of {the running coupling} on the Pomeron intercept in the anti-collinear regime is comparable to the effect of anti-collinear resummation: both lower the intercept significantly.

Finally in Sec.~\ref{sec:conclusions} we conclude with some remarks.
\\
%\new{The entire Intro mentions only 2 papers ...}

%% END OF INTRODUCTION %%%%%%%%%%%%%%%%%%%%%%%%%%%%%%%%%%%%%%%%%%%%%%%%%%%%%%%%%%%%%%%%%%%%

\section{Anti-collinear resummation in JIMWLK Hamiltonian with running coupling}\label{sec:AC-res}

\subsection{The resummation setup with running coupling}\label{sec:resumm-general}

The anti-collinearly resummed JIMWLK Hamiltonian  in the fixed-coupling approximation was derived in  Ref.~\cite{Kovner:2023vsy} and recapped in detail in  Ref.~\cite{Kovner:2026qfm}. Here we keep the general discussion brief and highlight only the details relevant for the inclusion of the running coupling effects.

The JIMWLK equation describes the evolution of the projectile operator ${\cal O}[S]$ composed of  (adjoint) Wilson lines $S^{ab}(\T{x})$  with respect to the ``projectile rapidity'' $Y\sim\ln P^+$ (this choice of the evolution variable is frequently referred to as the "+"-scheme):
\begin{equation}
    \frac{\partial}{\partial Y} {\cal O}_Y[S] = -\hat{H}_{\text{JIMWLK}}\, {\cal O}_Y[S]. \label{eq:JIMWLK} 
\end{equation}
The JIMWLK Hamiltonian with anti-collinear resummation~\cite{Kovner:2023vsy,Kovner:2026qfm} has the form:
\begin{eqnarray}
\hat{H}^{\text{(res.)}}_{\text{JIMWLK}}&=&  \int\limits_{\T{x},\T{y},\T{z}} \frac{\alpha_s(\mu^2_\star(\T{x},\T{y},\T{z}))}{2\pi^2} K(\T{x},\T{y},\T{z}) \bigg[ J_L^a(\T{x}) J_L^a(\T{y}) + J_R^a(\T{x}) J_R^a(\T{y}) \nonumber \\
&-& 2\Bbb{S}^{ab}(\T{z},Q_\star(\T{x},\T{y},\T{z})) J_L^a(\T{x}) J^b_R(\T{y})   \bigg], \label{eq:H-JIMWLK-res}
\end{eqnarray}
{where} $\Bbb{S}^{ab}(\T{x},Q)\equiv \Bbb{S}_Q^{ab}(\T{x})$ {is} the {\it dressed} Wilson line, which depends on the resolution scale %of the projectile 
$Q$. It is composed of the ``bare'' Wilson lines $S^{ab}(\T{x})$ {through} the solution of the corresponding evolution equations written below. The resolution scale of the anti-collinear resummation $Q_\star$ and that of the running coupling $\mu_\star$ are also defined below. The Weicz\"acker-Williams kernel in Eq.~(\ref{eq:H-JIMWLK-res}) is
\begin{eqnarray}
    && K(\T{x},\T{y},\T{z}) = \frac{\T{X}\cdot \T{Y}}{{X}^2 {Y}^2} = \frac{1}{2} \left( \frac{1}{{X}^2} + \frac{1}{{Y}^2} - K_D(\T{x},\T{y},\T{z})  \right), \\
    && K_D(\T{x},\T{y},\T{z})=\frac{(\T{x}-\T{y})^2}{(\T{x}-\T{z})^2 (\T{y}-\T{z})^2}, \label{eq:KD-LO-def}
\end{eqnarray}
with $\T{X}=\T{x}-\T{z}$, $\T{Y}=\T{y}-\T{z}$, $X^2=\T{X}^2=(\T{x}-\T{z})^2$ and $Y^2=\T{Y}^2=(\T{y}-\T{z})^2$.
Here $K_D({\bf x},{\bf y}, {\bf z})$, the so-called dipole kernel, is equal to the intensity of the gluon field of a color dipole with legs at ${\bf x}$ and ${\bf y}$.

 The left-right color rotation operators in Eq.~(\ref{eq:H-JIMWLK-res}) act on the ``bare'' eikonal gluon scattering matrix as \cite{Kovner:2005jc}:
\begin{eqnarray}
   && \hspace{-12mm}J_L^a(\T{y}) S^{bc}(\T{x}) = [T^a S (\T{x}) ]^{bc}  \delta^{(2)}(\T{x}-\T{y}), ~~~~ J_L^a(\T{y}) (S^\dagger(\T{x}))^{bc} = [- S^\dagger (\T{x}) T^a ]^{bc}  \delta^{(2)}(\T{x}-\T{y}), \label{eq:JL-def} \\
   && \hspace{-12mm} J_R^a(\T{y}) S^{bc}(\T{x}) = [S (\T{x}) T^a]^{bc}  \delta^{(2)}(\T{x}-\T{y}), ~~~~ J_R^a(\T{y}) (S^\dagger(\T{x}))^{bc} = [-T^a S^\dagger (\T{x})]^{bc}  \delta^{(2)}(\T{x}-\T{y}), \label{eq:JR-def} 
\end{eqnarray} 
where the $SU(N_c)$ generators in the adjoint representation are $T_{bc}^a=-if^{abc}$. 
%\old{Similar commutation relations hold between} 
 $J_{L(R)}$ act in a similar way on fundamental Wilson lines $V^{ij}(\T{x})$ which describe {the} eikonal propagation of quarks.
Another important property of the color rotation operators is
\begin{eqnarray}
    S^{ab}(\T{x}) J_{R}^{b}(\T{x}) = J^a_{L}(\T{x}),~~~~~
 S^{ab}(\T{x}) J_{L}^{a}(\T{x})   = J^b_{R}(\T{x}). \label{eq:bare-SJl-eqn}
\end{eqnarray}
%which in particular ensures the UV-finiteness of the Hamiltonian at $\T{z}\to \T{x}$ or $\T{z}\to\T{y}$. 
%\old{if one replaces $K\to -K_D/2$} in Eq. (\ref{eq:H-JIMWLK}).

 The resummation of {higher-order corrections in $\alpha_s$} to the JIMWLK Hamiltonian, introduced in  Ref.~\cite{Kovner:2023vsy} is relevant in the regime where the target, which is probed by the operator ${\cal O}_Y[S]$ has an intrinsic momentum scale $Q_T$ {that} is much greater than the scale $Q_P$, which characterizes the operator ${\cal O}_Y[S]$ itself. In the saturation regime $Q_T$ (and $Q_P$) has the meaning of saturation momentum, while in the weak-field perturbative regime it is the inverse correlation length of the color fields in the target (projectile).
 
 The next-to-leading order (NLO) correction to the JIMWLK Hamiltonian in QCD {was} first derived  in Refs.~\cite{Kovner:2013ona,Kovner:2014lca,Kovner:2014xia} using the results of Refs.~\cite{Balitsky:2007feb,Grabovsky:2013mba} and then re-derived independently {within} Light-Cone Perturbation Theory (LCPT) in Ref.~\cite{Lublinsky:2016meo}. By studying the NLO JIMWLK Hamiltonian it was shown in Ref.~\cite{Kovner:2023vsy} that the large logarithmic corrections  enhanced in the {anti-collinear} limit $Q_T\gg Q_P$ can be resummed by introducing the dressed Wilson lines $\Bbb{S}_Q^{ab}(\T{x})$ which evolve with the resolution scale $Q$ according to the following evolution equation:
\begin{eqnarray}
    && \frac{\partial}{\partial \ln Q^2} \Bbb{S}^{ab}(\T{x},Q) = -a_s(Q^2)\beta_0 \Bbb{S}^{ab}(\T{x},Q) \nonumber \\
    && - a_s(Q^2) \int\limits_0^1 d\xi\int\limits_0^{2\pi} \frac{d\phi}{2\pi} \bigg[  2N_c p_{gg}(\xi) \Bbb{D}^{ab}\left( \T{x} + (1-\xi) Q^{-1} \T{n}_\phi, \T{x} - \xi Q^{-1} \T{n}_\phi, Q \right) \nonumber \\
    && + 2T_F n_F p_{qg}(\xi) \Bbb{D}^{(F),ab}\left( \T{x} + (1-\xi) Q^{-1} \T{n}_\phi, \T{x} - \xi Q^{-1} \T{n}_\phi, Q \right) \bigg],\label{eq:eqn-SQ-quarks}
\end{eqnarray}
where we emphasize that the factors of the strong coupling $a_s$ {depend on} the evolution scale $Q^2$ as dictated by the leading logarithmic approximation (LLA) of DGLAP evolution. 
Equation \eqref{eq:eqn-SQ-quarks} is written for {a} theory with $n_F$ massless quark flavours. Here $a_s(Q^2)\equiv\alpha_s(Q^2)/(4\pi)$, $\beta_0=\beta_0^{(g)}+\beta_0^{(q)}$ with $\beta_0^{(g)}=11C_A/3$, $C_A=N_c$, $\beta_0^{(q)}=-4n_FT_F/3$, {and} $T_F=1/2$. 
At one loop,
\begin{eqnarray}
    \frac{\partial a_s(Q^2)}{\partial \ln Q^2} = -\beta_0 a_s^2(Q^2)+O(a_s^3)\quad 
  \Longrightarrow \quad a_s(Q^2)=\frac{a_s(\Lambda^2)}{1+\beta_0 a_s(\Lambda^2) \ln(Q^2/\Lambda^2)}.  \label{eq:as-Q2}
\end{eqnarray}
In eqn.~\eqref{eq:eqn-SQ-quarks},
  $\T{n}_\phi$ is the radial unit vector relative to {the} point $\T{x}$ in the transverse plane at azimuthal angle $\phi$.
  The  functions appearing in 
 {Eq.} (\ref{eq:eqn-SQ-quarks}) are:
\begin{eqnarray}
        && \Bbb{D}^{ab}(\T{x},\T{y}, Q)= \frac{1}{N_c} \tr\left[ T^a \Bbb{S}(\T{x},Q) T^b \Bbb{S}^\dagger (\T{y},Q) \right], \\
        &&  \Bbb{D}^{(F),ab}(\T{x},\T{y}, Q)=2\tr\left[ t^a \Bbb{V}(\T{x},Q) t^b \Bbb{V}^\dagger (\T{y},Q) \right], \\
        && p_{gg}(z)=\frac{g(z)}{z_+ (1-z)_+}, \\
        && g(z)= z^2+(1-z)^2+z^2(1-z)^2, \\
        && p_{qg}(z)=z^2+(1-z)^2,
\end{eqnarray}
where $T^a_{bc}$ and $t^a_{ij}$ are {the} $SU(N_c)$-generators in the adjoint and fundamental representations respectively. %The functions $p_{gg}(z)$ and $p_{qg}(z)$ are the usual DGLAP splitting functions $P_{gg}(z)$ and $P_{qg}(z)$ respectively. The $P_{gg}(z)$ is taken without the $\delta(1-z)$-term and with the high-energy $1/z$-pole subtracted. This subtraction avoids double-counting with the JIMWLK evolution.  
The two-sided plus distribution in $p_{gg}(z)$ is defined as: 
\begin{equation}
\int\limits_0^1 \frac{f(z) dz}{z_+(1-z)_+} = \int\limits_0^1 \bigg[\frac{f(z)}{z(1-z)} - \frac{f(0)}{z}-\frac{f(1)}{1-z}\bigg].
\end{equation}

 Equation~\eqref{eq:eqn-SQ-quarks} is accompanied by the evolution equation for the dressed fundamental Wilson line ($\Bbb{V}_Q^{ij}$)
\begin{eqnarray}
    && \frac{\partial}{\partial \ln Q^2} \Bbb{V}_{ij}(\T{x},Q) = -3a_s(Q^2) C_F   \Bbb{V}_{ij}(\T{x},Q) \nonumber \\
    && - 2a_s(Q^2) \int\limits_0^1 d\xi\int\limits_0^{2\pi} \frac{d\phi}{2\pi} p_{gq}(\xi) \Bbb{D}^{(FA)}_{ij}\left( \T{x} + (1-\xi) Q^{-1} \T{n}_\phi, \T{x} - \xi Q^{-1} \T{n}_\phi, Q \right) , \label{eq:eqn-VQ}
\end{eqnarray}
where $C_F=(N_c^2-1)/(2N_c)$ and
\begin{eqnarray}
        &&  \Bbb{D}^{(FA)}_{ij}(\T{x},\T{y}, Q)=\big[ t^a  \Bbb{V} (\T{y},Q) t^b \big]_{ij} \Bbb{S}^{ab}(\T{x},Q) , \\
        && p_{gq}(z)=\frac{1+(1-z)^2}{z_+}.
\end{eqnarray}
 Here  $p_{gq}(z)$ is the DGLAP splitting function $P_{gq}(z)$,  with its leading $1/z$-pole regularized by the standard plus-prescription. 

As was extensively discussed in Refs.~\cite{Kovner:2023vsy} and \cite{Kovner:2026qfm} the initial conditions for {Eqs.} (\ref{eq:eqn-SQ-quarks}) and (\ref{eq:eqn-VQ}) can be set at any scale $\Lambda>Q_T$ such that $$\Bbb{S}^{ab}_{\Lambda}(\T{x}) = S^{ab}(\T{x}) ~~~~ {\rm and} ~~~~  \Bbb{V}_{ij,\Lambda}(\T{x}) = V_{ij}(\T{x})\,.$$
Once the solution is found, one can  safely take the limit $\Lambda\to \infty$, as  will be done below.

{We  now} need to discuss the choice of scales $Q_\star$ and $\mu_\star$ in the Hamiltonian~(\ref{eq:H-JIMWLK-res}). The choice of the scale  $Q$ is dictated by the observation~\cite{Kovner:2023vsy} that the anti-collinear logarithmic region in the NLO JIMWLK Hamiltonian is cut-off by the position of the source $J_{L,R}$ nearest to the dressed Wilson line $\Bbb{S}_Q^{ab}(\T{z})$. The natural choice is therefore:
\begin{equation}
    Q^2_\star(\T{x},\T{y},\T{z})=\max(X^{-2},Y^{-2}). \label{eq:scale choice Q}
\end{equation}
 This scale choice is convenient for the analytic studies and we will  use it {extensively} below. However, in Ref.~\cite{Kovner:2026qfm} it was shown that the non-smooth behaviour of the function (\ref{eq:scale choice Q}) at the point $X=Y$ leads to unphysical artifacts, such as the appearance of the {spurious} pole {in} the BFKL characteristic function at $\gamma=1/2$. To avoid these artifacts it is sufficient to {promote} the scale choice (\ref{eq:scale choice Q}) {into} a smooth function e.g. as follows: 
\begin{equation}
    Q^2_{\star\lambda}(\T{x},\T{y},\T{z})=Y^{-2}\Theta_{\lambda}(X^{2},Y^{2}) + X^{-2}\Theta_{\lambda}(Y^{2},X^{2}), \label{eq:scale choice-smooth}
\end{equation}
where the function $\Theta_\lambda(x,y)$ is a smooth version of $\theta(x>y)$. It can be chosen for example as:
\begin{equation}
\Theta_\lambda(x,y)=\frac{1}{2}+\frac{1}{\pi}\arctan\big(\lambda \ln(x/y) \big),    
\end{equation}
where $\lambda$ is a smearing parameter and for $\lambda\to\infty$ the function $\Theta_\lambda(x,y)\to \theta(x>y)$.

\subsection{Setting the scale of the running coupling}\label{sec:muR-scale-setting}
We now  determine the scale of the running coupling $\mu_\star$ in (\ref{eq:H-JIMWLK-res}). It was argued in Ref.~\cite{Kovner:2023vsy} that the {appropriate choice} is
\begin{equation}
    \label{sub}\alpha_s(\mu^2_\star)\rightarrow \frac{g(X^{-2})g(Y^{-2})}{4\pi}.
\end{equation}
The {physical motivation} for this choice is that the gluon field at point $\T{z}$ is created in the amplitude by the charge at $\T{x}$ while in conjugate amplitude by the charge at $\T{y}$ and the scale of the coupling $g$ should correspond to the distance to the corresponding emitter.
We would now like to revisit this point. 
 The leading contribution in the anti-collinear regime comes from the regime of $X^2\simeq Y^2$, for which the argument above holds indeed. Yet, subleading terms arise from configurations when $X^2\gg Y^2$ (and vice versa), which get affected by large logarithms $\ln(X^2/Y^2)$. The substitution (\ref{sub}) fails for these subleading terms. This can be rectified by choosing the scale which eliminates all large {transverse} logarithms {in} the resummed JIMWLK Hamiltonian away from the strict anti-collinear regime as well.

First let us {briefly review} the argument {of} Ref.~\cite{Kovner:2023vsy}.
To resum the anti-collinear logarithms one introduces the resolution scale of the DGLAP cascade $Q^2$, and {splits} the LO+NLO Hamiltonian ~\cite{Kovner:2013ona,Kovner:2014lca,Kovner:2014xia}
into two pieces, 

\begin{equation}
    \hat{H}_{JIMWLK}^{NLO}=\hat{H}_Q^{JSJ}
    +\hat{H}_Q^{JSSJ}+\hat{H}_Q^{JVVJ} \,+\,\text{ ``JJJ'' terms}
\end{equation}
In the above, $\hat{H}_Q^{JVVJ}$ is the contribution from the quarks and
 ``JJJ''-terms represent all the non-linear in the projectile density terms. Those are irrelevant for the problem at hand  and are discarded from further discussion. Substituting 
$\mathbb{S}_Q$ for $S$, the two first parts of the Hamiltonian read (discarding all non-logarithmic terms):
\begin{eqnarray}\label{jsjq}
\hat{H}^{J\mathbb{S}J}_{ Q}
%\int_{\mathbf{x,y,z}}\frac{\alpha_{s}X\cdot Y}{2\pi^{2}X^{2}Y^{2}}\,\left(1+\frac{\alpha_s\beta_0}{4\pi}\,\left[\,\ln[X^{2}\mu^{2}]\,+\,\ln[Y^{2}\mu^2]\,-\ln\frac{\mu^2}{ Q^2}\right]\right)\\
%&\times&\left[\, J_{L}^{a}(\mathbf{x})\, J_{L}^{a}(\mathbf{y})\,+\, J_{R}^{a}(\mathbf{x})\, J_{R}^{a}(\mathbf{y})\,-\,2J_{L}^{a}(\mathbf{x})\, \mathbb{S}_{\bar Q}^{ab}(\mathbf{z})\, J_{R}^{b}(\mathbf{y})\,\right].\nonumber\\
&=&\int_{\mathbf{x,y,z}}\frac{\alpha_{s}(\mu^2)}{2\pi^{2}}\,
%\old{\frac{{\bf X\cdot Y}}{X^{2}Y^{2}}\,} 
{K(\T{x},\T{y},\T{z})}\,
\Bigg\{\left(1+\frac{\alpha_s(\mu^2)\beta_0}{8\pi}\left[\ln(X^{2}\mu^{2})+\,\ln(Y^{2}\mu^2)\right]\right)
\notag\\
&\times&\left[J_{L}^{a}(\mathbf{x}) J_{L}^{a}(\mathbf{y})+ J_{R}^{a}(\mathbf{x})\, J_{R}^{a}(\mathbf{y})-2J_{L}^{a}(\mathbf{x}) \mathbb{S}_{ Q}^{ab}(\mathbf{z}) J_{R}^{b}(\mathbf{y})\right]\\
&&\,+\,\frac{\alpha_s(\mu^2)\beta_0}{8\pi}\Big[(\ln (X^2 Q^2)+\ln (Y^2 Q^2))\, \left[J_{L}^{a}(\mathbf{x})\, J_{L}^{a}(\mathbf{y})\,+\, J_{R}^{a}(\mathbf{x})\, J_{R}^{a}(\mathbf{y})\right]\nonumber\\
&&\hspace{5cm}-2%\ln X^2\tilde Q^2+\ln Y^2 Q^2\right]
%\left[\, J_{L}^{a}(\mathbf{x})\, J_{L}^{a}(\mathbf{y})\,+\, J_{R}^{a}(\mathbf{x})\, J_{R}^{a}(\mathbf{y})\right]\,\nonumber\\
(\ln (X^2 Q^2)+\ln (Y^2 Q^2))J_{L}^{a}(\mathbf{x})\, \mathbb{S}_{Q}^{ab}(\mathbf{z})\, J_{R}^{b}(\mathbf{y})\,\Big]\Bigg\}\,.\nonumber
%J_{L}^{a}(\mathbf{x})\, \mathbb{S}_{ Q}^{ab}(\mathbf{z})\, J_{R}^{b}(\mathbf{y})\,\Bigg\}. 
\end{eqnarray}
and\footnote{A similar expression can be written for $\hat{H}_Q^{J\mathbb{VV}J}$. }
\begin{eqnarray}\label{jssjq}
 \hat{H}^{J\mathbb{SS}J}_Q &=&\int_{\mathbf{x},\, \mathbf{y},\mathbf{z}, \mathbf{z}^{\prime}}\,\hspace{-0.3cm} K_{JSSJ}(\mathbf{x},\mathbf{y},\mathbf{z}, \mathbf{z}^{\prime})J_{L}^{a}(\mathbf{x})\mathbb{D}_{ Q}^{ad}(\mathbf{z},\mathbf{z}^{\prime}) J_{R}^{d}(\mathbf{y}) \nonumber \\
  & -&\frac{\alpha^2_{s}(\mu^2)\beta^{(g)}_0}{4\pi^{3}}\int_{\mathbf{x},\, \mathbf{y},\mathbf{z}}\frac {X\cdot Y}{X^{2}Y^{2}}
 \mathbb{D}^{ab}_{ Q}(\mathbf {z})J_L^a(\mathbf{x})J_R^b(\mathbf{y}) -\bigg[\frac{N_c}{2}\int_{\mathbf{x},\, \mathbf{y},\mathbf{z}, \mathbf{z}^{\prime}}\hspace{-0.3cm} K_{JSSJ}(\mathbf{x},\mathbf{y},\mathbf{z}, \mathbf{z}^{\prime}) \nonumber \\
 & -&\frac{\alpha^2_{s}(\mu^2)\beta^{(g)}_0}{8\pi^{3}}\int_{\mathbf{x},\, \mathbf{y},\mathbf{z}} \frac{X\cdot Y}{X^{2}Y^{2}}\,\ln\frac{\mu^2}{Q^2}\bigg]\left[J_R^a(\mathbf{x})J_R^a(\mathbf{y})+J_L^a(\mathbf{x})J_L^a(\mathbf{y})\right]\,. 
\end{eqnarray}
where   $\mu$ is the UV cutoff scale.

The kernel $K_{JSSJ}$ %(and also $K_{JSJ}$) 
was introduced in~\cite{Kovner:2013ona,Kovner:2014lca,Kovner:2014xia} and is not quoted explicitly here. The important point is that the choice ($Q=Q_{\star}$) (see \eqref{eq:scale choice Q}) 
renders $\hat{H}^{J\mathbb{SS}J}_Q$ free of  large logarithms in the anti-collinear limit.

The $\hat{H}^{J\mathbb{S}J}_{ Q}$ part of the Hamiltonian has the same form as the leading order JIMWLK Hamiltonian with the eikonal Wilson line $S$ substituted by the resummed gluon amplitude $\mathbb{S}_Q$, and the kernel that contains the  $O(\alpha_s)$ correction 
\begin{equation}\begin{split}\label{kjsj}
 \frac{\alpha_{s}(\mu^2)}{2\pi^{2}} K\rightarrow  &\frac{\alpha_{s}(\mu^2)}{2\pi^{2}}\frac{X\cdot Y}{X^{2}Y^{2}}\,\left(1+\frac{\alpha_{s}(\mu^2)}{2\pi}\beta_0\,\left[\,\frac{1}{2}\ln[X^{2}\mu^{2}]\,+\,\frac{1}{2}\ln[Y^{2}\mu^2]\,-\frac{1}{2}\ln\frac{\mu^2}{Q^2}\right]\right).
 \end{split}\end{equation}
This expression clearly contains all the UV logarithms that have to be combined into the running coupling. 

With the choice of the resolution scale $Q=Q_\star$, the $Q_\star$ is the largest scale under the log (except for $\mu$). It is thus clear that the coupling should run from $\mu$ down to at least $Q_\star$.  Performing this (partial) resummation of UV logs we then have
\begin{equation}\begin{split}\label{kjsj1}
  \frac{\alpha_{s}(\mu^2)}{2\pi^{2}} K\rightarrow   &\frac{\alpha_{s}(Q_\star^2)}{2\pi^{2}}\frac{X\cdot Y}{X^{2}Y^{2}}\,
  \left(1+\frac{\alpha_{s}(Q_\star^2)}{2\pi}\beta_0\,\left[\,\frac{1}{2}\ln[X^{2}Q_\star^{2}]\,+\,\frac{1}{2}\ln[Y^{2}Q_\star^2]\right]\right).
 \end{split}\end{equation}

For $X\sim Y$ there are no large logarithms in this expression. In fact, {in anti-collinear regime} the leading contribution to the cross section  arises from emission of a gluon close to one of the sources in the wave function. {In this case,} {the emitting source}  in the amplitude and the conjugate amplitude {must either coincide or remain close to each other}  since otherwise the contribution is suppressed by a power of the ratio of the corresponding distances. This { situation} corresponds to $X\sim Y$, {for which} $\alpha(Q_\star)=\frac{g(X^{-2})g(Y^{-2})}{4\pi}$ 
with logarithmic accuracy.

One can fine tune the scale $\mu_\star$ so as to properly account for configurations where $X$ and $Y$ are parametrically  different, when the logarithms remaining in \eqref{kjsj1} {become} large and must be resummed. With our choice of $Q_\star$, the remaining logarithm corresponds precisely to the additional running of the coupling  between $Q_\star$ (which is the inverse of the smaller {of the two scales}  $X^2$ and $Y^2$) {and} the inverse of the larger between the two. Thus all  large {logarithms} in $H_{JIMWLK}$ are eliminated by running the coupling  from the UV scale down to the smallest momentum scale defined by the geometry of the emission:
\begin{equation}\begin{split}\label{kjsj11}
 %K_{JSJ}\rightarrow  %&\alpha_{s}\left(\min(X^{-2},Y^{-2})\right) \frac{{\bf %X\cdot Y}}{2\pi^{2}X^{2}Y^{2}}
 \alpha_{s}(\mu^2)K\rightarrow \alpha_{s}(\mu_\star^2)\,K\,,
 \qquad \mu_\star^2=\min(X^{-2},Y^{-2}).
 \end{split}\end{equation}

%{To illustrate this point }
Consider, for example the action of
this Hamiltonian on a %dressed 
dipole with legs at $\T{u}$ and $\T{v}$. 
%$\{\T{u},\T{v}\}=\tr\big[\Bbb{S}(\T{u}) \Bbb{S}^\dagger(\T{v}) \big]$, 
%which we will discuss in detail in the following sections. 
This action generates four terms. Two of {these} are ``diagonal terms'' which arise from the gluon emitted  off the same dipole leg in the amplitude and the conjugate amplitude,  $X=Y=U$ and $X=Y=V$. {The remaining two are} the interference terms where the gluon is emitted from different dipole legs {in the amplitude and conjugate amplitude}, {namely} $X=U,\ Y=V$ and $X=V,\ Y=U$. In the diagonal terms the scale of the coupling constant in \eqref{kjsj11} is {fixed} by the only scale present, and {is therefore} the same as proposed in Ref.~\cite{Kovner:2023vsy}. When $U$ is {parametrically} different from $V$ (anti-collinear regime), the total cross section is dominated by one of these terms since the other contributions are power suppressed. {Nevertheless,} the interference terms may {still} be important for {certain} observables, particularly the ones related to angular correlations\footnote{In terms of the BFKL characteristic function $\chi(n,\gamma)$ these correspond to $n\ne0$.}. 
In these, interference {contributions} the coupling constant is determined by the smaller of the two momentum scales (or { equivalently by the} larger of the the two distance scales). {At first sight,} {it may seem surprising that the coupling is controlled by} the largest distance scale. {Understanding the physical origin of this scale setting warrants a dedicated discussion, which we present in the next subsection.}

\subsubsection{Running coupling and the CSS evolution}\label{sec:scale-CSS}
To better understand the argument,  consider a slightly reorganized {version of \eqref{kjsj}},
\begin{equation}\begin{split}\label{kjsj2}
  &\frac{\alpha_{s}(\mu^2)}{2\pi^{2}}\frac{\bf X\cdot Y}{X^{2}Y^{2}}\,\left(1+\frac{\alpha_{s}(\mu^2)}{2\pi}\beta_0\,\bigg\{\,\frac{1}{2}\Big[\ln[X^{2}\mu^{2}]-\frac{1}{2}\ln\frac{\mu^2}{Q^2}\Big]\,+\,\frac{1}{2}\Big[\ln[Y^{2}\mu^2]\,-\frac{1}{2}\ln\frac{\mu^2}{Q^2}\Big]\bigg\}\right).
 \end{split}\end{equation}
If {one} could choose $Q^2=X^{-2}$ in the first square bracket, and $Q^2=Y^{-2}$ in the second, {the resulting} expression would give $g(X^{-2})g(Y^{-2})$, {corresponding to the prescription} (\ref{sub})  proposed in Ref.~\cite{Kovner:2023vsy}. As already mentioned, {this prescription correctly reproduces the} leading terms $X=Y$, but {fails for the} subleading ones where $X^2\gg Y^2$ (or $X^2\ll Y^2$). {The subleading terms we are concerned with are responsible for the angular dependence of the kernel are important for the coefficient of the left-most anti-collinear pole of the BFKL characteristic function at $n>0$, which is located at the point $\gamma=1+n/2$. } 

{For definiteness, consider the regime} $X^2\ll Y^2$. Recall that in this regime all large {logarithms} in $K_{JSSJ}$ are eliminated by the choice $Q^2=Q_\star^2=X^{-2}$. This is because the gluon DGLAP splittings in $K_{JSSJ}$ are regulated by the {distance to the nearest} source, and therefore the transverse resolution {scale} of the DGLAP dressed gluon is {set by} this distance. With this {choice of the} resolution scale, the first square bracket in (\ref{kjsj2}) indeed renormalizes one factor of the coupling constant to $g(X^{-2})$, however the second factor is {not} equal to $g(Y^{-2})$.

Recall that the NLO corrections in \eqref{kjsj2} originate both from the real and virtual diagrams (see e.g.~\cite{Lublinsky:2016meo}).
{For the sake of the argument},   consider {an} arbitrary {scale} $Q$,  not necessarily {equal to} $Q_\star$. First, note that  the two terms in the first square bracket in \eqref{kjsj2} {have distinct physical origin}. The term $\ln X^2\mu^2$ originates from the virtual diagram. This diagram can be {interpreted} as the correction to the classical field ${X_i}/{X^2}$ at {transverse} distance $X$ from the charged source. In this term the transverse momentum  in the {gluon} loop is cut off by $X^{-2}$ in IR, i.e. {by} the (inverse) distance at which the gluon is emitted from the source. On the other hand the second term, $\ln({\mu^2}/{Q^2})$ originates from the real diagram, {where} $Q^2$ {acts as} the IR cutoff on {the transverse} momentum of gluons in the final state. Physically this means that the dressed gluons in the real diagram are {probed} with transverse resolution $Q^2$. 

Now recall, that in QCD the running coupling $\alpha_s(\mu^2)$ is defined as the strength of the interaction between two sources at  a distance $\sim 1/\mu$, resolved with %due to {the} exchange of a dressed gluon defined with {the} 
transverse resolution $\mu$.

Thus in the standard definition of the running coupling, the same scale $\mu$ determines both, the transverse resolution of the {exchanged} dressed gluon, and the distance at which the gluon is emitted from the source. In other words, the momentum of the exchanged dressed gluon is of the  order of its transverse resolution. In our case, however, the resolution is {set by} $Q$, while the {emission} distance is $X$ in the amplitude and $Y$ in the conjugate amplitude. Choosing $Q^2=X^{-2}$, as we did above, indeed turns the renormalized coupling constant in the amplitude into $g(X^{-2})$. In the conjugate amplitude, however, the distance of the gluon from the emitter, $ Y$ is much greater than the inverse of the transverse resolution $Q^{-1}$. 

{This} situation is {strongly} reminiscent of the gluon TMD {evaluated} at the transverse momentum $\T{k}^2\simeq Y^{-2}$ defined  with high transverse resolution scale $\bar \mu^2=X^{-2}\gg \T{k}^2$. In fact the similarity here is more than just superficial. Indeed we are {effectively} calculating the magnitude of the gluon field, which {may be viewed, for all practical purposes, as a}  square root of the gluon TMD in the projectile.

To be more precise, consider the gluon TMD of a single color source at point $\T{y}$. At leading order in $\alpha_s$ in the eikonal approximation it is given by 

\begin{equation}\label{tmd}
    T_g(x, \T{k}^2) = \langle a^\dagger(x,\T{k}) a(x,\T{k})\rangle_{\T{y}} \sim 
    \int\limits_{\T{z},\T{z}'} e^{i\T{k}(\T{z}-\T{z}')} \alpha_s\, \frac{(\T{z}-\T{y})\cdot(\T{z}'-\T{y})}{(\T{z}-\T{y})^2(\T{z}'-\T{y})^2},
\end{equation}
where the dependence on the longitudinal momentum fraction $x$ is irrelevant for our discussion.
%\begin{equation}
 %   T_g(x, k^2; \bar\mu^2) = \langle a^\dagger(x,k) a(x,k)\rangle^P_{\bar\mu} \sim 
  %  \int_{z} e^{ikz}\int_{x,y} \bar\alpha_s(X^{-2},Y^{-2},\bar\mu^2) \frac{X Y}{X^2Y^2}
   %J_L(x)J_R(y)
%\end{equation}
%The last relation is the consequence of the eikonal nature of emission of soft gluons from the projectile charges, which in the JIMWLK representation are encoded by $J_{L,R}$. 
At leading order the TMD of course formally does not depend on the transverse resolution scale.
% Although we are not aware of fully consistent way to introduce transverse resolution in JIMWLK calculations,
However, since (at this order) the gluon emitted from the source itself is not further resolved (as it does not undergo any further splittings), it is clear that the resolution scale in the above relation should be identified  with the transverse momentum $\T{k}$ (see discussion in Refs.~\cite{Duan:2024nlr,Duan:2024qev}). Furthermore, for the gluon field in \eqref{tmd}, the momentum and the distance from the source are related by $|\T{k}|\approx 1/|\T{y}-\T{z}|\approx 1/|\T{y}-\T{z'}|$, so
we can write
\begin{equation}\label{tmd1}
    T_g(x, \T{k}^2; \bar\mu^2=\T{k}^2) \sim 
    \int\limits_{\T{z}, \T{z}'} e^{i\T{k}(\T{z}-\T{z}')} \alpha_s (Y^{-2})\frac{(\T{z}-\T{y})\cdot(\T{z}'-\T{y})}{(\T{z}-\T{y})^2(\T{z}'-\T{y})^2}.
\end{equation}

%Thus, the CSS equation can be regarded as an evolution in $\bar\mu^2$ of the effective coupling $\bar\alpha_s$ with the initial condition $\bar\alpha_s(k^2,k^2)=\alpha_s(k^2)$.

The dependence of TMD on the resolution scale $\bar{\mu}^2$ should be given by the solution of the CSS equation \cite{CollinsQCD}. It therefore follows that we should be able to use the CSS equation to calculate the dependence of the field intensity on the resolution scale $\bar\mu^2$. 
According to this logic, in order to calculate the intensity of the gluon field at distance $Y$ from the source with transverse resolution $X^{-2}$ we should evolve the TMD of a single source at $|\T{k}|\sim 1/Y$ from the initial transverse resolution  $\bar \mu_0^2=Y^{-2}$ 
up to the final high resolution $\bar\mu^2=X^{-2}$. 

At the initial {scale,} our amplitude ({the} square root of the TMD) is simply {given by} $g(Y^{-2})$ multiplied by the Weisz\"acker-Williams field as per Eq. \eqref{tmd1}\footnote{Assuming, as above that with logarithmic accuracy $k\approx 1/|\T{y}-\T{z}|\approx 1/|\T{y}-\T{z'}|$}. It then follows that in order to {reproduce} the correct scale {dependence} in \eqref{kjsj11}, the field at the end point of the evolution must evolve into $g^2(Y^{-2})/g(X^{-2})$. This is {rather puzzling}, {since it bears little resemblance to the expected solution of}  the CSS equation with initial condition \eqref{tmd1}. First, it does not exhibit any doubly logarithmic dependence on the resolution scale which {normally emerges} from  CSS evolution. Second, CSS evolution {with increasing} transverse resolution scale {generally suppresses}  TMD, as the {dominant effect}  is disappearance of the  {original parton through} DGLAP splittings {in}to parton pairs {carrying large} transverse momentum. {In contrast,}  in our case $g(Y^{-2})/g(X^{-2})>1$,
{which implies that the evolution must result in}
 an increase of the field amplitude from $g(Y^{-2})$ to $g^2(Y^{-2})/g(X^{-2})$.

{There is, however, an interesting twist to this argument which removes the  contradiction.} Recall that the  DGLAP cascade discussed in Refs.~\cite{Kovner:2023vsy, Kovner:2026qfm} does not involve the full {QCD} splitting function, but only the difference between the {full} splitting function and its small $x$ asymptotics. The reason is that the  small $x$ asymptotics {is already} included  in the JIMWLK evolution at leading order, and should therefore not be included in any further resummations in the JIMWLK Hamiltonian. It is {therefore} natural to expect that the same {logic} should apply to the CSS evolution discussed above, since (just like DGLAP resummation) it should only contain contributions not already included in the leading order JIMWLK. In other words, the {doubly} logarithmic term should be subtracted from the {CSS} kernel, and only the single logarithmic piece should be retained.

This indeed resolves the  puzzle {posed above}, since the single logarithmic contribution to the CSS kernel has  the opposite sign {compared} with the double logarithm, and thus leads to an {enhancement}, {rather than} a {suppression} of the TMD.

Indeed, consider the CSS evolution in the transverse resolution scale but with only the single logarithmic part in the kernel. For $\bar\mu^2\gg \T{k}^2$ it reads\footnote{The equation strictly speaking is local in coordinate space, but for very large resolution scale $\bar\mu^2$ the locality in momentum space holds as well, see for example Refs.~\cite{Duan:2024nlr,Duan:2024qev}.}
\begin{equation}
\frac{\partial}{\partial\ln {\bar\mu^2}}\ln {T_g(x, \T{k}^2;\bar\mu^2)}=\alpha_s(\bar\mu^2)
\frac{\beta_0}{4\pi}\,.
\end{equation}
{The solution is}
\begin{equation}\label{csssol}
    T_g(x,\T{k}^2;\bar\mu^2)=T_g(x,\T{k}^2;\T{k}^2)\exp\bigg[{\frac{\beta_0}{4\pi}\int\limits_{\T{k}^2}^{\bar\mu^2}\alpha_s(\bar\mu^{\prime2})\,d\ln{\bar\mu^{\prime2}}} \bigg].
\end{equation}
It is straightforward to {express} the exponential factor in \eqref{csssol} {in terms of} the running coupling.
Changing the integration variable in the exponent from $\ln \bar\mu^2$ to $\alpha_s(\bar\mu^2)$ using
\begin{equation}
    d\ln\bar\mu^{\prime2}=-\frac{4\pi}{\beta_0}\frac{d\alpha_s}{\alpha^2_s(\bar\mu^{\prime2})},
\end{equation}
the integral can be explicitly {evaluated, yielding} 
\begin{equation}\label{csssol2}
    T_g(x,\T{k}^2;\bar\mu^2)=T_g(x,\T{k}^2;\T{k}^2)\frac{\alpha_s(\T{k}^2)}{\alpha_s(\bar\mu^2)}\,.
    %e^{\frac{\beta_0}{4\pi}\int_{k^2}^{Q^2}\alpha_s(\mu^2)d\ln{\mu^2}}
\end{equation}
We {now} identify $\T{k}^2$ in \eqref{csssol2} with $Y^{-2}$ and $T_g(x,\T{k}^2;\T{k}^2)$
with ${\alpha_s(Y^{-2})}/{Y^2}$, {while  choosing $\bar\mu^2 = X^{-2}$}. {Taking the square root of both sides of the equation, we indeed find that the evolved  magnitude of the gluon field is proportional to the combination $g^2(Y^{-2})/g(X^{-2})$.} {Thus, the} additional scale factor on the RHS of \eqref{csssol2} {indeed} leads to desired evolution  
\begin{equation}
    g(Y^{-2})\rightarrow \frac{g^2(Y^{-2})}{g(X^{-2})}, 
    %\frac{\alpha_s(Y^{-2})}{g(X^{-2})}
\end{equation}
{and consequently} {produces} the factor $\alpha_s(Y^{-2})$ {multiplying} the WW kernel in the Hamiltonian, consistently with \eqref{kjsj11}.

We stress again the peculiarity of our CSS-like equation \eqref{csssol}, \eqref{csssol2} relative to the standard CSS equation. Due to the absence of the doubly logarithmic part of the kernel, it leads to growth, rather than {suppression} of TMD towards higher transverse resolution scale.
%In the standard CSS the Sudakov form factor suppresses the number of gluons. The physics of this suppression is that gluons are allowed to split into high transverse momentum pairs as the transverse resolution $\mu$ is increased. The Sudakov form factor there is dominated by the doubly logarithmic term. In \eqref{csssol} to the contrary, the analog of the Sudakov form factor is entirely due to the single logarithm, it has the opposite sign to the doubly logarithmic term and is positive. As a result the TMD grows with the increase of $\mu^2$. 

This is entirely consistent with the discussion of the DGLAP resummation within JIMWLK in \cite{Kovner:2023vsy}. %The resummation takes care of the remainder of the splitting function in addition to its low x asymptotics, which is taken into account in JIMWLK already in the leading order. This remainder is negative, since the leading order JIMWLK overestimates the total probability of gluon splitting. 
 If interpreted in terms of gluon cascade, the subtraction of the {small} $x$ asymptotics from the splitting function results in a DGLAP-like cascade with {\it negative} splitting probability. {It is precisely this effective}  ``negative probability'' of splitting that lead to the growth of {the} gluon TMD with increase of the resolution scale $\mu^2$ in the above discussion of CSS-like evolution. This {remarkable} inversion results in the scale of the running coupling in the asymmetric kinematics, $X\gg Y$, or $X\ll Y$ being set by the largest {transverse} distance scale.

Thus interestingly, the physics of the scale setting {of} the coupling  is closely intertwined {with}, and in a sense determined by, the resummation of DGLAP corrections \cite{Kovner:2023vsy}.

To summarize this discussion, we stipulate that the appropriate choice of the scale of the coupling constant in the DGLAP-resummed JIMWLK Hamiltonian is
\begin{equation}
    \mu_\star^2(\T{x},\T{y},\T{z})=\min(X^{-2},Y^{-2}). \label{eq: scale choice mu}
\end{equation}
We note that this choice is practically {equivalent to}  the one {proposed} in ~\cite{Balitsky:2007feb}.

We will use this scale choice in the analytic studies in this paper. {At the same time, we keep in mind that, in practical applications,}  in order to avoid unphysical artifacts away from the anti collinear region, it may be necessary to use a smooth version, for example:
\begin{equation}
    \mu_{\star\lambda}(\T{x},\T{y},\T{z}) = X^{-2} Y^{-2} Q^{-2}_{\star\lambda}(\T{x},\T{y},\T{z}).
\end{equation}

\subsection{Resummed BFKL kernel with running coupling}\label{sec:res-BFKL}

\subsubsection{Linearization procedure}

Following the computations in Ref.~\cite{Kovner:2026qfm}, we expand the resummed JIMWLK Hamiltonian in terms of the Reggeized gluon field $\alpha^a(\T{x})$, which is related with the bare Wilson lines as~\cite{Kovner:2005uw,Caron-Huot:2013fea}:
\begin{eqnarray}
S^{ab}(\T{x}) &=& \exp [ig T^c \alpha^c(\T{x})]^{ab}, %= \hat{1} + ig (T^{c_1})^{ab} \alpha^{c_1}(\T{x}) \nonumber \\
%&& + \frac{(ig)^2}{2} (T^{c_1} T^{c_2})^{ab} \alpha^{c_1}(\T{x}) \alpha^{c_2}(\T{x})+O(g^3),   
\label{eq:S-alpha-exp} \\
  V_{ij}(\T{x})&=&\exp[ i g t^c \alpha^c(\T{x}) ]_{ij}.\label{eq:V-alpha-exp}
\end{eqnarray}
%the color rotation operators ($J_{L,R}^a$) then can be defined by enforcing the properties (\ref{eq:JL-def}) and (\ref{eq:JR-def}) to all orders in $g$, leading to:
%\begin{eqnarray}
%  && (ig) J_{(L,R)}^a(\T{x}) = \frac{\delta}{\delta \alpha^a(\T{x})} \nonumber \\
%  && + \sum\limits_{n=1}^\infty g^n c_n^{(L,R)} \left( f^{a b_1 c_1} f^{b_1 b_2 c_2} \ldots f^{b_{n-1} b c_n} \right) \left[ \alpha^{c_1} (\T{x}) \ldots \alpha^{c_n}(\T{x}) \right] \frac{\delta}{\delta \alpha^b(\T{x})}, \label{eq:JLR-exp}
%\end{eqnarray}
%where $c_1^{(L)}=-c_1^{(R)} = +1/2$, $c_2^{(L)}=c_2^{(R)} = 1/12$, $c_3^{(L)}=c_3^{(R)} = 0$,  $c_4^{(L)}=c_4^{(R)} = -1/720$, $c_5^{(L)}=c_5^{(R)} = 0$, $c_6^{(L)}=c_6^{(R)}= 1/30240$ and so on. 
The color rotation operators can be {expressed} in terms of the fields $\alpha^a(\T{x})$ as~\cite{Altinoluk:2013rua}:
\begin{eqnarray}
    (ig) J^a_{(L,R)}(x) &=& \left[ F(\pm i T^c \alpha^c(\T{x})) \right]^{ab} \frac{\delta}{\delta \alpha^b(\T{x})}, \label{eq:JLR-alpha-repr}
\end{eqnarray}
with $F(x) = x/(1-e^{-x})$, where the $(+)$ sign is for $J_L$ and $(-)$ is for $J_R$.

For the dressed Wilson lines $\Bbb{S}^{ab}_Q$ and $\Bbb{V}_{ij,Q}$, the representations (\ref{eq:S-alpha-exp}) and (\ref{eq:V-alpha-exp}) are no longer valid since, as noted in Refs.~\cite{Kovner:2023vsy,Armesto:2025dwd}, the dressed Wilson line is not a unitary matrix. Therefore, following Ref.~\cite{Kovner:2026qfm}, we generalize the expansions (\ref{eq:S-alpha-exp}) and (\ref{eq:V-alpha-exp}) up to the second order in the Reggeized gluon fields as follows:
\begin{eqnarray}
  &&  \Bbb{S}^{ab}(\T{x},Q) = \delta^{ab} + ig T^{c_1}_{ab} \int\limits_{\T{z}_1} R_Q^{(1)}(\T{x}-\T{z}_1) \alpha^{c_1}(\T{z}_1) %+ g d^{c_1ab} \int\limits_{\T{z}_1} R_Q^{(1,d)}(\T{x}-\T{z}_1) \alpha^{c_1}(\T{z}_1)
  \nonumber \\ &&+ \frac{(ig)^2}{2} \sum\limits_{k=1}^{n_{\text{adj.}}} {\cal T}^{(k)c_1 c_2}_{ab} \int\limits_{\T{z}_1, \T{z}_2} R_Q^{(2,k)}(\T{x}-\T{z}_1,\T{x}-\T{z}_2) \alpha^{c_1}(\T{z}_1) \alpha^{c_2}(\T{z}_2) + O(g^3), \label{eq:Sbold-alpha-decomp} \\
   && \Bbb{V}_{ij}(\T{z}) = \delta_{ij} + ig t^{c_1}_{ij} \int\limits_{\T{z}_1} r_Q^{(1)}(\T{z}-\T{z}_1) \alpha^{c_1}(\T{z}_1) \nonumber \\
  && + \frac{(ig)^2}{2} \sum\limits_{k=1}^{n_{\text{fund.}}} {\cal T}^{(F,k)c_1 c_2}_{ij} \int\limits_{\T{z}_1 \T{z}_2} r_Q^{(2,k)}(\T{z}-\T{z}_1,\T{z}-\T{z}_2) \alpha^{c_1}(\T{z}_1) \alpha^{c_2}(\T{z}_2) + \ldots, \label{eq:Vbold-alpha-decomp}
\end{eqnarray}
where the basis of color structures ${\cal T}^{(k)c_1c_2}_{ab}$  spans the color-singlet subspace of the ${\bf 8}\otimes{\bf 8}\otimes {\bf 8}\otimes {\bf 8}$ and has $n_{\text{adj.}}=9$ elements for general $N_c$, while the basis ${\cal T}^{(F,k)c_1 c_2}_{ij}$  spans the color-singlet subspace of ${\bf 3}\otimes \bar{\bf 3} \otimes {\bf 8} \otimes {\bf 8}$ and has $n_{\text{fund.}}=3$ elements. We choose ${\cal T}^{(1)c_1c_2}_{ab}=(T^{c_1} T^{c_2})_{ab}$ and ${\cal T}^{(F,1)c_1 c_2}_{ij} = (t^{c_1} t^{c_2})_{ij}$ and the rest of the basis elements in such a way that their projection on the color singlet state of the pair of Reggeized gluons is zero:  $\delta_{c_1c_2} {\cal T}^{(k)c_1c_2}_{ab}=0$ and $\delta_{c_1 c_2} {\cal T}^{(F,1)c_1 c_2}_{ij} =0$ for $k>1$. The initial conditions for the resummation functions $R_Q^{(1)}$, $R_Q^{(2,k)}$, $r_Q^{(1)}$ and $r_Q^{(2,k)}$ at the starting scale of the evolution $Q=\Lambda$ are:
\begin{eqnarray}
  &&  R_\Lambda^{(1)}(\T{z})=\delta^{(2)}(\T{z}), \label{eq:R1-init-cond} \\
  && R_\Lambda^{(2,1)}(\T{z}_1,\T{z}_2) = \delta^{(2)}(\T{z}_1)\delta^{(2)}(\T{z}_2),\; ~~~~ R_\Lambda^{(2,k)}(\T{z}_1,\T{z}_2) = 0~~\text{ for }k>1,\label{eq:R2-init-cond} \\
%  && R_\Lambda^{(1,d)}(\T{z})=0. \label{eq:R1d-init-cond}
&&  r_\Lambda^{(1)}(\T{z})=\delta^{(2)}(\T{z}), \label{eq:r1-init-cond} \\
  && r_\Lambda^{(2,1)}(\T{z}_1,\T{z}_2) = \delta^{(2)}(\T{z}_1)\delta^{(2)}(\T{z}_2), \; ~~~~ r_\Lambda^{(2,k)}(\T{z}_1,\T{z}_2) = 0\text{ for }k>1.\label{eq:r2-init-cond}
\end{eqnarray}
The evolution equations 
for $R_Q^{(1)}$, $R_Q^{(2,k)}$, $r_Q^{(1)}$, $r_Q^{(2,k)}$ and the solutions will be presented below, in Sec. \ref{sec:evol-R2}.

\subsubsection{Resummed BFKL kernel with running coupling in coordinate space and coordinate-space characteristic function}\label{sec:res-BFKL-coord}
 Substituting  expressions (\ref{eq:JLR-alpha-repr}) and (\ref{eq:Sbold-alpha-decomp}) into the resummed Hamiltonian (\ref{eq:H-JIMWLK-res}) and expanding up to {the} second order in $\alpha$-fields, one obtains the resummed BFKL Hamiltonian:
\begin{eqnarray}
 \hat{H}_{\text{JIMWLK}}^{\text{(res. lin.)}}  &&  = 
 \int\limits_{\T{x},\T{y}}   \Bigg\{ \frac{f^{[c_1 a b_1} f^{c_2] a b_2}}{N_c} \bigg[ - \int\limits_{\T{z}}\bar{K}(\T{x},\T{y},\T{z}) \alpha^{c_1}(\T{x}) \alpha^{c_2}(\T{y})  \nonumber  \\
    &&- \int\limits_{\T{z}_{1},\T{z}_2}\bar{K}^{(2)}(\T{x},\T{y},\T{z}_1,\T{z}_2) \alpha^{c_1}(\T{z}_1) \alpha^{c_2}(\T{z}_2)  \label{eq:coll-res-K-Qt} \\
    && + \int\limits_{\T{z}}\bar{K}^{(1)}(\T{x},\T{y},\T{z}) \bigg( \alpha^{c_1}(\T{z}) \alpha^{c_2}(\T{x}) + \alpha^{c_1}(\T{z})\alpha^{c_2}(\T{y})  \biggr)   \bigg] \frac{{\delta}^2}{ \delta \alpha^{b_1}(\T{x}) \delta \alpha^{b_2}(\T{y})} \nonumber \\
    && +\delta^{(2)}(\T{y}-\T{x}) \int\limits_{\T{z}} [\bar{K}(\T{x},\T{x},\T{z}) \alpha^{b}(\T{x})-\bar{K}^{(1)}(\T{x},\T{x},\T{z}) \alpha^{b}(\T{z})] \frac{\delta}{ \delta \alpha^{b}(\T{x})} \Bigg\} , \nonumber
\end{eqnarray}
where $f^{[c_1 a b_1} f^{c_2] a b_2} = \big( f^{c_1 a b_1} f^{c_2 a b_2} + f^{c_2 a b_1} f^{c_1 a b_2} \big)/2$.  The kernels $\bar{K}$, $\bar{K}^{(1)}$ and $\bar{K}^{(2)}$ encapsulate the corresponding factors of $\alpha_s(\mu_{\star}^2)$ and resummation corrections with the scale choice (\ref{eq:scale choice Q}):
\begin{eqnarray}
&& \hspace{-1cm}\bar{K}(\T{x},\T{y},\T{z})= \frac{\alpha_s(\mu_{\star}^2(\T{x},\T{y},\T{z})) N_c}{2\pi^2} K(\T{x},\T{y},\T{z}) \\
 && \hspace{-1cm}\bar{K}^{(1)}(\T{x},\T{y},\T{z})= \int\limits_{\Tb{z}} \frac{\alpha_s(\mu_{\star}^2(\T{x},\T{y},\Tb{z})) N_c}{2\pi^2}  K(\T{x},\T{y},\Tb{z}) R^{(1)}_{Q_\star (\T{x},\T{y},\Tb{z})}(\Tb{z}-\T{z}), \label{eq:KQt1-def} \\
 &&\hspace{-1cm} \bar{K}^{(2)}(\T{x},\T{y},\T{z}_1,\T{z}_2)= \int\limits_{\Tb{z}} \frac{\alpha_s(\mu_{\star}^2(\T{x},\T{y},\Tb{z})) N_c}{2\pi^2} K(\T{x},\T{y},\Tb{z}) R^{(2,1)}_{Q_\star(\T{x},\T{y},\Tb{z})}(\Tb{z}-\T{z}_1,\Tb{z}-\T{z}_2) \label{eq:KQt2-def}\,.
\end{eqnarray} 
Acting with the linearised Hamiltonian (\ref{eq:coll-res-K-Qt}) on the {\it  ``BFKL Pomeron''} operator  $\alpha^a(\T{x}) \alpha^a(\T{y})$ one obtains the resummed BFKL kernel in coordinate space:
\begin{eqnarray}
  &&\langle \hat{H}_{\text{JIMWLK}}^{\text{(res. lin.)}} \ \alpha^a(\T{x}) \alpha^a(\T{y}) \rangle_Y = \int\limits_{\T{z}_1,\T{z}_2}  K^{\text{(res.)}}(\T{x},\T{y},\T{z}_1,\T{z}_2)  \langle \alpha^a(\T{z}_1) \alpha^a(\T{z}_2) \rangle_Y, \label{eq:H-alpha-alpha}
\end{eqnarray}
where $\langle {\cal O} \rangle_Y\equiv\bra{T} {\cal O} \ket{T}_Y$ denotes the average of the operator $\cal{O}$  over the target state $\ket{T}$ evolved to rapidity $Y$.  The kernel appearing in Eq.~(\ref{eq:H-alpha-alpha}) is
\begin{eqnarray}
   && K^{\text{(res.)}}(\T{x},\T{y},\T{z}_1,\T{z}_2) = \delta^{(2)}(\T{z}_1-\T{x})\delta^{(2)}(\T{z}_2-\T{y}) \bigg(\int\limits_{\T{z}} \bar{K}_D(\T{x},\T{y},\T{z})\bigg)\nonumber \\
   &&-2\bar{K}^{(2)}(\T{x},\T{y},\T{z}_1,\T{z}_2)  +\big( 2 \bar{K}^{(1)}(\T{x},\T{y},\T{z}_1) - \bar{K}^{(1)}(\T{y},\T{y},\T{z}_1) \big)\delta^{(2)}(\T{z}_2-\T{x}) \nonumber \\
   && + \big( 2\bar{K}^{(1)}(\T{x},\T{y},\T{z}_1) - \bar{K}^{(1)}(\T{x},\T{x},\T{z}_1) \big)\delta^{(2)}(\T{z}_2-\T{y}), \label{eq:K-res-def}
\end{eqnarray}
with %the kernel $\bar{K}_D$ being defined as:
\begin{eqnarray}
    \bar{K}_D(\T{x},\T{y},\T{z}) &=& \frac{N_c}{2\pi^2}\bigg[ -2\alpha_s(\mu^2_{\star}(\T{x},\T{y},\T{z})) K(\T{x},\T{y},\T{z}) \label{eq:KD-alphas} \\
    && + \alpha_s((\T{x}-\T{z})^{-2}) K(\T{x},\T{x},\T{z}) + \alpha_s((\T{y}-\T{z})^{-2}) K(\T{y},\T{y},\T{z}) \bigg]. \nonumber
\end{eqnarray}
We note that the three terms in the dipole kernel $\bar K_D$ come with different scales in the running of the coupling, as opposed to phenomenological prescriptions of factorizing a common factor of $\alpha_s$.

As we discussed in Ref.~\cite{Kovner:2026qfm}, in order to deal with IR-finite quantities, 
 instead of the Pomeron operator $\alpha^a(\T{x}) \alpha^a(\T{y})$, the Hamiltonian should be taken to act on the ``dipole'' operator $(\alpha^a(\T{x}) - \alpha^a(\T{y}))^2$: 
 \begin{eqnarray}
  &&\hspace{-10mm}\langle \hat{H}_{\text{JIMWLK}}^{\text{(res. lin.)}} \ (\alpha^a(\T{x}) - \alpha^a(\T{y}))^2 \rangle_Y = \int\limits_{\T{z}_1,\T{z}_2}  K_D^{\text{(res.)}}(\T{x},\T{y},\T{z}_1,\T{z}_2)  \langle (\alpha^a(\T{z}_1)- \alpha^a(\T{z}_2))^2 \rangle_Y, \label{eq:H-(alpha-alpha)^2}
\end{eqnarray}
where
\begin{eqnarray}
    && K_D^{\text{(res.)}}(\T{x},\T{y},\T{z}_1,\T{z}_2) = K^{\text{(res.)}}(\T{x},\T{y},\T{z}_1,\T{z}_2) +  \bar{K}^{(2)}(\T{x},\T{x},\T{z}_1,\T{z}_2) + \bar{K}^{(2)}(\T{y},\T{y},\T{z}_1,\T{z}_2) \nonumber \\
    && - \bar{K}^{(1)}(\T{x},\T{x},\T{z}_1)\delta^{(2)}(\T{z}_2-\T{x}) - \bar{K}^{(1)}(\T{y},\T{y},\T{z}_1) \delta^{(2)}(\T{z}_2-\T{y}) , \label{eq:K-Dip-res-coord}
\end{eqnarray}
and $K^{\text{(res.)}}(\T{x},\T{y},\T{z}_1,\T{z}_2)$ is given by eqn.~(\ref{eq:K-res-def}).

At  LO in the fixed-coupling approximation, the high-energy evolution is scale invariant, and therefore the eigenfunctions of the LO kernel are known exactly, %one of the main mathematical tools to study it is the, which can be defined via the action of t. The latter ones are known exactly, thanks to re-scaling symmetry of the LO kernel, and in coordinate space can be defined as:
\begin{equation}
    \big((\T{z}_1-\T{z}_2)^2\big)^{1-\gamma} e^{in\phi_{\T{z}_1\T{z}_2}}, \label{eq:EFs-coord}
\end{equation}
where $\phi_{\T{z}_1\T{z}_2}$ is the azimuthal angle of the vector $(\T{z}_1-\T{z}_2$). The action of the LO BFKL kernel on {its} eigenfunction defines the corresponding eigenvalue, or  {\it characteristic function} $\chi_0(n,\gamma)$,
\begin{equation}
  \chi_0(n,\gamma)= 2\psi(1)-\psi\left(\gamma+\frac{|n|}{2}\right) - \psi\left( 1-\gamma + \frac{|n|}{2} \right)\,, \label{eq:chi0}
\end{equation}
where the Euler's $\psi$-function $\psi(z)=\frac{d}{dz}\ln\Gamma(z)$.
The knowledge of the characteristic function allows one  to calculate evolution of the scattering amplitude.
%\new{The anti-collinear limit discussed in this paper corresponds to $\gamma\rightarrow 1$, for which the characteristic function  $\chi_0(n,\gamma\rightarrow 1)=1/(1-\gamma+n/2)$. }

Beyond strict high-energy LLA %in $\alpha_s$ for the kernel, 
the running coupling effects violate scale invariance and the functions (\ref{eq:EFs-coord}) are no longer eigenfunctions of the BFKL kernel. The proper eigenfunctions in this case %of the BFKL with higher-order corrections included is 
are known only up to NLO in $\alpha_s$~\cite{Chirilli:2013kca,Chirilli:2014dcb}. This is insufficient for our purposes, since our goal is to resum a series of higher-order corrections to the kernel, to all orders in $\alpha_s$. We therefore consider the next best {object} -- the {\it generalised characteristic function}.

 The {\it coordinate space generalised characteristic function} $\aleph^{\text{(res.)}}(n,\gamma,a_s(\mu_{xy}^2))$ is defined via the action of the kernel (\ref{eq:K-Dip-res-coord}) on the test functions (\ref{eq:EFs-coord}):
\begin{eqnarray}
   \int\limits_{\T{z}_{1,2}} %\left. 
   K_D^{\text{(res.)}}(\T{x},\T{y},\T{z}_1,\T{z}_2) %\right\vert_{\mu^2=(\T{x}-\T{y})^{-2}}
   &\times& \big((\T{z}_1-\T{z}_2)^2\big)^{1-\gamma} e^{in\phi_{\T{z}_1\T{z}_2}}  \\
  =&-&\frac{\alpha_s(\mu_{xy}^2) N_c}{\pi}\big((\T{x}-\T{y})^2\big)^{1-\gamma} e^{in\phi_{\T{z}_1\T{z}_2}} \aleph^{\text{(res.)}}(n,\gamma,a_s(\mu_{xy}^2)),\label{eq:chi_alpha-alpha_DEF}\nonumber
\end{eqnarray}
where $\mu^2_{xy}=c_E (\T{x}-\T{y})^{-2}$ with $c_E=4e^{-2\gamma_E}$ ($\gamma_E=0.57722$ being the Euler-Mascheroni constant). The generalised characteristic function has the following perturbative expansion:
\begin{eqnarray}
 &&   \aleph^{(\text{res.})}(n,\gamma,a_s) = \chi_0(n,\gamma) + \sum\limits_{m=1}^\infty \big(a_s N_c\big)^m \aleph^{(\text{res.})}_{m}(n,\gamma), \label{eq:aleph_exp} %\\
 %&&   \chi_0(n,\gamma)= %2\psi(1)-%\psi\left(\gamma+\frac{|n|}{2}\right) - %\psi\left( 1-\gamma + %\frac{|n|}{2} \right)\,. %\label{eq:chi0}
\end{eqnarray} 
Note that the higher order corrections to $\aleph^{(res)}(n,\gamma,a_s)$ depend on the scale $(\T{x}-\T{y})^2$ through the coupling $a_s(\mu_{xy}^2)$.  
{Although it is no longer, strictly speaking, an eigenvalue of the kernel}, the generalized characteristic function can {nevertheless} be used to calculate the scattering amplitude, as we demonstrate in Sec.~\ref{BFKLGF}.

\section{Momentum-space BFKL kernel}\label{sec:mom-BFKL}

\subsection{Momentum-space characteristic function}\label{sec:res-BFKL-mom}

 An alternative (but equivalent) formulation of the BFKL equation is obtained when one considers evolution of the target charge density correlator
 $\langle \rho_T^a(\T{x}) \rho_T^a(\T{y}) \rangle_Y$. 
 This correlator corresponds directly to the standard definition of the BKFL Green's function, which was  employed in the computations of NLO corrections to the BFKL kernel~\cite{Fadin:1998py,NLOCiafaloni1,NLOCiafaloni2}. Classically, the target charge density and Reggeon field operators are related as (see e.g.~\cite{Jalilian-Marian:1997qno}):
\begin{equation}
    \nabla^2 \alpha^a(\T{x}) = g \rho^a_T(\T{x}) + O(g^2\alpha^2), \label{eq:rhoT-alpha-rel}
\end{equation}
where by $O(g^2\alpha^2)$ we denote non-linear terms in the fields $\alpha^a(\T{x})$, which do not contribute to the linear BFKL equation. This relation leads to the following tree level relation between the correlators in the weak field limit:
\begin{equation}
    \nabla^2_{\T{x}} \nabla^2_{\T{y}} \langle \alpha^a(\T{x}) \alpha^a(\T{y}) \rangle_Y = g^2 \langle \rho_T^a(\T{x}) \rho_T^a(\T{y}) \rangle_Y. \label{eq:aa-rr-COORD}
\end{equation}
Beyond LO in $\alpha_s$ this relation receives loop corrections, which in particular lead to the running of the coupling $g(\mu^2)$ in \eqref{eq:aa-rr-COORD}. The choice of the scale of the coupling $\mu^2$ turns out to be important for  comparison with existing NLO BFKL results~\cite{Fadin:1998py,NLOCiafaloni1,NLOCiafaloni2,Kotikov:2000pm}. The scale choice which leads to agreement of the linearized BK and the NLO BFKL equations is most conveniently  given  in the momentum space version of eqn.~(\ref{eq:aa-rr-COORD}). For the case of the forward scattering amplitude it is:
\begin{equation}
    \T{k}^4 \langle \alpha^a(\T{k}) \alpha^a(-\T{k}) \rangle_Y = g^2(\T{k}^2) \langle \rho_T^a(\T{k}) \rho_T^a(-\T{k}) \rangle_Y. \label{eq:aa-rr-MOM}
\end{equation}
The rapidity evolution of the correlator $\langle \alpha^a(\T{x}) \alpha^a(\T{y}) \rangle_Y$ is governed by~(\ref{eq:H-alpha-alpha}). Thus with the help of the relation (\ref{eq:aa-rr-MOM}) we obtain the following momentum space BFKL equation for $\langle \rho_T^a(\T{k}) \rho_T^a(-\T{k}) \rangle_Y$: 
\begin{equation}
  \frac{\partial}{\partial Y} \langle \rho_T^a(\T{k}) \rho_T^a(-\T{k}) \rangle_Y = -\int\limits_{\T{q}} K_{\text{BFKL}}^{\text{(res.)}}(\T{k},\T{q})  \langle \rho_T^a(\T{q}) \rho_T^a(-\T{q}) \rangle_Y,  \label{eq:BFKL-mom}
\end{equation}
with the following definition of the momentum space BFKL kernel:
\begin{equation}
     K^{\text{(res.)}}_{\text{BFKL}}(\T{k},\T{q}) = \frac{\T{k}^4\alpha_s(\T{q}^2)}{\T{q}^4 \alpha_s(\T{k}^2)} \int\limits_{\T{x},\T{y},\T{z},\T{z}'}  \frac{e^{-i\T{k}(\T{x}-\T{y})}}{(2\pi)^{2}}   K^{\text{(res.)}}(\T{x},\T{y},\T{z},\T{z}') \frac{e^{i\T{q}(\T{z}-\T{z}')} + e^{-i\T{q}(\T{z}-\T{z}')} }{2 S_\perp}, \label{eq:Kres(k,q)-def}
\end{equation}
where $S_\perp=\int d^{2}\T{x}$. 
In momentum space, the anti-collinear resummation implies resummation of large logarithms $\ln \T{q}^2/\T{k}^2$, when $\T{q}^2\gg \T{k}^2$. The terms 
of the type $ \alpha_s^m \ln^m (\T{q}^2/\T{k}^2)$ correspond to 
the leading  logarithmic approximation (LLA), the abbreviation which will be used extensively below. 

A scale choice in the strong coupling factor in eqn.~(\ref{eq:aa-rr-MOM}) is equivalent to a choice of scale in the coupling of the Reggeized gluon to the color charge in the impact factor. The importance of this scale choice for the comparison of the linearised NLO BK equation with the NLO BFKL results was realised already in Ref.~\cite{Balitsky:2007feb} (Sec. VI). Our  eqn.~(\ref{eq:aa-rr-MOM}) is equivalent to eqn. (116) of Ref.~\cite{Balitsky:2007feb}\footnote{Note that our approach in this paper is somewhat different from that in Ref.~\cite{Balitsky:2007feb}. The comparison of the BFKL and BK equations at NLO in Ref.~\cite{Balitsky:2007feb} was done on the level of the dipole scattering amplitude. This involves convoluting the BFKL Green's function with the dipole impact factor. In the present paper, as an alternative approach,  we derive  the momentum space characteristic function, and thus compare the Green's functions directly. }.

 The {\it momentum-space generalized characteristic function} is defined using (\ref{eq:Kres(k,q)-def}) as
\begin{eqnarray}
    \int\limits_{\T{q}} %\left.
    K_{\text{BFKL}}^{\text{(res.)}}(\T{k},\T{q}) %\right\vert_{\old{\mu^2=\T{k}^2}}
    (\T{q}^2)^\gamma e^{in\phi_{\T{q}}} = -\frac{\alpha_s(\T{k}^2) N_c}{\pi} (\T{k}^2)^\gamma e^{in\phi_{\T{k}}} \chi^{\text{(res.)}}(n,\gamma, a_s(\T{k}^2)){\,.} \label{eq:chi_rho-rho_DEF} 
\end{eqnarray}

 Note that, beyond LO in $a_s$ the function
$\chi^{\text{(res.)}}(n,\gamma, a_s(\T{k}^2))$ differs 
from the coordinate-space characteristic function $\aleph^{\text{(res.)}}$ defined in eqn.~(\ref{eq:chi_alpha-alpha_DEF}). The definition \eqref{eq:chi_rho-rho_DEF}  corresponds to the action of the BFKL kernel on a set of test functions that are different from \eqref{eq:EFs-coord}, apart of the obvious Fourier transform, by a factor involving the ratio of two running couplings.  In Appendix~\ref{append:char-func-rel} we explore the relation between the coordinate space and momentum space characteristic functions. {In particular, in Appendix~\ref{sec:AC-res-NLO} we demonstrate that our scale choice (\ref{eq: scale choice mu}) correctly reproduces the $n$-dependence of the running-coupling contribution to the anti-collinear pole of the NLO BFKL characteristic function.} 

{Several approximations for the BFKL kernel with the anti-collinear resummation and/or running coupling will be considered in this paper. The approximation will be distinguished and labelled by the superscript ${}^\text{(L)}$.%, which meanwhile we keep general. 
The corresponding characteristic function carries the same superscript $\chi^{\text{(L)}}$.} 
We define the perturbative expansion of each of the functions $\chi^{\text{(L)}}$ in the same way as in eqn.~(\ref{eq:aleph_exp}):
\begin{eqnarray}
 &&   \chi^{(\text{L})}(n,\gamma,a_s) = \chi_0(n,\gamma) + \sum\limits_{m=1}^\infty \big(a_s N_c\big)^m \chi^{(\text{L})}_{m}(n,\gamma). \label{eq:chi_exp}
\end{eqnarray}
In the rest of the paper we will limit our analysis to the momentum-space computations, starting from computation of the resummed evolution kernel $K^{\text{(res.)}}_{\text{BFKL}}$ and then of the generalized characteristic function.

\subsection{Exact Fourier transform of the resummed BFKL kernel}\label{sec:K-BFKL-Fourier}

Similarly to what was done in Ref.~\cite{Kovner:2026qfm}, it is possible to compute the Fourier transform of the kernel (\ref{eq:K-res-def}) with the piecewise-smooth scale for anti-collinear resummation (\ref{eq:scale choice Q}) and running of the coupling (\ref{eq: scale choice mu}). The result is written in terms of the Fourier transformed resummation functions:
\begin{eqnarray}
 % \new{R^{(1)}(\T{p};Q)=}  
 R_Q^{(1)}(\T{p}) &=& \int\limits_{\T{z}} R_Q^{(1)}(\T{z}) e^{-i\T{p}\T{z}}, \label{eq:R1-q-def} \\ 
    % \new{R^{(2,1)}(\T{p};Q)=}
    R_Q^{(2,1)}(\T{p}) &=& \int\limits_{\T{z}_1,\T{z}_2} R^{(2,1)}_Q(\T{z}_1,\T{z}_2) e^{i\T{p}(\T{z}_1-\T{z}_2)}, \label{eq:R2-q-def}
\end{eqnarray} 
with initial conditions for these functions:
\begin{eqnarray}
    R_\Lambda^{(1)}(\T{p}) = R_\Lambda^{(2,1)}(\T{p})=1. \label{eq:R1-R21-mom-init-conds}
\end{eqnarray}
 In this section we only present the Fourier transformed kernel, while the details of its computation are given in Appendix~\ref{append:Fourier}:
\begin{eqnarray} 
K^{\text{(res.)}}_{\text{BFKL}}(\T{k},\T{q}) &=& \frac{N_c}{2\pi^2} \frac{\T{k}^4\alpha_s(\T{q}^2)}{\T{q}^4\alpha_s(\T{k}^2)} \bigg\{ \Delta K_{1}(\T{k},\T{q}) + \Delta K_{2}(\T{k},\T{q}) + \Delta K_{3}(\T{k},\T{q}) \label{eq:K_BFKL-exact} \\
 &+& K_\delta(\T{q}) \delta^{(2)}(\T{k}-\T{q}) + (\T{q}\to -\T{q}) \bigg\}. \nonumber
\end{eqnarray}
The kernel $K_\delta(\T{q})$ does not contribute in the anti-collinear limit $\T{q}^2\gg \T{k}^2$ and is not presented below (the expression for the kernel  is given in
Appendix~\ref{append:Fourier}). Following  Ref.~\cite{Kovner:2026qfm}, the kernels $\Delta K_{1,2}$ in ~(\ref{eq:K_BFKL-exact}) can be expressed as  integrals over an auxiliary scale $Q^2$:
\begin{eqnarray}
    \Delta K_{1}(\T{k},\T{q}) &=& \int\limits_0^{\infty} \frac{dQ^2}{Q^2}   \bigg[ \frac{\partial\alpha_s(Q^2)}{\partial \ln Q^2} \frac{J_0^2(|\T{q}-\T{k}|/Q)}{(\T{k}-\T{q})_+^2} + \frac{\partial [\alpha_s(Q^2) R_Q^{(2,1)}(\T{q})]}{\partial \ln Q^2}  \frac{J_0^2(|\T{k}|/Q)}{\T{k}^2} \nonumber \\ 
    &-&  \frac{2\T{k}\cdot (\T{k}-\T{q})}{\T{k}^2 (\T{k}-\T{q})^2} \frac{\partial[\alpha_s(Q^2) R^{(1)}_Q(\T{q})]}{\partial \ln Q^2}  J_0(|\T{k}|/Q) J_0(|\T{k}-\T{q}|/Q) \bigg] ,\label{eq:DK1-BFKL-res-ex}
\end{eqnarray}
where the $(+)$-distribution in transverse momentum space\footnote{%\old{When the expression \eqref{eq:DK1-BFKL-res-ex} is used in \eqref{eq:K_BFKL-exact}, the $+$ distribution is defined to act on all ${\bf q}$-dependent factors except for the Bessel functions \new{or $\alpha_s((\T{k}-\T{q})^2)$-factors in eqn.~(\ref{eq:DelK1-theta}) and below.} 
The $(+)$ prescription in
\eqref{eq:DK1-BFKL-res-ex} is defined with respect to any probe function the kernel
$\Delta K_1$ is assumed to act on. It is thus not affecting the factors present in the definition of the kernel itself, such as the Bessel functions.
} is defined as:
\begin{equation}
     \int\limits_{\T{q}} \frac{f(\T{q})}{(\T{k}-\T{q})^2_+} = \int\limits_{\T{q}} \frac{f(\T{q})-f(\T{k})\theta(|\T{k}-\T{q}|<|\T{k}|)}{(\T{k}-\T{q})^2}\,. \label{eq:1/(k-q)^2_+:DEF}
\end{equation}
The second term of the kernel can also be written with the single $Q^2$-integral:
\begin{eqnarray}
    \Delta K_{2}(\T{k},\T{q}) &=& 2 \int\limits_0^{\infty} \frac{dQ^2}{Q^2} \frac{\partial \alpha_s(Q^2)}{\partial \ln Q^2}\bigg[ \frac{\Delta(|\T{q}-\T{k}|,|\T{q}-\T{k}|,Q)}{(\T{k}-\T{q})_+^2}  + \frac{\Delta(|\T{k}|,|\T{k}|,Q)}{\T{k}^2} \label{eq:DK2-BFKL-res-ex} \\
  &-&  \frac{\T{k}\cdot (\T{k}-\T{q})}{\T{k}^2 (\T{k}-\T{q})^2} \bigg( \Delta(|\T{q}-\T{k}|,|\T{k}|,Q) + \Delta(|\T{k}|,|\T{q}-\T{k}|,Q)\bigg)\bigg],\nonumber 
\end{eqnarray}
where the function $\Delta(k_1,k_2,Q)\equiv J_0(k_1/Q)\big(1-J_0(k_2/Q) \big)$. 
Note that the kernel $\Delta K_{2}$
does not depend on any resummation function $R_Q^{(j)}$.

The  third kernel in (\ref{eq:K_BFKL-exact}) requires introduction of two auxiliary scales:   $Q_1$  for the resummation functions and  $Q_2$  for $\alpha_s$:
\begin{eqnarray}
    \Delta K_{3}(\T{k},\T{q}) =& -&2 \int\limits_0^{\infty} \frac{dQ_1^2}{Q_1^2}  \int\limits_0^{Q_1^2}\frac{dQ_2^2}{Q_2^2} \frac{\partial \alpha_s(Q_2^2)}{\partial \ln Q_2^2}\bigg[  \frac{\partial R_{Q_1}^{(2,1)}(\T{q}) }{\partial \ln Q_1^2} \frac{J_0(|\T{k}|/Q_1) J_0(|\T{k}|/Q_2)}{\T{k}^2}    \nonumber \\
  &-& \frac{\partial R^{(1)}_{Q_1}(\T{q})}{\partial \ln Q_1^2} \frac{\T{k}\cdot (\T{k}-\T{q})}{\T{k}^2 (\T{k}-\T{q})^2} \big( J_0(|\T{k}-\T{q}|/Q_1) J_0(|\T{k}|/Q_2) +  (Q_1\leftrightarrow Q_2)  \big) \bigg] \nonumber \\
 &-&2 \int\limits_0^{\infty} \frac{dQ^2}{Q^2} \frac{\partial \alpha_s(Q^2)}{\partial \ln Q^2}\bigg[ \big(R^{(2,1)}_Q(\T{q})-1\big) \frac{J^2_0(|\T{k}|/Q)}{\T{k}^2} \nonumber \\
 &-& \big(R^{(1)}_Q(\T{q})-1\big)\frac{2\T{k}\cdot (\T{k}-\T{q})}{\T{k}^2 (\T{k}-\T{q})^2} J_0(|\T{k}-\T{q}|/Q) J_0(|\T{k}|/Q) \bigg].  \label{eq:DK3-BFKL-res-ex} 
\end{eqnarray}
In the fixed coupling limit, only $\Delta K_1$ survives and we recover the result of Ref.~\cite{Kovner:2026qfm}. In the next subsection we extract the LLA terms for the three kernels introduced above.

\subsection{The LLA momentum-space BFKL kernel}\label{sec:LLA-BFKL-mom}

In the anti-collinear limit $\T{q}^2\gg \T{k}^2$, the exact momentum-space BFKL kernel (\ref{eq:K_BFKL-exact}) contains subleading terms, both  power suppressed (having extra factors of $\T{k}^2/\T{q}^2$) and leading power next-to-LLA logarithms (NLLs).  To isolate the LLA terms, we follow a two-step strategy introduced in Ref.~\cite{Kovner:2026qfm}. First, we approximate all the Bessel functions in eqn.~(\ref{eq:K_BFKL-exact}) by $\theta$-functions, following the  rule:
\begin{equation}
     J_0(x)\to \theta(1-x). \label{eq:J-theta-appr}
\end{equation}
That is  we neglect the decreasing tails of the Bessel functions at $x>1$ -- as demonstrated in ~\cite{Kovner:2026qfm},  these tails do not contribute in LLA. Indeed, eqns.~(\ref{eq:DK1-BFKL-res-ex}), (\ref{eq:DK2-BFKL-res-ex}) and (\ref{eq:DK3-BFKL-res-ex}) are organised {such that the logarithmic contribution from the integrations over the auxiliary scales $Q^2$ and $Q_{1,2}^2$}  {are cut-off by any deviation of the Bessel-function factors from a constant. The approximation (\ref{eq:J-theta-appr}) is  therefore not specific for the anti-collinear limit
and is valid also in the collinear regime of ($\T{k}^2 \gg \T{q}^2$), at LLA. The approximation (\ref{eq:J-theta-appr}) makes it possible  to compute all  the integrals over the auxiliary scales $Q^2$ (or $Q_{1,2}^2$) while preserving  its logarithmic structure. {The resulting} expression  still {contains certain}  NLL and subleading-power {contributions}. However, as a second step, {these terms} can be easily identified {and discarded}, {yielding} {the anti-collinearly resummed BFKL kernel in the LLA,  in momentum space}.

Let us now perform the first step and apply the approximation (\ref{eq:J-theta-appr}) to eqns.~(\ref{eq:DK1-BFKL-res-ex}), (\ref{eq:DK2-BFKL-res-ex}), and (\ref{eq:DK3-BFKL-res-ex}). 
 Two scales will emerge below:
\begin{equation}\label{2scales}
{\cal Q}\equiv \max(|\T{k}|,|\T{k}-\T{q}|)\,;
\qquad\qquad 
\bar{\cal Q}\equiv \min(|\T{k}|,|\T{k}-\T{q}|).
\end{equation}
Within the approximation (\ref{eq:J-theta-appr}), Eq. (\ref{eq:DK1-BFKL-res-ex})  for $\Delta K_1$ reads:
\footnote{As noted in the previous footnote, the $+$ prescription here does not apply to $\alpha_s((\T{q}-\T{k})^2)$.}
\begin{equation}
     \Delta K^{\text{($\theta$-appr.)}}_{1}(\T{k},\T{q}) =-\frac{\alpha_s((\T{q}-\T{k})^2)}{(\T{k}-\T{q})_+^2} - \frac{\alpha_s(\T{k}^2) R_{|\T{k}|}^{(2,1)}(\T{q})}{\T{k}^2}  +\frac{2\T{k}\cdot (\T{k}-\T{q})}{\T{k}^2(\T{k}-\T{q})^2}\alpha_s({\cal Q}^2) R_{\cal Q}^{(1)}(\T{q})\,. \label{eq:DelK1-theta}
\end{equation}
By the superscript ``$\theta$-appr.'' we have explicitly indicated that the
kernel is derived using the approximation (\ref{eq:J-theta-appr}).
%\begin{eqnarray}
%&&\old{   \Delta K^{\text{($\theta$-appr.)}}_{1}(\T{k},\T{q}) =-\frac{\alpha_s((\T{q}-\T{k})^2)}{(\T{k}-\T{q})_+^2} - \frac{\alpha_s(\T{k}^2) R^{(2,1)}(\T{q},|\T{k}|)}{\T{k}^2} } \\
%&&+\old{    \frac{2\T{k}\cdot (\T{k}-\T{q})}{\T{k}^2(\T{k}-\T{q})^2}\alpha_s(\max[\T{k}^2,(\T{k}-\T{q})^2]) R^{(1)}(\T{q},\max[|\T{k}|,|\T{k}-\T{q}|]),\label{eq:DelK1-theta}.\nonumber}
%\end{eqnarray}
%\old{where we have used the notation $R^{(j)}(\T{p},Q)\equiv R^{(j)}_Q(\T{p})$.} 

Turning to $\Delta K_2$ in eqn.~(\ref{eq:DK2-BFKL-res-ex}), the approximation (\ref{eq:J-theta-appr}) implies that $\Delta(k_1,k_2,Q)\simeq \theta(k_1<Q<k_2)\theta(k_2>k_1)$ and $\Delta(k_1,k_1,Q)\simeq 0$.
Hence
\begin{eqnarray}
  \hspace{-5mm}\Delta K^{\text{($\theta$-appr.)}}_{2}(\T{k},\T{q}) =- \frac{2\T{k}\cdot (\T{k}-\T{q})}{\T{k}^2 (\T{k}-\T{q})^2} \big( \alpha_s(\bar {\cal Q}^2) - \alpha_s({\cal Q}^2) \big).
\end{eqnarray}
%\begin{eqnarray}
%\old{  \hspace{-5mm}\Delta K^{\text{($\theta$-appr.)}}_{2}(\T{k},\T{q}) =- \frac{2\T{k}\cdot (\T{k}-\T{q})}{\T{k}^2 (\T{k}-\T{q})^2} \big( \alpha_s(\min[\T{k}^2,(\T{k}-\T{q})^2]) - \alpha_s(\max[\T{k}^2,(\T{k}-\T{q})^2]) \big).}
%\end{eqnarray}

The remaining kernel, $\Delta K_3$ (eq. ~(\ref{eq:DK3-BFKL-res-ex})), can be put into the following form:
\begin{eqnarray}
    \Delta K^{\text{($\theta$-appr.)}}_{3}(\T{k},\T{q}) = - \frac{2\T{k}\cdot (\T{k}-\T{q})}{\T{k}^2 (\T{k}-\T{q})^2} \big(  \alpha_s({\cal Q}^2)-\alpha_s(\bar{\cal Q}^2)\big) \big(R_{\cal Q}^{(1)}(\T{q})-1\big). 
\end{eqnarray}
%\begin{eqnarray}&&
%   \old{  \Delta K^{\text{($\theta$-appr.)}}_{3}(\T{k},\T{q}) = - \frac{2\T{k}\cdot (\T{k}-\T{q})}{\T{k}^2 (\T{k}-\T{q})^2} \big(  \alpha_s(\max[\T{k}^2,(\T{k}-\T{q})^2])-\alpha_s(\min[\T{k}^2,(\T{k}-\T{q})^2])\big)
%   }\nonumber \\
%    &&
%    \old{\times \big(R^{(1)}(\T{q},\max[|\T{k}|,|\T{k}-\T{q}|])-1\big). }
%\end{eqnarray}
%\old{Finally % eqn.~(\ref{eq:K_BFKL-exact})
%one obtains:}
%\begin{eqnarray}
%   &&\old{   K^{\text{(res., $\theta$-appr.)}}_{\text{BFKL}}(\T{k},\T{q}) = -\frac{N_c}{2\pi^2} \frac{\T{k}^4\alpha_s(\T{q}^2)}{\T{q}^4\alpha_s(\T{k}^2)}} \nonumber \\
%     &&
%     \old{\times\bigg\{ \frac{\alpha_s((\T{q}-\T{k})^2)}{(\T{k}-\T{q})_+^2}  + \frac{\alpha_s(\T{k}^2)}{\T{k}^2} -\frac{2\T{k}\cdot (\T{k}-\T{q})}{\T{k}^2 (\T{k}-\T{q})^2}\alpha_s(\min[\T{k}^2,(\T{k}-\T{q})^2]) }\nonumber \\
%     &&
%     \old{+ \frac{\alpha_s(\T{k}^2) }{\T{k}^2} \big(R^{(2,1)}(\T{q},|\T{k}|) -1 \big) }\nonumber \\
%     &&
%     \old{-\frac{2\T{k}\cdot (\T{k}-\T{q})}{\T{k}^2(\T{k}-\T{q})^2}\alpha_s(\min[\T{k}^2,(\T{k}-\T{q})^2]) \big( R^{(1)}(\T{q},\max[|\T{k}|,|\T{k}-\T{q}|]) -1 \big) }\nonumber \\
% &&\old{+ \text{(terms $\propto \delta^{(2)}(\T{q}-\T{k})$)} + (\T{q}\to -\T{q}) \bigg\}.\label{eq:K-BFKL-res-theta}}
%\end{eqnarray}
 Substituting all the  kernels $\Delta K_{1,2,3}^{\text{$\theta$-appr.}}$ into  \eqref{eq:K_BFKL-exact}, we arrive at the ``$\theta$-approximated'' resummed BFKL kernel,
\begin{eqnarray}
    &&  K^{\text{(res., $\theta$-appr.)}}_{\text{BFKL}}(\T{k},\T{q}) = -\frac{N_c}{2\pi^2} \frac{\T{k}^4\alpha_s(\T{q}^2)}{\T{q}^4\alpha_s(\T{k}^2)} \nonumber \\
     &&
     \times\bigg\{ \frac{\alpha_s((\T{q}-\T{k})^2)}{(\T{k}-\T{q})_+^2}  + \frac{\alpha_s(\T{k}^2)}{\T{k}^2} -\frac{2\T{k}\cdot (\T{k}-\T{q})}{\T{k}^2 (\T{k}-\T{q})^2}\alpha_s(\bar{\cal Q}^2) +\frac{\alpha_s(\T{k}^2) }{\T{k}^2} \big(R_{|\T{k}|}^{(2,1)}(\T{q}) -1 \big) \nonumber \\
     &&
     -\frac{2\T{k}\cdot (\T{k}-\T{q})}{\T{k}^2(\T{k}-\T{q})^2}\alpha_s(\bar{
     \cal Q}^2) \big( R_{\cal Q}^{(1)}(\T{q}) -1 \big) +\text{(terms $\propto \delta^{(2)}(\T{q}-\T{k})$)} + (\T{q}\to -\T{q}) \bigg\}.\label{eq:K-BFKL-res-theta}
\end{eqnarray}
As mentioned, the kernel $K^{\text{(res., $\theta$-appr.)}}_{\text{BFKL}}$ is correct both in the collinear and anti-collinear regimes. 
We now consider the latter one, that is the limit $\T{q}^2\gg \T{k}^2$. In this limit, ${\cal Q}\simeq |\T{q}|$ while 
$\bar{\cal Q}= |\T{k}|$. Since
$ R_{|\T{q}|}^{(1)}(\T{q})\simeq 1$, the first term in the last line
can be neglected ~\cite{Kovner:2026qfm}.
The first term in the curly brackets is formally power suppressed too. Yet, we opt to
keep this term, modulo replacing $\alpha_s(\T{q}^2)$ with $\alpha_s(\T{k}^2)$ (which is safe to do because the term is subleading anyway). It is convenient to keep this term, since together with the second and third terms it combines into the full LO BFKL kernel, thus explicitly separating the effect of the resummation.
Discarding the irrelevant $\delta$-functional terms, 
the final result for the LLA
resummed BFKL kernel is
\begin{eqnarray}
 \hspace{-6mm}   K^{\text{(res., LLA)}}_{\text{BFKL}}(\T{k},\T{q}) = \frac{\alpha_s(\T{q}^2) N_c}{2\pi^2} \frac{\T{k}^4}{\T{q}^4} \bigg[ -\frac{\T{q}^2}{\T{k}^2 (\T{k}-\T{q})^2_+} + \frac{1}{\T{k}^2} \big( 1 - R^{(2,1)}_{|\T{k}|}(\T{q}) \big) + (\T{q}\leftrightarrow -\T{q}) \bigg] \nonumber\\
 \equiv \alpha_s(\T{q}^2)\,{\tilde K^{\text{(res., LLA)}}_{\text{BFKL}}}.
 \label{eq:K-BFKL-res-LLA}
\end{eqnarray}
The result coincides with the eqn.~(2.58) of Ref.~\cite{Kovner:2026qfm}, up to running of $\alpha_s$. Note, that the scale for $\alpha_s$ in the momentum space BFKL kernel (\ref{eq:K-BFKL-res-LLA}) turned-out to be in accordance with the natural expectation in the anti-collinear limit $\T{q}^2\gg \T{k}^2$.  
This is in sharp contrast with
 the coordinate space   where the scale was  chosen in (\ref{eq: scale choice mu}) to be  the lowest one among the available scales. This inversion of the scales is entirely due to the definition of the BFKL kernel (\ref{eq:Kres(k,q)-def}), {which} contains an additional factor $\alpha_s(\T{q}^2)/\alpha_s(\T{k}^2)$. 

\subsection{LLA evolution for $R^{(2,1)}_Q$ with running coupling}\label{sec:evol-R2}

As in the fixed-coupling case~\cite{Kovner:2026qfm}, the LLA result for the resummed BFKL kernel (\ref{eq:K-BFKL-res-LLA}) depends  on the resummation function $R^{(2,1)}_Q$ only. As opposed to  the previous subsection, where we discussed running of the coupling in the BFKL kernel, we will now address the effect of the running in the ``DGLAP'' resummation equations for $R^{(2,1)}_Q$. The coordinate space evolution equations for the functions $R_Q^{(j)}$ are obtained from the evolution equations derived in  Sec. 2.3.1 of Ref.~\cite{Kovner:2026qfm} via  replacement $a_s\to a_s(Q^2)$. 
The  momentum-space evolution equations for  $R_Q^{(2,1)}$ and its quark counterpart $r_Q^{(2,1)}$ are obtained via the same substitution: 
\begin{equation}
    \frac{\partial}{\partial \ln Q^2} {\cal R}_Q^{(2,1)}(\T{p}) = -a_s(Q^2) \bigg[ \Pi_2 {\cal R}_Q^{(2,1)}(\T{p}) + J_0\left(\frac{|\T{p}|}{Q} \right) \rho_Q^{(1)}(\T{p}) \bigg],\label{eq:R2-eqn-mom}
\end{equation}
where %two dimensional vectors are introduced,
\begin{equation}
{\cal R}^{(2,1)}_Q(\T{p}) \equiv \left( \begin{array}{c}
         R^{(2,1)}_Q(\T{p}) \\
         r^{(2,1)}_Q(\T{p}) 
    \end{array} \right)\,,\qquad
     \rho_Q^{(1)}(\T{p}) \equiv \left( \begin{array}{c}
         \frac{11}{3}N_c [R_Q^{(1)}(\T{p})]^2 + \frac{2n_F}{3N_c^2} [r_Q^{(1)}(\T{p})]^2  \\
         3N_c R_Q^{(1)}(\T{p}) r_Q^{(1)}(\T{p}) 
    \end{array}\right), \label{eq:inhomog-part-quarks}
\end{equation}
The source term $\rho_Q^{(1)}$ depends on the first-order resummation functions $R^{(1)}_Q(\T{p})$ and $r^{(1)}_Q(\T{p})$.
The constant matrix $\Pi_2$ is %(setting $T_F=1/2$):
\begin{equation}
    \Pi_2 = \left( \begin{array}{cc}
       \beta_0-2N_c\frac{11}{3}  & \frac{4C_F n_F}{3N_c}  \\
        -3N_c & 0
    \end{array} \right) , \label{eq:Pi2-matr}
\end{equation}
The basis of normalised eigenvectors ${\cal E}_{\pm}$ of the matrix $\Pi_2$:
\begin{equation}
{\cal E}_{\pm}=\frac{1}{\sqrt{9N_c^2+\lambda_{\pm}^2}} \left( \begin{array}{c}
         -\lambda_\pm \\
          3N_c
    \end{array} \right).
    \label{eq:Pi2-eigen-basis} 
\end{equation}
The corresponding eigenvalues of the matrix (\ref{eq:Pi2-matr}) 
\begin{eqnarray}
     \lambda_{\pm} = \frac{\beta_0}{2}-\frac{11}{3}N_c \pm \sqrt{\left( \frac{\beta_0}{2}-\frac{11}{3}N_c \right)^2 - 4C_Fn_F} \; . \label{eq:lambda-PM-expr}
\end{eqnarray}

Following Ref.~\cite{Kovner:2026qfm}, we expand the solutions of Eq.~(\ref{eq:R2-eqn-mom}) in this basis,
\begin{equation}
    {\cal R}_Q^{(2,1)}(\T{p}) = \bar{R}_{Q,+}^{(2,1)}(\T{p}) {\cal E}_+ +  \bar{R}_{Q,-}^{(2,1)}(\T{p}) {\cal E}_- . \label{eq:R2-sol-EB-exp}
\end{equation}
Taking into account running of the coupling \eqref{eq:as-Q2}, 
%\begin{eqnarray}
 %   \frac{\partial a_s(Q^2)}{\partial \ln Q^2} = -\beta_0 a_s^2(Q^2)+O(a_s^3)\quad 
  %\Longrightarrow \quad a_s(Q^2)=\frac{a_s(\Lambda^2)}{1+\beta_0 a_s(\Lambda^2) \ln(Q^2/\Lambda^2)},  \label{eq:as-Q2}
%\end{eqnarray}
we arrive at the exact solution of \eqref{eq:R2-eqn-mom}, expressed in terms of 
 the expansion coefficients $\bar{R}_{Q,+}^{(2,1)}(\T{p})$ and $ \bar{R}_{Q,-}^{(2,1)}(\T{p})$:
\begin{eqnarray}
    \bar{R}_{Q,\pm}^{(2,1)}(\T{p}) &=& \left( \frac{a_s(Q^2)}{a_s(\Lambda^2)} \right)^{\lambda_{\pm}/\beta_0} \bigg[ \bar{R}_{\Lambda,\pm}^{(2,1)}(\T{p}) -  \int\limits_{\Lambda^2}^{Q^2} \frac{dq^2}{q^2} a_s(q^2) \left( \frac{a_s(q^2)}{a_s(\Lambda^2)} \right)^{-\lambda_{\pm}/\beta_0} J_0\left( \frac{|\T{p}|}{q} \right) \bar{\rho}_{q,\pm}^{(1)}(\T{p}) \bigg]. \nonumber \\\label{eq:sol-eigen}
\end{eqnarray}
 Here $\bar{R}_{\Lambda,\pm}^{(2,1)}(\T{p})$ is the vector of  the initial conditions at the scale $\Lambda$ and $\bar{\rho}_{q,\pm}^{(1)}(\T{p})$ 
 is the source vector (\ref{eq:inhomog-part-quarks}), decomposed in the basis (\ref{eq:Pi2-eigen-basis}).  

The observation made in Ref.~\cite{Kovner:2026qfm} holds true: the LLA-terms in the solution (\ref{eq:sol-eigen}) in the region $q<|\T{p}|$  do not depend on the detailed behaviour of the first-order resummation functions $R^{(1)}_q(\T{p})$ and $r^{(1)}_q(\T{p})$. This is because the contribution of this region to the integral in (\ref{eq:sol-eigen}) is suppressed by oscillations of the Bessel function. Therefore, to obtain the LLA solution we can first use the approximation (\ref{eq:J-theta-appr}) and then replace the functions $\bar{\rho}_{q,\pm}^{(1)}(\T{p})$ with the corresponding initial conditions $\bar{\rho}^{(1)}_{\Lambda,\pm}$, which is valid 
for $q>|\T{p}|$. {Proceeding} this way, one obtains the LLA solution in the eigenbasis (\ref{eq:Pi2-eigen-basis}). It leads to  the following LLA result for  $R^{(2,1)}_Q$:
\begin{equation}
    R^{(2,1),\text{LLA}}_{Q}(\T{p}) = h(\lambda_+,\lambda_-) \left( \frac{a_s(\max [Q^2, \T{p}^2])}{a_s(Q^2)} \right)^{-\lambda_-/\beta_0} + (\lambda_+ \leftrightarrow \lambda_-), \label{eq:R2-LLA}
\end{equation}
where
\begin{equation}
    h(\lambda_+,\lambda_-) = \frac{3\beta_0-11N_c^3-11N_c-3N_c^2\lambda_+}{3N_c^2(\lambda_- - \lambda_+)}.\label{eq:h(lam+,lam-)-DEF}
\end{equation}
The fixed coupling limit of the factors governing the scale dependence of eqn.~(\ref{eq:R2-LLA}) is found by using  eqn.~(\ref{eq:as-Q2}):
\begin{equation}
  \beta_0\to 0:\;\;  \left( \frac{a_s(\max [Q^2, \T{p}^2])}{a_s(Q^2)} \right)^{-\lambda/\beta_0} \to \left( \frac{\max [Q^2, \T{p}^2]}{Q^2} \right)^{a_s \lambda},\label{eq:FC-limit}
\end{equation}
for fixed $\lambda$. With the help of the substitution (\ref{eq:FC-limit}) one recovers the fixed-coupling solution, obtained in Ref.~\cite{Kovner:2026qfm}:
\begin{equation}
    R^{(2,1),\text{FC-LLA}}_{Q}(\T{p}) = h(\lambda_+,\lambda_-) \left( \frac{\max [Q^2, \T{p}^2]}{Q^2} \right)^{a_s \lambda_-} + (\lambda_+ \leftrightarrow \lambda_-). \label{eq:R2-LLA-FC}
\end{equation}
Here the superscript ``FC''
stands for Fixed Coupling. We note, that as in the case of the usual DGLAP equation, both the fixed coupling (\ref{eq:R2-LLA-FC}) and  running coupling (\ref{eq:R2-LLA}) solutions resum LLA terms $\propto \alpha_s^n(Q^2)\ln^{n}(\T{p}^2/Q^2)$. Yet, only the running coupling solution provides resummation of the  complete LLA series.

\section{Running coupling effects in the generalized characteristic function}\label{sec:chi-RC-main}

In this Section we are going to investigate the resummed BFKL kernel through the lens of its generalized characteristic function. As discussed above, the coupling constant enters in two distinct places. The first one is in the kernel of the resummed JIMWLK/BFKL equation \eqref{eq:H-alpha-alpha}. The second place is inside the
DGLAP resummation  \eqref{eq:R2-eqn-mom}. In the anti-collinear limit the running of  these couplings is equally important. In this limit, both the JIMWLK and  DGLAP cascades  are dominated by short distance emissions, and hence give rise to identical transverse logarithm.  Nevertheless it is both interesting and important to disentangle the effects of the two couplings. Therefore we will  first discuss  the running of $\alpha_s$ in the resummed BFKL kernel, and then include the running in the DGLAP resummation equation.

\subsection{Running of the coupling in the BFKL kernel only}\label{sec:RC-kernel}

Our goal now is to understand how the scale-dependence of the overall factor $\alpha_s(\T{q}^2)$ in the LLA-resummed BFKL kernel (\ref{eq:K-BFKL-res-LLA})  manifests itself in the generalised characteristic function (\ref{eq:chi_rho-rho_DEF}):
%  while temporarily ignoring the running  in the resummation equations for $R_{Q}^{(j)}$.
% From (\ref{eq:chi_rho-rho_DEF}),  the generalized characteristic function is
\begin{eqnarray}
    \chi^{\text{(res.)}}(n,\gamma,a_s)&=& -\frac{\pi}{N_c \alpha_s(\T{k}^2)} \int\limits_{\T{q}} \widetilde{K}_{\text{BFKL}}^{\text{(res.)}}(\T{k},\T{q}) \alpha_s(\T{q}^2) \bigg( \frac{\T{q}^2}{\T{k}^2}\bigg)^\gamma e^{in(\phi_{\T{q}}-\phi_{\T{k}})} \nonumber \\
    &=& \big[1+a_s(\T{k}^2)\beta_0 \partial_\gamma \big]^{-1} \widetilde{\chi}^{\text{(res.)}}(n,\gamma,a_s), \label{eq:chi_as(q)_oper}
\end{eqnarray}
where  the kernel  $\widetilde{K}_{\text{BFKL}}^{\text{(res.)}}$ is defined in (\ref{eq:K-BFKL-res-LLA}) and $\widetilde{\chi}^{\text{(res.)}}(n,\gamma,a_s)$ is the corresponding characteristic function, 
\begin{eqnarray}
   \label{chtilde}
   \widetilde{\chi}^{\text{(res.)}}(n,\gamma,a_s)= -\frac{\pi}{N_c } \int\limits_{\T{q}} \widetilde{K}_{\text{BFKL}}^{\text{(res.)}}(\T{k},\T{q}) \bigg( \frac{\T{q}^2}{\T{k}^2}\bigg)^\gamma e^{in(\phi_{\T{q}}-\phi_{\T{k}})}. 
\end{eqnarray}
In the second line of \eqref{eq:chi_as(q)_oper} we  used the one-loop formula \eqref{eq:as-Q2}} for $\alpha_s(\T{q}^2)$  with $\Lambda^2\to \T{k}^2$, and  represented the factors of $\ln (\T{q}^2/\T{k}^2)$ in the operator form.
 
%In the second line of eqn.~(\ref{eq:chi_as(q)_oper}) we have expressed the factor $\alpha_s(\T{q}^2)$ in the operator form (\ref{eq:RC-op-action}). 

{It is} useful to introduce a Cauchy-type integral representation for  the action of the running coupling operator {of} eqn.~(\ref{eq:chi_as(q)_oper}) on an arbitrary function $f(\gamma)$:
\begin{equation}
    \big[ 1 + a_s \beta_0 \partial_\gamma \big]^{-1} f(\gamma) = \int\limits_{C_{\bar{\gamma}}} \frac{d\bar{\gamma}}{2\pi i}\, \Gtil(0,\bar{\gamma}-\gamma,a_s) f(\bar{\gamma}), \label{eq:RC-op-action}
\end{equation}
{where}
%\old{with the following function, which will appear several times in the results of the present paper:}
\begin{eqnarray}
     \Gtil(s,\gamma_0,a_s) &=&  \int\limits_0^{\infty} dL\; e^{-\gamma_0 L} \big(1+a_s\beta_0 L \big)^{s-1} \label{eq:Gtil-int-def} = \gamma_0^{-s}(a_s\beta_0)^{s-1} e^{\gamma_0/(a_s \beta_0)} \Gamma\bigg(s,\frac{\gamma_0}{a_s\beta_0} \bigg)\,. \label{eq:Gtil-inc-Gamma}
\end{eqnarray}
{Here} $\Gamma(s,t_0)=\int\limits_{t_0}^\infty dt\; t^{s-1} e^{-t}$  {is} an incomplete $\Gamma$-function. The contour $C_{\bar{\gamma}}$ in eqn.~(\ref{eq:RC-op-action}) encircles the {branch} cut of the function $\Gtil(0,\bar{\gamma}-\gamma,a_s)$ in the $\bar{\gamma}$-plane. The cut originates at the branch point $\bar{\gamma}=\gamma$ and extends towards negative values of $\bar{\gamma}$. {Provided that the analytic structure of $f(\bar{\gamma})$ permits it}, the {contour} $C_{\bar{\gamma}}$ can be deformed into a  Mellin contour {running} parallel to the imaginary axis.  To prove the equality (\ref{eq:RC-op-action}) one can use the following asymptotic (but Borel-summable) expansion of the function (\ref{eq:Gtil-int-def}) in {powers of} $a_s$, 
\begin{equation}
    \Gtil(s,\gamma_0,a_s)=\frac{1}{\gamma_0}\sum\limits_{n=0}^\infty \bigg( \frac{-a_s \beta_0}{\gamma_0}\bigg)^n \frac{\Gamma(1-s+n)}{\Gamma(1-s)}. \label{eq:Gtil-series}
\end{equation}
%The representation (\ref{eq:Gtil-int-def}) also allows one to prove\footnote{The identity can be easily proven if one assumes the straight contour $C_{\bar{\gamma}}$, parallel to the imaginary axis, uses the integral representation (\ref{eq:Gtil-int-def}) and integrates-out $\bar{\gamma}$ first, to obtain the delta function in $L$. } the following useful identity for the functions $\Gtil$:
%\begin{equation}
%    \int\limits_{C_{\bar{\gamma}}} \frac{d\bar{\gamma}}{2\pi i} \Gtil(s_1,\gamma_1-\bar{\gamma},a_s) \Gtil(s_2,\bar{\gamma}-\gamma_2,a_s)  = \Gtil(s_1+s_2-1,\gamma_1-\gamma_2). \label{eq:Gtil-property}
%\end{equation}
The value of  $\Gtil$ at $\gamma_0=0$,
\begin{equation}
  \Gtil(s,0,a_s)=-\frac{1}{a_s \beta_0 s}, \label{eq:Gtil_gamma0=0}  \,
\end{equation}
 is useful for evaluation of the characteristic function at $\gamma=1$.

 The characteristic function $\widetilde\chi$ 
has two contributions: the LO $\chi_0$ and 
the term due to the resummation.
In the anti-collinear limit, the LO BFKL characteristic function 
 $$\chi_0(n,\gamma\rightarrow 1+n/2)
 \rightarrow \frac{1}{(1-\gamma+n/2)}.$$ 
The operator (\ref{eq:RC-op-action}) acts on a simple pole function as follows: 
\begin{eqnarray}
    \big[1+a_s \beta_0 \partial_\gamma \big]^{-1} \frac{1}{\gamma-\gamma_0} &=& -\Gtil(0,\gamma_0-\gamma,a_s) \label{eq:as(q2)-action-pole}\\
    &=&  \frac{1}{a_s\beta_0}\bigg(\gamma_E - \ln\frac{a_s\beta_0}{\gamma_0-\gamma} \bigg) + O((\gamma_0-\gamma)^0),\label{eq:as(q2)-action-pole-01}
\end{eqnarray}
where in the second line we have expanded the function near $\gamma=\gamma_0$. 
%The effect of the  running of the coupling on the left-most anti-collinear pole of $\chi_0$, in the anti-collinear limit, can be thus expressed in the following form,
%\begin{eqnarray}
%\big[1+a_s \beta_0 \partial_\gamma \big]^{-1}\chi_0(n,\gamma)\simeq \chi_0(n,\gamma) - \frac{1}{1-\gamma+n/2} +\Gtil(0,1-\gamma+n/2,a_s) \equiv \overline{\chi}_0(n,\gamma,a_s). \nonumber \\ \label{eq:chi0+as(q)}
%\end{eqnarray}
The anti-collinear poles of the function $\chi_0(n,\gamma)$ located at $\gamma=l+n/2$, for all integer $l>1$
 correspond to ``higher twist'',   power suppressed terms in the kernel $K_{\text{BFKL}}\sim (\T{k}^2/\T{q}^2)^{n/2+l}/\T{q}^2$   at $\T{q}^2\gg \T{k}^2$. In our derivation of eqn.~(\ref{eq:K-BFKL-res-LLA}) we  systematically neglect{ed} such  terms. Hence a resummation for these poles in the characteristic function would exceed {the} accuracy {of our calculation}. Consequently in the expressions below we only modify the leftmost ($\gamma\rightarrow 1+n/2$) anti-collinear pole {of} the LO BFKL characteristic function.

As long as the running of $\alpha_s$ in the resummation equation is {neglected},  the solution for $R^{(2,1)}_Q$ {is given by}  eqn.~(\ref{eq:R2-LLA-FC}), and the corresponding characteristic function $\widetilde{\chi}$ coincides with the function $\chi^{\text{(anti-coll,LLA)}}_{+}(n,\gamma)$  obtained in  eqn.~(3.6) of Ref.~\cite{Kovner:2026qfm}. {Using} eqn.~(\ref{eq:as(q2)-action-pole}) one can {construct} the corresponding resummed expression {that} accounts for the running of the coupling in the BFKL kernel. For $n=0$ we obtain
\begin{eqnarray}
     \chi_+^{\text{(FC-LLA+$\alpha_s$)}}(0,\gamma,a_s) &=&%\new{ \big[1+a_s \beta_0 \partial_\gamma \big]^{-1} \tilde\chi} = 
     %\overline{\chi}_0(0,\gamma)-\Gtil(0,1-\gamma,a_s)   \nonumber \\
     \chi_0(0,\gamma)-\frac{1}{1-\gamma} \nonumber \\
    &+& h(\lambda_+,\lambda_-) \Gtil(0,1-a_s\lambda_- -\gamma,a_s) + (\lambda_-\leftrightarrow\lambda_+). \label{eq:chi+_FC-LLA+as(q)}
\end{eqnarray}
Here the superscript ``FC-LLA''
is for fixed coupling LLA DGLAP resummation,
while ``+$\alpha_s$'' indicate running in the BFKL kernel. The subscript ``+'' refers to the "+"-scheme. The value of the characteristic function in (\ref{eq:chi+_FC-LLA+as(q)}) at $\gamma=1$ for $n_F=0$
\begin{equation}
    \chi_+^{\text{(FC-LLA+$\alpha_s$)}}(0,1,a_s) = \frac{\pi}{\alpha_s N_c} \frac{12}{11}e\Gamma(0,1), \label{eq:chi_FC-LLA+as(q)_gamma=1}
\end{equation}
where $e\Gamma(0,1)\simeq 0.596$. 

The value in \eqref{eq:chi_FC-LLA+as(q)_gamma=1} is smaller compared to the value obtained 
in  Ref.~\cite{Kovner:2026qfm} in the fixed coupling case. This behaviour is expected. 
As discussed  in Ref.~\cite{Kovner:2026qfm}, the resummation shifts the pole of the leading order characteristic function from  $\gamma=1$ to  $\gamma=1+O(\alpha_s)$, thus making its  value  at $\gamma=1$ finite.  When running of the coupling in the kernel is turned on, it converts the pole into a weaker  {(logarithmic)} singularity (see eqn.~(\ref{eq:as(q2)-action-pole-01})),  further suppressing the value of the characteristic function at $\gamma=1$. The behaviour of $ \chi_+^{\text{(FC-LLA+$\alpha_s$)}}(0,\gamma,a_s)$ as a function of $\gamma$ is displayed in
Fig. \ref{fig:chi-res_plot}.

For $n>0$ the anti-collinear resummation does not have an effect ~\cite{Kovner:2026qfm} and we take into account only the effect of the running coupling on the left-most anti-collinear pole. This leads to:
\begin{eqnarray}
    \chi_+^{\text{(FC-LLA+$\alpha_s$)}}(n>0,\gamma,a_s) = %\overline{\chi}_0(n,\gamma,a_s).
    \chi_0(n,\gamma) - \frac{1}{1-\gamma+n/2} +\Gtil(0,1-\gamma+n/2,a_s) . \label{eq:chi0+as(q)}
\end{eqnarray}

\subsection{Including running of the coupling in the DGLAP resummation equation}\label{sec:RC-resumm}
 We now include the running of the coupling also  in the DGLAP resummation, and investigate its effect on the generalized characteristic function.
To this end  %obtain the characteristic function, which takes into account the running of $\alpha_s$ both in the BFKL kernel and in the anti-collinear resummation, 
we substitute the running coupling solution (\ref{eq:R2-LLA}) into the kernel (\ref{eq:K-BFKL-res-LLA}) and then use the definition of the momentum-space characteristic function (\ref{eq:chi_rho-rho_DEF}). The computation involves the following integral:
\begin{eqnarray}
     \int\limits_{\T{k}^2}^{\infty} \frac{d\T{q}^2}{(\T{k}^2)^{\gamma-1}} \left( \frac{a_s(\T{k}^2)}{a_s(\T{q}^2)} \right)^{\lambda/\beta_0-1 } (\T{q}^2)^{\gamma-2} = \Gtil (\lambda/\beta_0,1-\gamma,a_s(\T{k}^2)),
\end{eqnarray}
whith the function $\Gtil$ defined in (\ref{eq:Gtil-int-def}). 
The final result for the characteristic function reads:
\begin{eqnarray}
     \chi_+^{\text{(LLA+$\alpha_s$)}}(0,\gamma,a_s)%= \int\limits_{C_{\bar{\gamma}}} \frac{d\bar{\gamma}}{2\pi i} \Gtil(0,\bar{\gamma}-\gamma,a_s) \, \widetilde{\chi}_+^{\text{(LLA)}}(0,\bar{\gamma},a_s) \nonumber \\
   %&&
   &=& %\overline{\chi}_0(0,\gamma,a_s)-\Gtil(0,1-\gamma,a_s) \nonumber \\
   \chi_0(0,\gamma)-\frac{1}{1-\gamma} \nonumber \\
   &+& h(\lambda_+,\lambda_-)\Gtil(\lambda_-/\beta_0,1-\gamma,a_s) + (\lambda_+ \leftrightarrow \lambda_-). \label{eq:chi_LLA+as(q)}
\end{eqnarray}

Remarkably, computing the  characteristic function (\ref{eq:chi_LLA+as(q)}) at $\gamma=1$ using  eqn.~(\ref{eq:Gtil_gamma0=0}),  we obtain exactly the same values as found in Ref.~\cite{Kovner:2026qfm}, in the fixed-coupling case:
\begin{eqnarray}
    \chi_+^{\text{(LLA+$\alpha_s$)}}(0,1,a_s)=\frac{\pi}{\alpha_s N_c}\left\{\begin{array}{cc}
       4/3  & \text{for }n_F\neq 0, \\
       12/11  & \text{for }n_F=0.
    \end{array}  \right. \label{eq:chi_LLA+as(q)_gamma=1}
\end{eqnarray}
The general trend here {follows our naive expectations}: switching on the running of the coupling in the DGLAP evolution slows it down and therefore weakens the effect of the resummation. That is the running in the DGLAP effectively works in the direction opposite to the one in the BFKL.
However the exact cancellation between the running coupling effects in the BFKL equation and in DGLAP resummation is quite surprising. It remains unclear to us whether this cancellation is purely accidental or  there exists a deeper reason behind it. We note that the exact cancellation happens  at $\gamma=1$ only, while at $\gamma<1$ the  cancellation is partial, 
as will be demonstrated below.

The perturbative expansion of eqn.~(\ref{eq:chi_LLA+as(q)}) has the form:
\begin{eqnarray}
   \chi_+^{\text{(LLA+$\alpha_s$)}}(0,\gamma,a_s)&=&\chi_0(0,\gamma) -(a_sN_c)\frac{2 \left(11 N_c^3-N_c^2 n_F+n_F\right)}{3 (\gamma -1)^2
   N_c^3} \label{eq:chi_LLA+as(q)_exp-as} \\
   &+&(a_sN_c)^2\frac{ \left(-726 N_c^4-8 \left(N_c^2-1\right)
   n_F^2+2 \left(75 N_c^2-53\right) N_c n_F\right)}{9 (\gamma -1)^3
   N_c^4}+O(a_s^3)\,. \nonumber
\end{eqnarray}
The $O(\alpha_s)$-term completely agrees with the $1/(1-\gamma)^2$-term of the NLO BFKL characteristic function in the ``+''-scheme (see, e.g. eqns.~(3.8) -- (3.10) of Ref.~\cite{Kovner:2026qfm}). The $O(\alpha_s^2)$-term in eqn.~(\ref{eq:chi_LLA+as(q)_exp-as}) constitutes the prediction of our formalism for the $1/(1-\gamma)^3$-term of the NNLO BFKL characteristic function in the ``+''-scheme. Yet, as we will further comment in Sec.~\ref{sec:comp-Salam-Stasto}, we do not expect this prediction to be accurate as this calculation misses certain effects, which are not taken into account in our present formalism.

\begin{figure}
    \centering
    \includegraphics[width=0.65\linewidth]{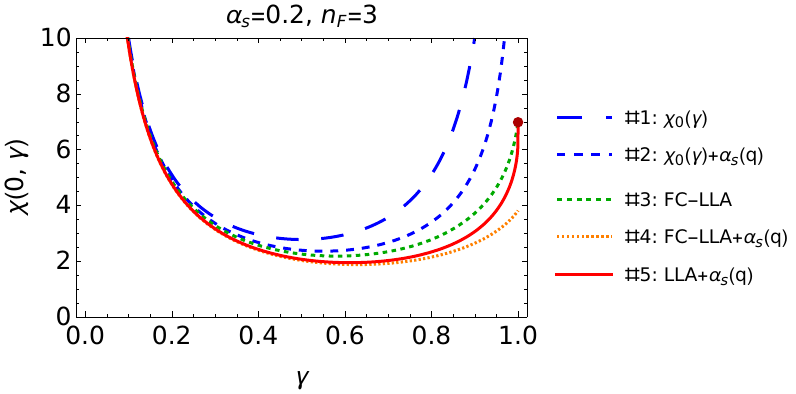}
    \caption{Momentum-space characteristic functions for $n=0$ with the anti-collinear resummation and running coupling,  $\alpha_s(\T{k}^2)=0.2$ and $n_F=3$. The curves: 1) eqn.~(\ref{eq:chi0});  2) eqn.~(\ref{eq:chi0+as(q)}) taken at $n=0$; 3) eqn.~(3.6) of Ref.~\cite{Kovner:2026qfm}; 4) eqn.~(\ref{eq:chi+_FC-LLA+as(q)}); 5) the compete running coupling + anti-collinear resummation result of eqn.~(\ref{eq:chi_LLA+as(q)}). The value $4\pi/(3\alpha_s N_c)$ at $\gamma=1$ is marked by the dot.}
    \label{fig:chi-res_plot}
\end{figure}

The influence of various types of anti-collinear LLA corrections on the BFKL characteristic function is illustrated  in Fig.~\ref{fig:chi-res_plot}. {Curve \#1} shows the LO BFKL characteristic function $\chi_0$ {and serves as a} reference. Comparing the curves \#1 and \#2 illustrates the softening of the anti-collinear pole singularity due to the inclusion of the factor $\alpha_s(\T{q}^2)$ in the BFKL kernel. The curve \#3 is our fixed-coupling anti-collinearly resummed result {from} Ref.~\cite{Kovner:2026qfm}. The curve \#4 is obtained from {curve \#3}  by taking into account the effect of $\alpha_s(\T{q}^2)$ in the BKFL kernel. Finally, the curve \#5 includes both the effect of $\alpha_s(\T{q}^2)$ in the kernel and  of $\alpha_s(Q^2)$ in the anti-collinear resummation equations (\ref{eq:eqn-SQ-quarks}) and (\ref{eq:eqn-VQ}). The dot in the plot of Fig.~\ref{fig:chi-res_plot} corresponds to the value (\ref{eq:chi_FC-LLA+as(q)_gamma=1}) of the characteristic function at $\gamma=1$, which is {common to} the curves \#3 and \#5.  

\subsection{Comparison with the CCSS RG-improved approach}\label{sec:comp-Salam-Stasto}

In Ref.~\cite{Ciafaloni:2003rd} (eqn. 15) the following expression for the symmetric-scheme BFKL kernel with the collinear ($\T{k}^2\gg \T{q}^2$) and anti-collinear ($\T{q}^2\gg \T{k}^2$) resummation was proposed by Ciafaloni, Colferai, Salam and Stasto (CCSS):
\begin{eqnarray}
 \label{eq:Salam-Stasto}
 K_{S}(\T{k},\T{q},\omega) &=& -\frac{\alpha_s(\T{k}^2) N_c}{\pi^2} \frac{\T{k}^2}{\T{q}^2}\bigg[\frac{1}{\T{k}^2} \left(\frac{|\T{q}|}{|\T{k}|}\right)^{\omega}
 \left(\frac{\alpha_s(\T{k}^2)}{\alpha_s(\T{q}^2)}\right)^{-\frac{4 A_1(\omega)}{\beta_0}} \theta(\T{k}^2-\T{q}^2)
\;  + \nonumber \\
&&  + \; \frac{1}{\T{q}^2} \left(\frac{|\T{k}|}{|\T{q}|}\right)^{\omega}
 \left(\frac{\alpha_s(\T{k}^2)}{\alpha_s(\T{q}^2)}\right)^{\frac{4 A_1(\omega)}{\beta_0}-1} \theta(\T{q}^2-\T{k}^2) \bigg] \,,
\end{eqnarray}
where $\omega$ is the Mellin variable conjugate to the rapidity $Y$. 
Here
\begin{eqnarray}
      A_1(\omega)&=&\frac{1}{2}\gamma_{gg}(\omega)- \frac{N_c}{\omega} \label{eq:A1-Salam-def},\, %\quad \text{ for }\,n_F=0,
      \label{eq:A1-Salam-exp}
\end{eqnarray}
where $\gamma_{gg}(\omega)$ is  the standard DGLAP anomalous dimension of the gluon:
\begin{equation}
     \gamma_{gg}(\omega)=2N_c\bigg(\frac{1}{\omega}-\frac{1}{\omega+1} +\frac{1}{\omega+2} - \frac{1}{\omega+3} - H_{\omega+1}\bigg)+\frac{\beta_0}{2}\,. \label{eq:gamma_gg-00}
 \end{equation}
 Where $H_n=\sum\limits_{k=1}^n 1/k={\psi(n+1)-\psi(1)}$ is the harmonic sum.  
For $n_F=0$, $A_1$ at small $\omega$ is expressible in terms of the eigenvalue $\lambda_-$ defined in ~(\ref{eq:lambda-PM-expr}):
\begin{equation}
    A_1(\omega, n_F=0)= -\frac{11 N_c}{12} + O(\omega)= \frac{\lambda_-}{4}+O(\omega).
\end{equation}
A distinguishing feature of the approach of Ref.~\cite{Ciafaloni:2003rd} is the dependence of the kernel on the variable $\omega$, in contrast to the standard BFKL formalism, where the kernel is independent of rapidity. This $\omega$-dependence of the kernel can be understood as encoding an infinite resummation of higher-order corrections to the standard BFKL kernel, which was the original motivation behind the construction of Ref.~\cite{Ciafaloni:2003rd}.

There are two sources of $\omega$-dependence in eqn.~(\ref{eq:Salam-Stasto}){, each} with {a distinct} physical meaning. The first one is in the factors $(|\T{k}|/|\T{q}|)^{\pm \omega}$, which are responsible for the resummation of the leading (double-logarithmic) series of the collinear and anti-collinear poles of the symmetric-scheme characteristic function $\chi_S(0,\gamma)$ ($\propto \alpha_s^n/\gamma^{2n+1}$ and $\propto \alpha_s^n/(1-\gamma)^{2n+1}$){. This resummation follows from }  the relation between ``$\pm$'' and symmetric rapidity-factorisation schemes (see ~\cite{Salam98} and also Appendix D of Ref.~\cite{Kovner:2026qfm}). 

The second source  is the $\omega$-dependence of $A_1(\omega)$. One can show that{, starting from NNLO,}  this $\omega$-dependence  is important for the resummation of the collinear ($\propto \alpha_s^n/\gamma^{n+1}$) and anti-collinear ($\propto \alpha_s^n/(1-\gamma)^{n+1}$) ``DGLAP'' poles. 
These contributions to the characteristic function contain terms of highest transcendental weight $\propto \alpha_s^n \zeta(n)$ (See the Sec. 3 of Ref.~\cite{Deak:2019wms}). These terms originate from a physical mechanism which is not accounted for by our resummation of Sec.~\ref{sec:resumm-general}, particularly they are missing in the NNLO result (\ref{eq:chi_LLA+as(q)_exp-as}). The $\omega$-dependence of $A_1$ indicates that the resummed BFKL equation is {no longer} a simple {local} differential equation in rapidity, but {instead} contains some effects of retardation. We believe that these {effects originate from} {the finite rapidity extent of} the DGLAP cascade. We intend to study this issue in a separate publication. 

Nevertheless even {after} setting $\omega=0$, eqn.~(\ref{eq:Salam-Stasto}) constitutes a non-trivial resummation. It can {therefore} be {meaningfully} compared with our result. The second term in ~(\ref{eq:Salam-Stasto}) corresponding to the anti-collinear limit, for $n_F=0$  can be rewritten as:
\begin{equation}
  K_{+}^{\text{(anti-coll.)}}(\T{k},\T{q},0)= -\frac{\alpha_s(\T{q}^2)N_c}{\pi^2} \frac{\T{k}^2}{\T{q}^4} \bigg(\frac{a_s(\T{q}^2)}{a_s(\T{k}^2)} \bigg)^{-\lambda_-/\beta_0} = -\frac{\alpha_s(\T{q}^2)N_c}{\pi^2} \frac{\T{k}^2}{\T{q}^4} R_{|\T{k}|}^{(2,1)}(\T{q}), \label{eq:K-anti-coll-from-RG}  
\end{equation}
where the subscript $(+)$ indicates the "+"-scheme: that is, compared to (\ref{eq:Salam-Stasto}), the factor $(|\T{k}|/|\T{q}|)^{\omega}$ 
is dropped. In eqn.~(\ref{eq:K-anti-coll-from-RG}) one recognizes the contribution of the resummation  function $R^{(2,1)}_Q(\T{p})$ to the resummed BFKL kernel (\ref{eq:K-BFKL-res-LLA}). Thus for $n_F=0$ our resummation formalism {reproduces precisely the $\omega=0$}  contribution in the formalism of Ref.~\cite{Ciafaloni:2003rd}. For $n_F\ne 0$, we believe that in our formalism the $n_F$-dependence of the resummation is under better control than that in Ref.~\cite{Ciafaloni:2003rd}: in our resummation the $n_F$ dependence  does not simply {reduces to the fermionic}  contribution  to the $\beta$-function {alone,} but {also} includes the genuine {dynamics} of fermion{-induced} DGLAP splittings.

So far we have explored the effect of resummation and running coupling on the generalized characteristic function.  As discussed above,
 the generalized characteristic function $\chi(n,\gamma)$ is obtained by acting with the evolution kernel on a class of test functions. While
these test functions are eigenfunctions of the LO BFKL kernel at fixed coupling, they are no longer  eigenfunctions when the coupling {runs}. The question then arises {of} how {to} calculate physical observables (cross sections) from the knowledge of $\chi(n,\gamma)$. We address this question in the next section.

\section{BFKL Green's function with running coupling and anti-collinear resummation}
\label{BFKLGF}

A classic approach to computing physical cross sections involves a convolution of process-dependent impact factors with the BFKL Green's function. The goal of this Section is to construct a practically useful and, to the best of our knowledge, new representation of the Green's function, which takes into account both  the running of the coupling and anti-collinear resummation.  

In  Sec.~\ref{sec:GF-solution} we compute the BFKL Green's function, starting from the generalised characteristic function $\chi^{\text{(L)}}(n,\gamma,a_s)$ defined in eqn.~(\ref{eq:chi_rho-rho_DEF}), including the running coupling effects. Then in Sec.~\ref{sec:saddle}, using a saddle point approximation we explore 
how the anti-collinear resummation of Sec.~\ref{sec:resumm-general}  affects the result. 

\subsection{BFKL Green's function from the generalized characteristic function}\label{sec:GF-solution}

The BFKL equation with running coupling has been  well studied (see e.g. Ref.~\cite{Forshaw:1997dc} for an early review on the subject). For example, in the discrete Pomeron approach~\cite{Lipatov:1985uk,Ellis:2008yp,Kowalski:2010ue,Kowalski:2015paa,Kowalski:2017umu} 
 one expresses the Green's function as a sum over a discrete set of eigenfunctions of the operator $\alpha_s(\T{k}^2)\chi_0(0,\partial/\partial\ln\T{k}^2)$.  While well developed, this approach has not been studied in conjunction with the (anti-)collinear resummations.

In the present paper,  we have focused so far on  computation of the generalized characteristic function defined by eqn.~(\ref{eq:chi_rho-rho_DEF}). It takes into account both the anti-collinear resummation and running coupling effects. When the coupling runs, there exists no straightforward formalism, unlike in the fixed coupling case,  to construct the Green's function from the characteristic function. Below we report on a progress in building such a formalism.

The BFKL Green's function $G$ is defined in terms of the solution of the BFKL equation (\ref{eq:BFKL-mom}) for the $\langle \rho \rho \rangle$-correlator, rescaled by a factor of transverse momentum squared\footnote{Conventionally one defines $G(Y,\T{k};\T{p})\equiv \frac{1}{\T{k}^2}\langle\rho^a(\T{k})\rho^a(-\T{k})\rangle$, where $\T{p}$ refers to a transverse scale in the initial condition.}:
\begin{equation}
    \frac{\partial}{\partial Y} G(Y,\T{k}; \T{p}) = -\int\limits_{\T{q}} K^{\text{(L)}}_{\text{BFKL}}(\T{k},\T{q}) \frac{\T{q}^2}{\T{k}^2} G(Y,\T{q}; \T{p}),\label{eq:BFKL-eqn-G}
\end{equation}
with the initial condition at $Y=0$: $G(0,\T{k};\T{p})=\delta^{(2)}(\T{k}-\T{p})$.  In this subsection, our discussion is going to be pretty general, so that it is valid for any approximation for the BFKL kernel considered in the present paper (indicated by the superscript ${}^{\text{(L)}}$). To make a connection with our physical picture in the JIMWLK case, we will refer to the momentum $\T{p}$ as a scale in the target ($Q_T$). %The rapidity $Y\sim \ln P^+$ is increasing from the target to the projectile. 

Even though the eigenfunctions of the LO BFKL equation are not eigenfunctions of the kernel with the running coupling, we can still use them as a basis for expansion of the Green's function:
\begin{eqnarray}
    G(Y,\T{k};\T{p})&=& \frac{1}{\T{k}^2}\sum\limits_{n=-\infty}^{\infty} \int\limits_{1/2-i\infty}^{1/2+i\infty} \frac{d\gamma}{2\pi^2 i} \bigg(\frac{\T{k}^2}{\T{p}^2} \bigg)^\gamma e^{in\phi_{\T{k}\T{p}}} \, G_{n}(Y,\gamma,\alpha_s(\T{p}^2)), \label{eq:G-conf-exp_(-)}
\end{eqnarray}
%Note that $\T{p}^2$ is the external scale which enters the RHS of \eqref{eq:G-conf-exp_(-)} via the initial condition.

 At $Y=0$ the initial condition is  $G_{n}(0,\gamma,\alpha_s)=1$.  Substituting eqn.~(\ref{eq:G-conf-exp_(-)}) into eqn.~(\ref{eq:BFKL-eqn-G}) and using the definition of the generalised characteristic function (\ref{eq:chi_rho-rho_DEF}), one shows that $G_n(Y,\gamma,\alpha_s(\T{p}^2))$ satisfies:
\begin{equation}
    \frac{\partial}{\partial Y}G_n(Y,\gamma,\alpha_s(\T{p}^2)) = \frac{\alpha_s(-\partial_\gamma,\T{p}^2) N_c}{\pi} \chi^{\text{(L)}}(n,\gamma,a_s(-\partial_\gamma,\T{p}^2)) G_n(Y,\gamma,\alpha_s(\T{p}^2))\, \label{eq:BFKL_G-gamma:targ}
\end{equation}
where we have again expressed $\alpha_s$ in the operator form using  the one-loop expression \eqref{eq:as-Q2}:
\begin{equation}
    \alpha_s(-\partial_\gamma,\T{p}^2) \equiv \alpha_s(\T{p}^2) \big[1-a_s(\T{p}^2)\beta_0\partial_\gamma \big]^{-1}. \label{eq:alpha_s-op-def}
\end{equation}
In (\ref{eq:BFKL_G-gamma:targ}), $\chi^{\text{(L)}}$ has been promoted into 
a function of the operator $a_s(-\partial_\gamma,\T{p}^2)\equiv \alpha_s(-\partial_\gamma,\T{p}^2)/(4\pi)$. It should be understood through the expansion (\ref{eq:chi_exp}), with the same ordering of the factors $a_s(-\partial_\gamma,\T{p}^2)$ and $\chi_m^{\text{(L)}}(n,\gamma)$.  
  
A formal solution of  eqn.~(\ref{eq:BFKL_G-gamma:targ}) reads
\begin{equation}
    G_n(Y,\gamma,\alpha_{s}(\T{p}^2)) = \exp\bigg[ Y \frac{\alpha_s(- \partial_\gamma,\T{p}^2) N_c}{\pi} \chi^{\text{(L)}}(n,\gamma,a_s(-\partial_\gamma,\T{p}^2)) \bigg] 1,\label{eq:G(+-)-sol-00}
\end{equation}
where the right-most factor of $1$ is annihilated by any $\partial_\gamma$-derivatives acting on it. 

In eqns. \eqref{eq:RC-op-action}
and \eqref{eq:Gtil-inc-Gamma}, we have 
shown how the operator $a_s(-\partial_\gamma,\T{p}^2)$ acts on an 
arbitrary function $f(\gamma)$. Now we provide an alternative representation of this operation and also generalize it for
powers of $a_s(-\partial_\gamma,\T{p}^2)$, which will be useful later (for the reminder of this section, $\alpha_s=\alpha_s(\T{p}^2)$ if not stated otherwise):
\begin{eqnarray}
    \big( \alpha_s(-\partial_\gamma,\T{p}^2) \big)^n f(\gamma) &=& \int\limits_0^{\infty} d\tau\, \frac{\tau^{n-1}}{(n-1)!} \exp[-\tau (1-a_s\beta_0\partial_\gamma)] f(\gamma) \nonumber \\
    &=&\int\limits_0^{\infty} d\tau\,e^{-\tau} \frac{\tau^{n-1}}{(n-1)!}  f(\gamma+\tau a_s\beta_0 ). 
    \label{an}
\end{eqnarray}
 With the help of this representation the solution (\ref{eq:G(+-)-sol-00}) can be rewritten in the following form:
\begin{eqnarray}
    G_n(Y,\gamma,\alpha_{s}) &=& \exp\bigg[ Y\frac{\alpha_s N_c }{\pi} \int\limits_0^{\infty}d\tau\, e^{-\tau} e^{\tau a_s \beta_0\partial_\gamma} \overline{\chi}^{\text{(L)}}(n,\gamma,a_s\tau) \bigg]\, 1 \label{eq:Gn(Y,gamma)-sol-00}   \\
    &=& \sum\limits_{k=0}^{\infty} \frac{1}{k!} \bigg(\frac{Y \alpha_s N_c }{\pi} \bigg)^k \int\limits_0^{\infty} d\tau_1\ldots  d\tau_k\, e^{-\tau_1-\ldots-\tau_k} \, \overline{\chi}^{\text{(L)}}(n,\gamma+a_s\beta_0\tau_1,a_s\tau_1)% \label{eq:Gn(Y,gamma)-sol-000}
    \nonumber  \\
    & \times & \overline{\chi}^{\text{(L)}}(n,\gamma+a_s\beta_0(\tau_1+\tau_2),a_s\tau_2)\ldots \overline{\chi}^{\text{(L)}}(n,\gamma+a_s\beta_0(\tau_1+\ldots +\tau_k),a_s\tau_k),\nonumber \\
    & = &\sum\limits_{k=0}^{\infty} \frac{1}{k!} \bigg(\frac{Y \alpha_s N_c }{\pi} \bigg)^k \int\limits_0^{\infty} d\tau_k\int\limits_0^{\tau_k}d\tau_{k-1}\ldots \int\limits_0^{\tau_2} d\tau_1\, e^{-\tau_k} \, \overline{\chi}^{\text{(L)}}(n,\gamma+a_s\beta_0\tau_1,a_s\tau_1) \nonumber \\
    & \times  &\overline{\chi}^{\text{(L)}}(n,\gamma+a_s\beta_0\tau_2,a_s(\tau_2-\tau_1))\ldots \overline{\chi}^{\text{(L)}}(n,\gamma+a_s\beta_0\tau_k,a_s(\tau_k-\tau_{k-1})). \nonumber
\end{eqnarray}
Where we have introduced a Borel transform $  \overline{\chi}^{\text{(L)}}(n,\gamma,a)$ of the perturbative expansion (\ref{eq:chi_exp}):
\begin{equation}
    \overline{\chi}^{\text{(L)}}(n,\gamma,a) \equiv\text{BT}[{\chi}^{\text{(L)}}(n,\gamma,a)]
    \equiv
    \chi_0(n,\gamma) + \sum\limits_{m=1}^{\infty} \frac{(a N_c)^m}{m!}\chi_m^{\text{(L)}}(n,\gamma)\,.\label{eq:chi-Borel}
\end{equation}
Note that in eqn.~(\ref{eq:Gn(Y,gamma)-sol-00}), $a=a_s\tau_i$ or $a_s(\tau_i-\tau_{i-1})$.

 We note that the Borel transform has appeared here not as a tool to sum a divergent series,  but rather as a means to extract the operator $a_s(-\partial_\gamma,\T{p}^2)$ from the argument of the function $\chi^{\text{(L)}}$ in eqn.~(\ref{eq:G(+-)-sol-00}).  Eqn. (\ref{eq:Gn(Y,gamma)-sol-00}) is applicable to the main results of the present paper, particularly to (\ref{eq:chi_LLA+as(q)})
 for the generalized characteristic function. The latter  contains the function $\Gtil$ whose Borel transform is:
 \begin{equation}
    \bar{g}(s,\gamma_0,a)\equiv \text{BT}[{g}(s,\gamma_0,a)]=\gamma_0^{-s}\big(\gamma_0+a\beta_0 \big)^{s-1}.
\end{equation}

In the following  subsections we derive two approximate, closed-form expressions for the series (\ref{eq:Gn(Y,gamma)-sol-00}). In Sec.~\ref{sec:Gn-closed-form-no-as-running-in-HO}, as a warm-up, we take $\chi^{\text{(L)}} = \chi_0$ and obtain a closed-form solution of the LO BFKL equation, only with the running coupling $\alpha_s(\T{k}^2)$ in the kernel. In  Sec.~\ref{sec:anti-coll-LLA-G} we derive  an expression for the series (\ref{eq:Gn(Y,gamma)-sol-00}) with the characteristic function given in eqn.~(\ref{eq:chi_LLA+as(q)}),  applicable in the anti-collinear regime. For $n=0$, its Borel transform   takes the form:
\begin{eqnarray}
    \overline{\chi}_{+}^{\text{(LLA+$\alpha_s$)}}(0,\gamma,a)&=& \chi_0(0,\gamma) - \frac{1}{1-\gamma} + \Delta \overline{\chi}^{\text{(LLA+$\alpha_s$)}}(\gamma,a), \label{barchi} \\
\Delta \overline{\chi}^{\text{(LLA+$\alpha_s$)}}(\gamma,a) &=& h(\lambda_+,\lambda_-) \bar{g}(\lambda_-/\beta_0,1-\gamma,a) + (\lambda_+\leftrightarrow \lambda_-).\label{eq:DeltaChi-check-res+as}
\end{eqnarray}

\subsubsection{{LO BFKL Green's function with running coupling $\alpha_s(\T{k}^2)$}
}\label{sec:Gn-closed-form-no-as-running-in-HO}

 {Eqn.}~(\ref{eq:Gn(Y,gamma)-sol-00}) 
 expresses the BFKL Green's function in terms of the Borel transformed generalized characteristic function. Yet it is impractical to use without further approximations. One such approximation leading  to a closed-form expression for the Green's function is to substitute $ \chi^{\text{(L)}}(n,\gamma,a_s)\to \chi_0(n,\gamma)  $, in  eqn.~(\ref{eq:G(+-)-sol-00}). This characteristic function corresponds to the BFKL kernel $\alpha_s(\T{k}^2) \widetilde{K}_{\text{BFKL}}^{\text{(LO)}}(\T{k},\T{q})$. Consequently, $\overline{\chi}^{(L)}= \chi_0$ 
 and the solution ~(\ref{eq:Gn(Y,gamma)-sol-00}) for the Green's function  becomes
 \begin{eqnarray}
     G_n^{\text{(LO+$\alpha_s$)}}(Y,\gamma,\alpha_{s})  &=& \sum\limits_{k=0}^{\infty} \frac{1}{k!} \bigg(\frac{Y \alpha_s N_c }{\pi} \bigg)^k \int\limits_0^{\infty} d\tau_k\int\limits_0^{\tau_k}d\tau_{k-1}\ldots \int\limits_0^{\tau_2} d\tau_1\, e^{-\tau_k}\nonumber \\
    & \times &\chi_0(n,\gamma+a_s\beta_0\tau_1) \chi_0(n,\gamma+a_s\beta_0\tau_2)\ldots \chi_0(n,\gamma+a_s\beta_0\tau_k) \label{eq:Gn(Y,gamma)-sol-01} \\
     &=& \sum\limits_{k=0}^{\infty} \frac{1}{(k!)^2} \bigg(\frac{Y \alpha_s N_c }{\pi} \bigg)^k \int\limits_0^{\infty} d\tau_k\int\limits_0^{\infty}d\tau_{k-1}\ldots \int\limits_0^{\infty} d\tau_1\, e^{-\max(\tau_1,\ldots,\tau_k)} \nonumber\\
    & \times& \chi_0(n,\gamma+a_s\beta_0\tau_1) \chi_0(n,\gamma+a_s\beta_0\tau_2)\ldots \chi_0(n,\gamma+a_s\beta_0\tau_k). \label{eq:Gn(Y,gamma)-sol-02} 
\end{eqnarray}
In the second equality we have used the symmetry of the integrand w.r.t. permutations of $\tau_1,\ldots,\tau_k$.
 The last equation can be simplified using the identity:
\begin{eqnarray}
    \int\limits_{0}^{\infty} d\tau_1 \ldots \int\limits_0^{\infty} d\tau_k \int\limits_{\max(\tau_1,\ldots,\tau_k)}^\infty d\tau\; e^{-\tau} f(\tau_1)\ldots f(\tau_k) = \int\limits_0^{\infty} d\tau\, e^{-\tau} \bigg( \int\limits_0^{\tau} d\tau'\, f(\tau') \bigg)^k,
\end{eqnarray}
leading to the compact closed-form expression for the series (\ref{eq:Gn(Y,gamma)-sol-02}):
\begin{equation}
    G_n^{\text{(LO+$\alpha_s$)}}(Y,\gamma,\alpha_s) = \int\limits_0^{\infty}d\tau\, e^{-\tau} {\cal I}_0\bigg[\frac{Y\alpha_s N_c}{\pi} \int\limits_0^{\tau} d\tau'\,  \chi_0(n,\gamma+a_s\beta_0\tau')  \bigg].
    \label{eq:Gn(Y,gamma)-appr-sol}
\end{equation}
The function ${\cal I}_0(x)\equiv I_0(2\sqrt{x})$ is expressed in terms of the modified Bessel function $I_0(x)$.

\subsubsection{The BFKL Green's function
with anti-collinear resummation: a closed-form solution in gluodynamics}\label{sec:anti-coll-LLA-G}

The general expression for the Green's function is given in (\ref{eq:Gn(Y,gamma)-sol-00}). In the rest of this section we will limit the discussion to $n=0$ case, that is we will compute $G_0$ only. The corresponding Borel transformed characteristic function  $\overline{\chi}^{(L)}=\overline{\chi}^{\text{(LLA+$\alpha_s$)}}_+$  given by \eqref{barchi}.
In the anti-collinear LLA, $\chi_0(0,\gamma)=1/(1-\gamma)$. 
Hence, in this approximation, $\overline{\chi}^{\text{(LLA+$\alpha_s$)}}_+= \Delta \overline{\chi}^{\text{(LLA+$\alpha_s$)}}$
introduced in eqn.~(\ref{eq:DeltaChi-check-res+as}), which in $G_0$ resums all the terms of the type $\alpha_s^m/(1-\gamma)^{m+1}$.

It turns out that  a closed form expression for $G_0$ can be found in a theory with $n_F=0$. In gluodynamics,  $\lambda_-=-\beta_0^{(g)}=-11N_c/3$ and $\lambda_+=0$, such that eqn.~(\ref{eq:DeltaChi-check-res+as}) reduces to
\begin{equation}
    \Delta \overline{\chi}^{\text{(LLA+$\alpha_s$)}}_{n_F=0}(\gamma,a) = \bar{g}(-1,1-\gamma,a).
\end{equation}
Without higher-order perturbative corrections
($a=0$), this function recovers  the anti-collinear pole of the LO characteristic function: $\bar{g}(-1,1-\gamma,0)=1/(1-\gamma)$. 
Substituting the explicit expression for $\bar{g}(-1,1-\gamma,a)$ from (\ref{eq:DeltaChi-check-res+as}), 
\begin{eqnarray}
     \hspace{-0.5cm}\Delta \overline{\chi}^{\text{(LLA+$\alpha_s$)}}_{n_F=0}(\gamma+a_s\beta_0^{(g)}\tau_i,a_s(\tau_i-\tau_{i-1})) 
     %&&= \big(1-\gamma-a_s\beta_0^{(g)}\tau_i \big) \big(1-\gamma - \cancel{a_s\beta_0^{(g)}\tau_i} + a_s\beta_0^{(g)}(\cancel{\tau_i}-\tau_{i-1}) \big)^{-2} \nonumber \\     
     =\big(1-\gamma-a_s\beta_0^{(g)}\tau_i \big) \big(1-\gamma - a_s\beta_0^{(g)}\tau_{i-1}  \big)^{-2}.
\end{eqnarray}
Hence, the product of the characteristic functions in eqn.~(\ref{eq:Gn(Y,gamma)-sol-00}) re-factorises,
\begin{eqnarray}
    &&\Delta \overline{\chi}^{\text{(LLA+$\alpha_s$)}}_{n_F=0}(\gamma+a_s\beta_0^{(g)}\tau_1,a_s\tau_1) \prod\limits_{i=2}^{k} \Delta \overline{\chi}^{\text{(LLA+$\alpha_s$)}}_{n_F=0}(\gamma+a_s\beta_0^{(g)}\tau_i,a_s(\tau_i-\tau_{i-1})) \nonumber \\
    &&= \frac{1-\gamma-a_s\beta_0^{(g)}\tau_k}{(1-\gamma)^2} \prod_{i=1}^{k-1} \big( 1-\gamma-a_s\beta_0^{(g)}\tau_i \big)^{-1}.  
\end{eqnarray}
Using the symmetry of the integrand   w.r.t. permutations of $\tau_1,\ldots,\tau_{k-1}$,
\begin{eqnarray}
  &&  \int\limits_0^{\infty} d\tau_k\int\limits_0^{\tau_k}d\tau_{k-1}\ldots \int\limits_0^{\tau_2} d\tau_1\, e^{-\tau_k}\Delta \overline{\chi}^{\text{(LLA+$\alpha_s$)}}_{n_F=0}(\gamma+a_s\beta_0^{(g)}\tau_1,a_s\tau_1)  \nonumber \\
  && \times  \prod\limits_{i=2}^{k} \Delta \overline{\chi}^{\text{(LLA+$\alpha_s$)}}_{n_F=0}(\gamma+a_s\beta_0^{(g)}\tau_i,a_s(\tau_i-\tau_{i-1})) \nonumber \\
  && =\int\limits_0^{\infty} \frac{d\tau\, e^{-\tau}}{(k-1)!}  \frac{1-\gamma-a_s\beta_0^{(g)}\tau}{(1-\gamma)^2} \bigg( \int\limits_0^{\tau} \frac{d\tau'}{1-\gamma-a_s\beta_0^{(g)}\tau'} \bigg)^{k-1}. 
\end{eqnarray}
Finally, substituting this result into ~(\ref{eq:Gn(Y,gamma)-sol-00}) and summing  over $k$, we obtain the following closed-form result for $G_0$,  valid for  $\gamma\simeq 1$:
\begin{eqnarray}
     G_{0,n_F=0}^{\text{(LLA+$\alpha_s$)}}(Y,\gamma,\alpha_s) &=& 1 \label{eq:G0-sol-nF=0-anti-coll} \\
    &&\hspace*{-2cm}+ \frac{\alpha_s N_c Y}{\pi(1-\gamma)} \int\limits_0^{\infty} d\tau\, e^{-\tau} \bigg( 1-\frac{a_s\beta_0^{(g)}\tau}{1-\gamma} \bigg) {\cal I}_1\bigg[ -\frac{4 N_c Y}{\beta_0^{(g)}} \ln\bigg(1-\frac{a_s\beta_0^{(g)}\tau}{{1-\gamma}} \bigg)\bigg],  \nonumber
\end{eqnarray}
where the function ${\cal I}_1(x)\equiv I_1(2\sqrt{x})/\sqrt{x}$ and $I_1(x)$ is the modified Bessel function. This closed-form expression contains all running-coupling and anti-collinear LLA effects in the approximation of the present paper.

In the next subsection we will glue the two approximations discussed above and extract a value for the Pomeron intercept.

\subsection{Saddle-point approximation and Pomeron intercept}\label{sec:saddle}

Our starting point is an approximation for  $G_0$: %(we drop all the super/sub-scripts):
\begin{equation}
    G_{0}(Y,\gamma,\alpha_s) = G_{0,n_F=0}^{\text{(LLA+$\alpha_s$)}}(Y,\gamma,\alpha_s) + G_{0,n_F=0}^{\text{(LO+$\alpha_s$)}}(Y,\gamma,\alpha_s)-1, \label{eq:G0-matched-nF=0}
\end{equation}
where the first term is given by eqn.~(\ref{eq:G0-sol-nF=0-anti-coll}) and contains all the LLA anti-collinear terms $\sim\alpha_s^m/(1-\gamma)^m$. The second term ($G_{0,n_F=0}^{\text{(LO+$\alpha_s$)}}$), whose role  is to improve the approximation away from $\gamma=1$, 
is given by eqn.~(\ref{eq:Gn(Y,gamma)-appr-sol}) with the replacement:
\begin{equation}
    \chi_0(0,\gamma)\to \chi_0(0,\gamma)-\frac{1}{1-\gamma}.
\end{equation}
The subtraction is required to prevent  a double-counting of the LLA anti-collinear terms. 

The next step is to evaluate the $\gamma$ integral (\ref{eq:G-conf-exp_(-)}). Define $L_T\equiv\ln(\T{k}^2/\T{p}^2)$ ($L_T>0$ corresponds to the collinear and $L_T<0$ -- to the anti-collinear regimes). We will consider only the Green's function averaged oner the azimuthal angle, which determines the $Y$-scaling of the total cross section. In this case, only the contribution of $n=0$ in eqn.~(\ref{eq:G-conf-exp_(-)}) is relevant,
\begin{equation}
   \bar{G}(Y,\T{k}^2;\T{p}^2)\equiv \T{k}^2\langle G(Y,\T{k};\T{p}) \rangle_{\phi_{\T{k}\T{p}}} =  \int\limits_{1/2-i\infty}^{1/2+i\infty} \frac{d\gamma}{2\pi^2 i} \exp \big[ V(\gamma; Y,L_T) \big],\label{eq:Gbar(Y)-def}
\end{equation}
with the ``potential'' $V(\gamma; Y,L_T)$  defined as 
\begin{equation}
    V(\gamma; Y, L_T) = \gamma L_T + \ln G_0(Y,\gamma,\alpha_s).\label{eq:pot-V-def}
\end{equation}
In the limit of large $Y$ or $L_T$, the integral is dominated by saddle points of the potential in $\gamma$-plane. The complex-valued saddle points will lead to unphysical oscillations of the integral as a function of $Y$ or $L_T$. The physically consistent approximation for the BFKL kernel should have  minimum of the potential at a real value of $\gamma=\gamma_S$. That
is at $\gamma=\gamma_S$ the derivative of the potential vanishes,
\begin{equation}
    V'(\gamma_S;Y,L_T)=0.
\end{equation}
The saddle-point approximation for the integral (\ref{eq:Gbar(Y)-def}) then becomes:
\begin{equation}
    \bar{G}_{\text{SP}}(Y,\T{k}^2;\T{p}^2)= \frac{e^{V(\gamma_{S};Y,L_T)}}{\sqrt{2\pi V''(\gamma_{S};Y,L_T)}}. 
\end{equation}
The potential (\ref{eq:pot-V-def}) is a quantity closely related to the characteristic function. 
In the fixed-coupling approximation the potential is:
\begin{equation}
    V^{\text{(fix. $\alpha_s$)}}(\gamma; Y,L_T,\alpha_s) = \gamma L_T + \frac{\alpha_s N_c}{\pi} Y \chi^{\text{(fix. $\alpha_s$)}}(0,\gamma,a_s).\label{eq:V(gamma)-fix-coupl-DEF}
\end{equation}
Here $\chi^{\text{(fix. $\alpha_s$)}}$ refers to any characteristic function computed at fixed coupling. In practice, we will consider two: the LO $\chi_0$ 
and the anti-collinearly resummed characteristic function in the fixed-coupling approximation, derived in Ref.~\cite{Kovner:2026qfm}. 

In the running-coupling case one substitutes eqn.~(\ref{eq:Gn(Y,gamma)-sol-00}) for $G_0$ into eqn.~(\ref{eq:pot-V-def}).
We will use the matched LO running coupling + anti-collinear LLA approximation (\ref{eq:G0-matched-nF=0}) for $G_0$ with $n_F=0$ in the numerical computations in this section.

 Pomeron intercept is defined as an effective exponent characterising the growth of the cross section with $Y$. For the BFKL Pomeron it is defined by approximating $e^{V(\gamma_S;Y,L_T)}\simeq e^{\omega_P Y}$ or equivalently
\begin{eqnarray}
\omega_P(L_T,Y_0)&=&\frac{dV(\gamma_S;Y_0,L_T)}{dY_0} = \underbrace{V'(\gamma_S;Y_0,L_T)}_{0} \frac{\partial\gamma_S}{\partial Y_0} + \frac{\partial V(\gamma_S;Y_0,L_T)}{\partial Y_0} = \frac{\partial V(\gamma_S;Y_0,L_T)}{\partial Y_0}, \nonumber \\
\label{eq:OmegaP-DEF-00}
\end{eqnarray}
%or in terms of the function $G_0$:
%\begin{eqnarray}
%    \omega_P(L_T,Y_0)= \frac{1}{G_0(Y,\gamma_S(L_T,Y_0),\alpha_s)} \frac{\partial G_0(Y_0,\gamma_S(L_T,Y_0),\alpha_s)}{\partial Y_0}.\label{eq:OmegaP-RC}
%\end{eqnarray}
The intercept $\omega_P$ might still have a mild dependence on the ``probe rapidity'' $Y_0$, because $\gamma_S$ depends on $Y_0$. One however typically expects a much stronger dependence on $L_T$, which we will explore below.

In the fixed-coupling case, the intercept is proportional to the value of the characteristic function at the saddle point:
\begin{eqnarray}
    \omega_P^{\text{(fix. $\alpha_s$)}}(L_T,Y_0)= \frac{\alpha_s N_c}{\pi} \chi^{\text{(fix. $\alpha_s$)}}(0,\gamma_S(L_T,Y_0),a_s),\label{eq:OmegaP-fix-coupl}
\end{eqnarray}
while in the running-coupling case the potential is computed from the corresponding approximation for $G_0$ by eqn.~(\ref{eq:pot-V-def}).

%\old{We are turning to numerical analysis, which requires additional, phenomenological inputs. First, we will freeze $\alpha_s(k^2)$ at the  lowest {\it projectile} scale $k^2$, $\alpha_s(k^2)=0.2$. Second, the integrations over $\tau$ in (\ref{eq:Gn(Y,gamma)-appr-sol}) and (\ref{eq:G0-sol-nF=0-anti-coll}) run into  IR for $\tau\gtrsim  1/\alpha_s$. This is the typical infrared diffusion of the BFKL ~\cite{Bartels:1993hh,Bartels:1995yk}. The diffusion to IR manifests itself as crossing of the integrand's singularities  by the $\tau$-integration contour. In practice, we will cut the $\tau$-integration at the value $\tau_{\max}=(1-\gamma)/(\beta_0^{(g)} a_s(\T{p}^2))$ to avoid crossing the logarithmic branch-point in eqn.~(\ref{eq:G0-sol-nF=0-anti-coll}). Thus our computation corresponds to taking into account only the perturbative contributions while disregarding any non-perturbative contributions to the Green’s function.}\CommentMy{MN:I like my old paragraph better.}

\begin{figure}
    \centering
    \includegraphics[width=0.8\linewidth]{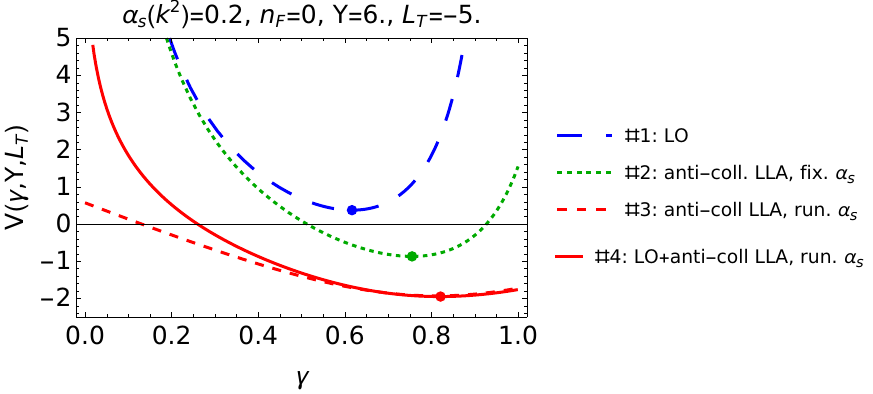}
    \caption{Effective potential (\ref{eq:pot-V-def}) for several approximations for  $G_0$: \#1 -- fixed-coupling approximation (\ref{eq:V(gamma)-fix-coupl-DEF}) with LO characteristic function $\chi_0(0,\gamma)$; \#2 -- fixed-coupling approximation (\ref{eq:V(gamma)-fix-coupl-DEF}) with the anti-collinearly resummed characteristic function of Ref.~\cite{Kovner:2026qfm}; \#3 -- running-coupling $G_{0,n_F=0}^{\text{(LLA+$\alpha_s$)}}$ of eqn.~(\ref{eq:G0-sol-nF=0-anti-coll}); \#4 -- running-coupling LO + anti-collinear LLA approximation of eqn.~(\ref{eq:G0-matched-nF=0}). Positions of the corresponding saddle-points are marked by the dots on the curves.}
    \label{fig:V-plots}
\end{figure}

\begin{figure}
    \centering
    \includegraphics[width=0.8\linewidth]{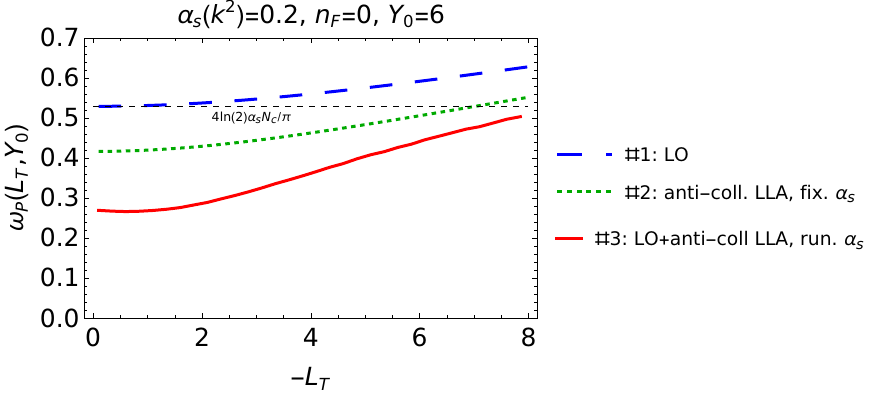}
    \caption{Plots of the Pomeron intercept (\ref{eq:OmegaP-DEF-00}), corresponding to the different approximations for the Green's function: \#1 -- fixed-coupling approximation (\ref{eq:OmegaP-fix-coupl}) with $\chi_0(0,\gamma)$; \#2 -- fixed-coupling approximation (\ref{eq:OmegaP-fix-coupl}) with the anti-collinearly resummed characteristic function of Ref.~\cite{Kovner:2026qfm}; \#3 -- running-coupling intercept (\ref{eq:OmegaP-DEF-00}) with LO running-coupling + anti-collinearly resummed $G_0$ of eqn.~(\ref{eq:G0-matched-nF=0}). }
    \label{fig:OmegaP-plots}
\end{figure}

We will study the behaviour of the Green's function in the anti-collinear regime $L_T<0$. To avoid crossing the Landau pole while going deeper into this regime, we fix the $\alpha_s(\T{k}^2)$ at the lower {\it projectile} scale $\T{k}^2$ to a given value, which is indicated in Figs.~\ref{fig:V-plots} and \ref{fig:OmegaP-plots}. We then evolve $\alpha_s$ towards the higher target scale $\T{p}^2$ and use thus obtained value of $\alpha_s(\T{p}^2)$ in the formulas (\ref{eq:Gn(Y,gamma)-appr-sol}) or (\ref{eq:G0-sol-nF=0-anti-coll}). 

Even with this precaution, $G_0$ evaluated via eqns.~(\ref{eq:Gn(Y,gamma)-appr-sol}) and (\ref{eq:G0-sol-nF=0-anti-coll}) is not free from infrared contribution, associated with the phenomenon of the diffusion to infra red, typical for the solutions of BFKL equation~\cite{Bartels:1993hh,Bartels:1995yk}. In eqns.~(\ref{eq:Gn(Y,gamma)-appr-sol}) and (\ref{eq:G0-sol-nF=0-anti-coll}) the diffusion to IR manifests itself as crossing of the singularities of the integrand by the $\tau$-integration contour at large values of $\tau\sim 1/\alpha_s$.  In the complete phenomenological computation one perhaps would need to limit the integration in $\tau$ such that only the perturbative terms are included, parametrise the non-perturbative contribution from large-$\tau$ region and fit the resulting model for the Green's function to experimental data, similarly to how it is done for the non-perturbative contribution to the Collins-Soper kernel in TMD-factorisation~\cite{CollinsQCD,Scimemi:2019cmh}. Such a complete phenomenological analysis is far beyond the scope of the present paper. Instead, to access the importance of these nonpertubative effects we simply cut the $\tau$-integration at the value $\tau_{\max}=(1-\gamma)/(\beta_0^{(g)} a_s(\T{p}^2))$ to avoid crossing the logarithmic branch-point in eqn.~(\ref{eq:G0-sol-nF=0-anti-coll}) and generation of unphysical imaginary part of the Green's function. Thus our computation corresponds to taking into account only the perturbative contributions and disregarding any non-perturbative contributions to the Green's function. We thus consider the numerical results of this section as semi-quantitative indications of the influence of the different resummations considered here on the ``effective potential'' and the Pomeron intercept.

%\CommentMy{What is the difference between the two plots on Fig 2? Is it due to cutting off the region of integration in $\tau$? The effect seems to be dramatic for $\gamma\rightarrow 1$, precisely the region where we pretend to provide a reliable result. I am totally lost }\CommentMy{MN: I left the old plots where I just took $Re G_0$ for comparison with the new ones (with a cut on $\tau$), because I have shown the old plots to you before. Later I realised that just taking the real part of $G_0$ still leaves some non-perturbative terms in the series, so the cut on $\tau$ is the safest option. I wanted to discuss this with you and A, but we had more interesting topics...}
In Fig.~\ref{fig:V-plots} several plots of the effective potential are shown for  $L_T=-5$, deeply in the anti-collinear regime. The curve \#1 is a reference curve of the fixed-coupling approximation (\ref{eq:V(gamma)-fix-coupl-DEF}) with $\chi_0(0,\gamma)$.
One can see that the potential with the fixed-coupling anti-collinear resummation of Ref.~\cite{Kovner:2026qfm}  has a shape (curve \#2) qualitatively different from the one with the resummation and running coupling (curve \#4). One can also note that at this particular value of $L_T$, the potential around the minimum is completely dominated by the anti-collinear contribution (\ref{eq:G0-sol-nF=0-anti-coll}) which is plotted separately (curve \#3).

In Fig.~\ref{fig:OmegaP-plots},  plots of the $L_T$-dependence of the Pomeron intercept for a fixed value of $Y_0$ are shown for various approximations for the Green's function. The intercept decreases with consecutive inclusion of anti collinear fixed-coupling (curve \#2) and running-coupling (curve \#3) corrections in comparison to the LO intercept (curve \#1). Interestingly, the intercept with inclusion of the running-coupling and anti-collinear corrections reaches values as low as $\omega_P\sim 0.3$, which are phenomenologically reasonable for the hard Pomeron, see e.g. Fig.~2 in Ref.~\cite{ChVeraRGI2}.

The effect of excluding the nonperturbative region on the effective potential is very small to the left of the saddle point. However  the potential becomes rather flat in the vicinity of the saddle point, and thus the effect on its exact location and the Pomeron intercept is quite significant.

We note that the running coupling effects are very important in the calculations of this section. For example, the reduction of the Pomeron intercept at values of $Y_0$ and $L$ considered for fig.~\ref{fig:OmegaP-plots}  due to the running coupling roughly doubles the reduction due to the anti collinear resummation at fixed coupling. This is in contradistinction to the situation discussed in the previous section, where the running of the coupling had no effect on the value of characteristic function at $\gamma=1$.

%In the complete phenomenological computation one perhaps would need to limit the integration in $\tau$ such that only the perturbative terms are included, parametrise the non-perturbative contribution from large-$\tau$ region and fit the resulting model for the Green's function to experimental data, similarly to how it is done for the non-perturbative contribution to the Collins-Soper kernel in TMD-factorisation~\cite{CollinsQCD,Scimemi:2019cmh}. Such a complete phenomenological analysis is far beyond the scope of the present paper.

\section{Conclusions and outlook}\label{sec:conclusions}

In this paper we have {investigated the} effects of running of the QCD coupling  $\alpha_s$ on the anti-collinearly resummed JIMWLK evolution in the linear {(BFKL)} regime. 
We incorporated the running of the coupling {both in}  the JIWMLK kernel {and}  in the resummation (DGLAP) equation which determines the resummation functions  entering the kernel.  

We have elucidated the {appropriate} choice of scale in the running coupling in the JIMWLK kernel. Interestingly this choice of scale is intimately related with the anti-collinear resummation itself. It can be understood as the naive choice proposed in \cite{Kovner:2023vsy} improved by the CSS evolution with respect to the transverse resolution scale. In the {phenomenologically} important kinematic regions, where the {emitted} gluon is  either far from both sources in the amplitude and conjugate amplitude, or much closer to one of the sources than to the other, the {resulting} scale {coincides with the prescription}   proposed by Balitsky {and Chirilli} in Ref.~\cite{Balitsky:2007feb}.

We have derived the resummed BFKL equation with the running coupling and  calculated {its} generalized characteristic function. We followed here the conventional definition of the generalized characteristic function $\chi(\gamma)$, even though this object is not an eigenvalue of the BFKL equation once the running of the coupling  is included. This however facilitates straightforward comparisons with existing literature. 

In general, both the anti-collinear resummation and the running of the coupling  slow down  the BFKL evolution. {Their combined effect} leads to an additional suppression of $\chi(\gamma)$ of order of $10\%$ for $0.6<\gamma<0.9$ relative to the fixed-coupling case. However interestingly enough, for $\gamma=1$ the running coupling effect completely disappears, and the resulting value of $\chi(\gamma=1)$ is not changed relative to the fixed coupling case. The disappearance of the effect of the running at $\gamma=1$ is quite intricate: it happens due to {an} exact cancellation between the effect of running of the coupling in the BFKL kernel, and the running of the coupling in the resummation (DGLAP) equation. {If either of these effects is switched off, } the value of $\chi(\gamma=1)$ changes.

We have also {investigated} the question of how  the generalized characteristic function {can be used} to calculate physical observables. In particular we have found an approximate expression for the BFKL Green's function which includes the anti-collinear resummation with running coupling, and have used the saddle-point approximation to calculate the Pomeron intercept in gluodynamics. Here, as in the {conventional} BFKL analysis, {the calculation is affected} by the diffusion into IR. Nevertheless our results show  the {reduction of the} intercept {due to the resummation effects}, {with values approaching the}  phenomenologically reasonable value close to $0.3$ in {certain} kinematical regimes. {We emphasize that these estimates should be regarded as}  semi-quantitative.

{Our result for $\chi(\gamma=1)$ raises an interesting question.} {On one hand,} in Ref.~\cite{Kovner:2026qfm}, we found that, for fixed coupling, $\chi(\gamma=1)$ is given by \eqref{eq:chi_LLA+as(q)_gamma=1}, and, in the present paper, we have confirmed that  running of $\alpha_s$ does not affect this value. On the other hand, momentum conservation in conformal theory (e.g. ${\cal N}=4$ Yang-Mills) requires $\chi(\gamma=1)=\pi/(\alpha_sN_c)$. {Although } the discrepancy between the two is not {large}, {it is}  significant {enough to}  require an explanation. 

At present we believe that at heart of this discrepancy is the fact that the DGLAP cascade in QCD has a finite extent in rapidity. In our current calculation on the other hand it is treated as point-like, %\old{and the whole cascade of DGLAP splittings develops at a single point of JIMWLK evolution} 
{with the entire cascade of DGLAP splittings occurring at a single step of the JIMWLK evolution}. 
%\CommentMy{ML: I dont understand the previous sentence}
We believe that the same {approximation} is behind the discrepancy between our results and prediction of the RG improvement of Ref.~\cite{Ciafaloni:2003rd}, beyond $\omega=0$. That is the dependence of $A_1$ on $\omega$ in \eqref{eq:Salam-Stasto} also stems from a finite extent of the DGLAP cascade in $Y$. We are currently extending the anti-collinear resummation {framework} to account for this effect. 

{Nevertheless we emphasize that our current approximation captures most of the physics and the main numerical effect of the anti-collinear resummation. {This is illustrated by the numerical closeness of our value of $\chi(\gamma=1)$, eqn.~(\ref{eq:chi_LLA+as(q)_gamma=1}), to the CFT expectation $\pi/(\alpha_s N_c)$.} It is valuable due to its relative simplicity in comparison with the resummation that would fully incorporate all the DGLAP effects, since the latter has to introduce some kind of retardation effects in JIMWLK evolution.}

Another interesting {open} question is how to extend the current approach to resum collinear logarithms. In principle, one could try to use the dense-dilute duality~\cite{Kovner:2005en,Kovner:2005uw} to {interchange the roles of} the projectile and target. {The} collinear and anti-collinear regimes {map onto each other} under this duality transformation. It is {nevertheless important}  to understand how both regimes {emerge within}  the evolution of the wave function of the same object, i.e. projectile in our approach. {Work along this lines is currently in progress.} 

\acknowledgments

%This is the most common positions for acknowledgments. A macro is available to maintain the same layout and spelling of the heading.

%\paragraph{Note added.} This is also a good position for notes added after the paper has been written.

This research was supported by   Binational Science Foundation grants \#2022132,
\#2024818. This work is also supported by the U.S. Department of Energy, Office of Science, Office of Nuclear Physics, within the framework of the Saturated Glue (SURGE) Topical Theory Collaboration.
 The work of  A.K. is supported by the NSF Nuclear Theory grant \#2514546.
The work of ML and MN is supported by the ISF grant \#910/23. The work of V.S. is supported by the U.S. Department of Energy, Office of
Nuclear Physics through contract DE-SC0020081. \\
We  thank Physics Departments of the Ben Gurion University of the Negev and University of Connecticut for hospitality during mutual visits. \\

\appendix

\section{Relating coordinate space and momentum space characteristic functions}\label{append:char-func-rel}

In this Appendix, we {investigate} the relation between the generalized characteristic function in the coordinate and momentum space.

\subsection{All-order relation between $\aleph$ and $\chi$}\label{sec:chi_aa-chi_rr-relation}

The relation between the coordinate-space characteristic function $\aleph^{{\text{(L)}}}$ and momentum-space characteristic function $\chi^{{\text{(L)}}}$ can be derived using their definitions, Eqs.~(\ref{eq:chi_alpha-alpha_DEF}) and (\ref{eq:chi_rho-rho_DEF}) 
and the definition of the Fourier-transform of the kernel (\ref{eq:Kres(k,q)-def})\footnote{{Here again the superscript ${(L)}$ indicates no reference to any specific approximation.}}. {We first note that }
\begin{eqnarray}
  && \hspace{-1cm} \alpha_s(\T{q}^2) (\T{q}^2)^\gamma = \frac{\alpha_s(\T{k}^2) (\T{q}^2)^\gamma}{ 1 + a_s(\T{k}^2)\beta_0\ln(\T{q}^2/\T{k}^2) } = \alpha_s(\T{k}^2)\big[ 1 + a_s(\T{k}^2)\beta_0\big(\partial_\gamma - \ln \T{k}^2 \big) \big]^{-1} (\T{q}^2)^\gamma, \label{eq:RC-OP-01} \\ 
  && 
  \nonumber \\ 
  && 
  \hspace{-1cm} \alpha_s(c_E(\T{x}-\T{y})^{-2}) ((\T{x}-\T{y})^2)^{1-\gamma} \nonumber \\
  && =\alpha_s(\T{k}^2)\big[ 1 + a_s(\T{k}^2)\beta_0\big(\ln c_E +\partial_\gamma - \ln \T{k}^2 \big) \big]^{-1} ((\T{x}-\T{y})^2)^{1-\gamma}. \label{eq:RC-OP-02}
\end{eqnarray}
 Here we have used the one-loop {running coupling formula} \eqref{eq:as-Q2} and  represented factors of $\ln {\bf q}^2$ and $\ln [({\bf x}-{\bf y})^{-2}]$  as differential operators acting on the corresponding momentum- and coordinate-space eigenfunctions.
 
{Retaining} only the terms proportional to $\beta_0$ is sufficient to study the anti-collinear resummation {of the} BFKL {characteristic function} in LLA. In this approximation, the all-order relation between the two characteristic functions takes the form:
\begin{eqnarray}
    \chi(n,\gamma) = \big[ 1 + a_s(\T{k}^2)\beta_0\partial_\gamma\big]^{-1} \frac{\aleph(n,\gamma)}{F(n,\gamma)} \big[ 1 + a_s(\T{k}^2)\beta_0\big(\ln c_E +\partial_\gamma \big) \big]^{-1} F(n,\gamma). \label{eq:chi_alpha-alpha_rho-rho_REL}
\end{eqnarray}
%\CommentMy{ML: Is it a "+" scheme or general? MN: This is general. Just changing the $\alpha_s$-argument.}
where  $F(n,\gamma)=\frac{\Gamma(2-\gamma +n/2) }{4^{\gamma} \Gamma(\gamma+n/2-1)}$ arises from the Fourier-transform of the test function (\ref{eq:EFs-coord}) to momentum space. Here { $\chi(n,\gamma)\equiv\chi^{{(L)}}(n,\gamma,a_s(\T{k}^2))$} and 
$\aleph(n,\gamma)\equiv\aleph^{{(L)}}(n,\gamma,a_s(\T{k}^2))$; i.e. the coupling {entering} the higher order corrections {is evaluated at a fixed scale $\T{k}^2$}.  

Expanding  eqn.~(\ref{eq:chi_alpha-alpha_rho-rho_REL}) up to NLO, one obtains the following relation between NLO corrections to the characteristic function, defined in eqns.~(\ref{eq:aleph_exp}) and (\ref{eq:chi_exp}):
\begin{eqnarray}
    \chi_{1}(n,\gamma) = \aleph_{1}(n,\gamma) + \frac{\beta_0}{N_c} \bigg(-\big(\chi_0'(n,\gamma) +\chi_0^2(n,\gamma)\big) + \frac{2(1-\gamma)\chi_0(n,\gamma)}{(1-\gamma)^2-n^2/4} \bigg),  \label{eq:NLO_chi-chi_relation}
\end{eqnarray}
%\CommentMy{ML: Is it a "+" scheme or general?}
where $\chi_0'(n,\gamma)=\partial_\gamma \chi_0(n,\gamma)$.

%\begin{equation}
%    \int\limits_{C_{\bar{\gamma}}} \frac{d\bar{\gamma}}{2\pi i} \Gtil(s_1,\gamma_1-\bar{\gamma},a_s) \Gtil(s_2,\bar{\gamma}-\gamma_2,a_s)  = \Gtil(s_1+s_2-1,\gamma_1-\gamma_2). \label{eq:Gtil-property}
%\end{equation}
%We also note the value of the function (\ref{eq:Gtil-int-def}) at $\gamma_0=0$:
%\begin{equation}
%  \Gtil(s,0,a_s)=-\frac{1}{a_s \beta_0 s}, \label{eq:Gtil_gamma0=0}  
%\end{equation}
%which is useful for evaluation of the characteristic functions at $\gamma=1$.

\subsection{Testing the relation at NLO}\label{sec:rel-testing-NLO}

%\CommentMy{ML: Worth saying that our procedure is inverse to that  of [22]. MN: We explain the relation with [22] in Sec.~\ref{sec:res-BFKL-mom}. }

The relation (\ref{eq:NLO_chi-chi_relation}) can be tested by comparing the NLO BK/JIMWLK and BFKL results. {To this end, consider the} contribution to the NLO BK kernel~\cite{Balitsky:2007feb}, proportional to  $\beta_0$:
\begin{eqnarray}
  && \hspace{-10mm} K_{D}^{\text{($\beta_0$-NLO)}}(\T{x},\T{y},\T{z},\mu^2) % =  -2K_{\beta_0}^{\text{(NLO)}}(\T{x},\T{y},\T{z}) + K_{\beta_0}^{\text{(NLO)}}(\T{x},\T{x},\T{z}) + K_{\beta_0}^{\text{(NLO)}}(\T{y},\T{y},\T{z}) \nonumber \\
  %&& 
  = \frac{\alpha_s^2(\mu)N_c\beta_0}{8\pi^3}\frac{(\T{x}-\T{y})^2}{X^2 Y^2} \bigg[ \ln[(\T{x}-\T{y})^2\mu^2] - \frac{Y^2-X^2}{(\T{X}-\T{Y})^2} \ln\frac{Y^2}{X^2} \bigg] \,.\label{eq:beta0-D-logs}
\end{eqnarray}
{Upon} linearisation of the NLO BK equation, this kernel governs the evolution of the  operator $(\alpha^a(\T{x})-\alpha^a(\T{y}))^2$, eqn. (\ref{eq:H-(alpha-alpha)^2}). 

The coordinate-space characteristic function corresponding to the kernel (\ref{eq:beta0-D-logs}){, as defined by} eqn.~(\ref{eq:chi_alpha-alpha_DEF}), was computed  in Ref.~\cite{Balitsky:2007feb} {and reads}
\begin{eqnarray}
    \aleph_{1}^{\text{($\beta_0$)}}(n,\gamma) = \frac{\beta_0}{N_c}\bigg[ \frac{1}{2}\big(\chi_0^2(n,\gamma) + \chi_0'(n,\gamma) \big) - \frac{2(1-\gamma)\chi_0(n,\gamma)}{(1-\gamma)^2 - n^2/4} \bigg] .  \label{eq:chi1-Bal-beta0}
\end{eqnarray}
%\CommentMy{ML: Is it a "+" scheme or general?}
Substituting this result to  eqn.~(\ref{eq:NLO_chi-chi_relation}), one obtains
\begin{equation}
     \chi_{1}^{\text{($\beta_0$)}}(n,\gamma) = -\frac{\beta_0}{2N_c} \big( \chi_0 ^{2}(n,\gamma ) + \chi_0'(n,\gamma) \big). \label{eq:chi1-beta0-BFKL}
\end{equation}
%\CommentMy{ML: Is it a "+" scheme or general?}
{This result  coincides exactly, for arbitrary $\gamma$ and $n$, with the $\beta_0$-dependent part of the NLO correction to the generalized BFKL characteristic function computed in Ref.~\cite{Kotikov:2000pm}; see Eq.~(27) therein.
}

%This agreement serves for us as a motivation for the scale choice in eqns.~(\ref{eq:aa-rr-MOM}), which we from now on assume to be valid to all orders.  

\subsection{Relating to the anti-collinear behaviour at NLO}\label{sec:AC-res-NLO}
The leading anti-collinear (and collinear) structure of the NLO characteristic function {in the ``+''-scheme}~\cite{Kotikov:2000pm,Kovner:2026qfm} is given by
\begin{eqnarray}
    \chi_{{+}1}(n,\gamma)&=& -\frac{{4}}{\left( \gamma+n/2 \right)^3} - \frac{\delta_{n,0}}{\gamma^2} \left( \frac{11}{3}+\frac{2n_F}{3N_c^3} \right) + 2\frac{\psi(1)-\psi(n+1)}{\big( \gamma+\frac{n}{2} \big)^2}   \label{eq:NLO-chi-exp-poles} \\
    && -\frac{\delta_{n,0}}{(1-\gamma)^2} \left( \frac{11}{3}+\frac{2n_F}{3N_c^3} \right) + \frac{\uwave{-\beta_0/N_c} + 2\big( \psi(1)-\psi(n+1) \big) }{\big( 1-\gamma+\frac{n}{2} \big)^2} \nonumber \\
    && + O\big( (\gamma+\frac{n}{2})^{-1}, (1-\gamma+\frac{n}{2})^{-1}   \big).\nonumber 
\end{eqnarray}

In Ref.~\cite{Kovner:2026qfm} we {explained the origin of all contributions to} the anti-collinear $1/(1-\gamma)^2$-pole of $\chi_{+1}(n,\gamma)$ for $n=0$, except {for the underlined term} proportional to $\beta_0/N_c$. 
We can now {identify the origin of this term in the anti-collinear regime as a running-coupling effect and demonstrate its consistency with} the resummation formalism described in Sec.~\ref{sec:resumm-general}.

{The $\beta_0/N_c$ term can be traced directly to the contribution (\ref{eq:chi1-beta0-BFKL}) to the NLO BFKL characteristic function.}
For general $n$ the anti-collinear expansion of ~(\ref{eq:chi1-beta0-BFKL}) is:
\begin{equation}
   \chi_{+1}^{\text{($\beta_0$)}}(n,\gamma) = -\frac{\beta_0}{N_c} \frac{1}{\big(1-\gamma+n/2\big)^2} + O\big((1-\gamma+n/2)^{-1}\big). \label{eq:beta0-AC-NLO-BFKL}
\end{equation}
{Thus, after transforming to the momentum-space characteristic function, the NLO BK contribution in Eq.~(\ref{eq:beta0-D-logs}) reproduces the correct NLO BFKL term, Eqs.~(\ref{eq:chi1-beta0-BFKL}) and~(\ref{eq:beta0-AC-NLO-BFKL}).}
Therefore our goal is to show that the {\bf coordinate space} characteristic function itself  contains no additional anti-collinear poles {beyond} those found in \cite{Kovner:2026qfm} in the fixed coupling approximation. 

{Because the effects of DGLAP resummation and running coupling enter additively at NLO,}
we do not need to {re}consider the  anti-collinear resummation {itself}. {It is sufficient} to study the consequences of the renormalisation scale choice (\ref{eq: scale choice mu}) for the coordinate space NLO characteristic function. To  this {end}, in eqn.~(\ref{eq:K-Dip-res-coord}) we replace the resummation functions $R^{(1)}_Q(\T{z}_1)$ and $R^{(2,1)}_Q(\T{z}_1,\T{z}_2)$ with their initial conditions (\ref{eq:R1-init-cond}) and (\ref{eq:R2-init-cond}){. The resulting expression}  reduces it to the dipole kernel with running coupling {in Eq.~}(\ref{eq:KD-alphas}):
\begin{eqnarray}
  &&  \left.  K_D^{\text{(res.)}}(\T{x},\T{y},\T{z}_1,\T{z}_2)  \right\vert_{Q_\star\to\Lambda} = \delta^{(2)}(\T{z}_1-\T{x})\delta^{(2)}(\T{z}_2-\T{y}) \bigg(\int\limits_{\T{z}} \bar{K}_D(\T{x},\T{y},\T{z})\bigg)\nonumber \\
  && - \bar{K}_D(\T{x},\T{y},\T{z}_1)\big( \delta^{(2)}(\T{z}_2-\T{x}) +\delta^{(2)}(\T{z}_2-\T{y}) \big). \label{eq:KDres_Q-to-Lambda}
\end{eqnarray}
Expanding the kernel (\ref{eq:KD-alphas}) in $\alpha_s((\T{x}-\T{y})^{-2})$ and taking into account the scale choice (\ref{eq: scale choice mu}) one finds:  
\begin{eqnarray}
    \bar{K}_D(\T{x},\T{y},\T{z})&=&\frac{\alpha_s((\T{x}-\T{y})^{-2}) N_c}{2\pi^2} \bigg[ K_D(\T{x},\T{y},\T{z}) + a_s((\T{x}-\T{y})^{-2}) \beta_0  \bigg( \hspace{-0.1cm}2\frac{X\cdot Y}{X^2 Y^2} \ln\big({{\mu_\star^2}(\T{x}-\T{y})^{2}}\big)  \nonumber \\
    &&+ \frac{1}{X^2}\ln\frac{X^{2}}{(\T{x}-\T{y})^{2}} + \frac{1}{Y^{2}} \ln\frac{Y^{2}}{(\T{x}-\T{y})^{2}} \bigg) +O(a_s^2) \bigg] \nonumber \\
    &=&  \frac{\alpha_s((\T{x}-\T{y})^{-2}) N_c}{2\pi^2} K_D(\T{x},\T{y},\T{z}) + K_D^{\text{($\beta_0$-NLO)}}(\T{x},\T{y},\T{z},(\T{x}-\T{y})^{-2}) \nonumber \\
    && + \Delta K_D^{\text{($\mu_\star$-NLO)}}(\T{x},\T{y},\T{z})+O(\alpha_s^3),
\end{eqnarray}
where $\mu_\star^2=\min(X^{-2},Y^{-2})$.
In the last two lines we have separated  the $\beta_0$-{dependent} contribution {to} NLO BK kernel (\ref{eq:beta0-D-logs}) {from the remaining}  difference term:
\begin{eqnarray}
 \hspace{-10mm} \Delta K_D^{\text{($\mu_\star$-NLO)}}(\T{x},\T{y},\T{z}) = \frac{\alpha_s^2 N_c \beta_0}{8\pi^3} \bigg[  \frac{(\T{x}-\T{y})^2}{X^2 Y^2} \ln\frac{\max[X^2,Y^2]}{(X-Y)^2} - \frac{1}{\max[X^2,Y^2]} \bigg\vert \ln\frac{X^2}{Y^2} \bigg\vert \bigg]. \label{eq:DeltaKD-b0}
\end{eqnarray}

{Using  (\ref{eq:chi_alpha-alpha_DEF}) and (\ref{eq:KDres_Q-to-Lambda}), the contribution of the term in (\ref{eq:DeltaKD-b0}) to the coordinate space characteristic function is}
\begin{eqnarray}
    \Delta \aleph_1^{\text{($\mu_\star$-NLO)}}(n,\gamma) &=& - \frac{4\pi^3}{\alpha_s^2N_c^2} \int\limits_{\T{z}} \frac{\Delta K_D^{\text{($\mu_\star$-NLO)}}(\T{x},\T{y},\T{z})}{((\T{x}-\T{y})^2)^{1-\gamma} e^{in\phi_{\T{x}\T{y}}}} \bigg( ((\T{x}-\T{y})^2)^{1-\gamma} e^{in\phi_{\T{x}\T{y}}} \nonumber \\
    && - ((\T{x}-\T{z})^2)^{1-\gamma} e^{in\phi_{\T{x}\T{z}}} - ((\T{y}-\T{z})^2)^{1-\gamma} e^{in\phi_{\T{y}\T{z}}} \bigg). \label{eq:DeltaAleph1_mustar_DEF}
\end{eqnarray}
The anti-collinear behaviour of this function at $\gamma\to 1+n/2$ is determined by the UV limit $\T{z}\to \T{x}$ (or $\T{z}\to \T{y}$). The kernel $\Delta K_D^{\text{($\mu_\star$-NLO)}}(\T{x},\T{y},\T{z})$ is growing at most logarithmically in this limit and therefore does not lead a pole at $\gamma\to 1$ for $n=0$. 

To study the case $n>0${, we} consider, for definiteness, the limit $\T{z}\to \T{x}$. At each order of the Taylor-expansion of the kernel in $\T{\delta}=\T{z}-\T{x}$, the term with the highest power of $\cos\phi$, {where $\phi$ is} an angle between $\T{\delta}$ and $\T{\Delta}=\T{y}-\T{x}$, determines the residue of the left-most anti-collinear pole for given $n$. For the kernel (\ref{eq:DeltaKD-b0}) those terms are:
\begin{eqnarray}
    \Delta K_D^{\text{($\mu_\star$-NLO)}}(\T{x},\T{y},\T{z}) = -\frac{\alpha_s^2 N_c \beta_0}{8\pi^3}\frac{1}{|\T{\delta}| |\T{\Delta}|} \sum\limits_{k=0}^\infty \frac{|\T{\delta}|^k}{|\T{\Delta}|^k}\bigg( c_k  \cos^{k+1}\phi + O(\cos^{k}\phi) \bigg), \label{eq:DeltaKD_mustar_exp-delta} 
\end{eqnarray}
with coefficients $c_k=2^{k+1}H_{k+1}$. %\old{Here $H_n=\sum\limits_{k=1}^n 1/k=\psi(n+1)-\psi(1)$ {is} the harmonic sum}.
There are no factors of $\ln(\T{\delta}^2/\T{\Delta}^2)$ in the terms explicitly written in eqn.~(\ref{eq:DeltaKD_mustar_exp-delta}), therefore their $\T{\delta}$-dependence leads merely to the first-order poles at the point $\gamma=1+(k+1)/2$, while the angular integration with $e^{in\phi}$ fixes $k=n-1$. Thus the corresponding left-most anti-collinear pole of the characteristic function (\ref{eq:DeltaAleph1_mustar_DEF}) for given $n$ is:
\begin{eqnarray}
    \Delta \aleph_1^{\text{($\mu_\star$-NLO)}}(n,\gamma) = -\frac{\beta_0}{N_c} \frac{\psi(n+1)-\psi(1)}{1-\gamma+n/2}+O((1-\gamma+n/2)^0).
\end{eqnarray}

{We conclude that} this contribution to the NLO kernel (\ref{eq:DeltaKD-b0}) does not introduce any second order pole. Hence the residue of the second order pole in the coordinate space characteristic function is  the same as the one originating from the kernel (\ref{eq:beta0-D-logs}). Therefore the momentum space characteristic function has the residue as in eqn.~(\ref{eq:beta0-AC-NLO-BFKL}) which is indeed the same as in the NLO BFKL result~\cite{Kotikov:2000pm}. 

To reiterate, the particular structure of the residue for $n\ne 0$ pole in \eqref{eq:NLO-chi-exp-poles} has nothing to do with the anti-collinear resummation, but is entirely due to the choice of the ``test function'' on which we decide to act with the JIMWLK kernel, and the choice of the scale in the coupling constant in JIMWLK Hamiltonian.

The computation performed in this Appendix shows, that the choice of the scale for the strong coupling constant in the kernel (\ref{eq: scale choice mu}) is indeed consistent with the NLO BFKL results.
%The goal of the rest of this paper will be to derive the all-order in $\alpha_s$-results for the momentum-space BFKL kernel and the characteristic function in the anti-collinear limit, following from the anti-collinear resummation with running coupling effects (Sec.~\ref{sec:resumm-general}), combined with the scale-choice (\ref{eq: scale choice mu}) for the coupling in the JIMWLK kernel.

\section{Fourier transform of the kernel to momentum space}\label{append:Fourier}

In this Appendix we perform the exact Fourier-transform of the resummed coordinate-space BFKL kernel (\ref{eq:K-res-def}) 
{to the momentum space}{. We use} the definition (\ref{eq:Kres(k,q)-def}) and the piecewise-smooth scale choices (\ref{eq:scale choice Q}) and (\ref{eq: scale choice mu})  for $Q_\star$ and $\mu_\star$. {Then we derive}   eqn.~(\ref{eq:K_BFKL-exact}). {We} first re-shuffle a little the terms in the eqn.~(\ref{eq:K-res-def}) as follows:
\begin{eqnarray}
   && K^{\text{(res.)}}(\T{x},\T{y},\T{z}_1,\T{z}_2) = K_{I}^{\text{(res.)}}(\T{x},\T{y},\T{z}_1,\T{z}_2) + K_{II}^{\text{(res.)}}(\T{x},\T{y},\T{z}_1,\T{z}_2). \label{eq:K-res-coord-00} \\
   && K_I^{\text{(res.)}}(\T{x},\T{y},\T{z}_1,\T{z}_2)= -\int\limits_{\bar{z}} \frac{\alpha_s(\mu_{\star}^2(\T{x},\T{y},\Tb{z}))N_c}{\pi^2} K(\T{x},\T{y},\Tb{z}) \nonumber \\
   &&\times\bigg\{ \delta^{(2)}(\T{z}_1-\T{x})\delta^{(2)}(\T{z}_2-\T{y}) +R^{(2,1)}_{Q_\star(\T{x},\T{y},\Tb{z})}(\Tb{z}-\T{z}_1,\Tb{z}-\T{z}_2) \nonumber \\
   && -  R^{(1)}_{Q_\star(\T{x},\T{y},\Tb{z})}(\Tb{z}-\T{z}_1) \big( \delta^{(2)}(\T{z}_2-\T{x}) + \delta^{(2)}(\T{z}_2-\T{y})\big) \bigg\}, \label{eq:K-res-coord-I} \\
   && K_{II}^{\text{(res.)}}(\T{x},\T{y},\T{z}_1,\T{z}_2)= \frac{N_c}{2\pi^2}\int\limits_{\bar{z}} \bigg\{ \alpha_s(\mu_{\star}^2(\T{x},\T{x},\Tb{z})) K(\T{x},\T{x},\Tb{z}) \nonumber \\
   && \times \bigg[ \delta^{(2)}(\T{z}_1-\T{x}) - R^{(1)}_{Q_\star(\T{x},\T{x},\Tb{z})}(\Tb{z}-\T{z}_1) \bigg]\delta^{(2)}(\T{z}_2-\T{y}) + (\T{x}\leftrightarrow \T{y},\T{z}_1\leftrightarrow \T{z}_2) \bigg\}.\label{eq:K-res-coord-II}
\end{eqnarray}
so that $K_I^{\text{(res.)}}$ corresponds to emission from the source at point ${\bf x}$ in the amplitude, and from ${\bf y}$ in the conjugate amplitude, and $K_{II}^{\text{(res.)}}$  to the emission from the same source in the amplitude and conjugate amplitude.

We will use the technique {applied} in Ref.~\cite{Kovner:2026qfm} and introduce the integration over two auxiliary scale variables: $Q_1$ for the scale in the resummation functions $R^{(j)}_{Q_1}$ and $Q_2$ for the scale in $\alpha_s$. Then, the Fourier transform (\ref{eq:Kres(k,q)-def}) of the kernel $K_I^{\text{(res.)}}$ is expressed in terms of the function
\begin{eqnarray}
 &&    \kappa (\T{K},\T{P},Q_1,Q_2) = \frac{1}{(2\pi)^2} \int\limits_{\T{x},\T{z}} \frac{\T{x}\cdot \T{z}}{\T{x}^2 \T{z}^2} e^{i\T{x}\T{K}} e^{i\T{z}\T{P}} \theta(\T{z}^2>\T{x}^2) \delta(Q_1^2-\T{x}^{-2}) \delta(Q_2^2-\T{z}^{-2}) \nonumber \\
 && = - \frac{\theta(Q_1>Q_2)}{Q_1^2 Q_2^2} \frac{\T{K}\cdot \T{P}}{\T{K}^2 \T{P}^2} \frac{\partial^2}{\partial \ln Q_1^2 \, \partial \ln Q_2^2} \big( J_0(|\T{K}|/Q_1) J_0(|\T{P}|/Q_2) \big),    
\end{eqnarray}
as follows
\begin{eqnarray}
    && \bigg( \frac{\T{k}^4\alpha_s(\T{q}^2)}{\T{q}^4\alpha_s(\T{k}^2)} \bigg)^{-1} K^{\text{(res.)}}_{I}(\T{k},\T{q}) =  \int\limits_0^{\infty} dQ_1^2 \int\limits_0^{\infty}dQ_2^2 \frac{\alpha_s(Q_2^2)N_c}{\pi^2}\bigg[ \kappa(\T{q}-\T{k},\T{q}-\T{k},Q_1,Q_2) \nonumber \\
    && + R^{(2,1)}(\T{q},Q_1) \kappa(\T{k}^2,\T{k}^2,Q_1,Q_2) \nonumber \\
  && + R^{(1)}(\T{q},Q_1)\big( \kappa(\T{k}-\T{q},-\T{k},Q_1,Q_2) + \kappa(\T{k},\T{q}-\T{k},Q_1,Q_2) \big) \bigg]  + (\T{q}\to -\T{q}). \label{eq:Kres-I-FT-00}
\end{eqnarray}
The first term in the square brackets in this equation requires additional care, because for $\T{q}=\T{k}$ it contains an IR-divergence which cancels the IR-divergence in the Fourier-transform of the kernel $K_{II}^{\text{(res.)}}$. 

Let's write-down the Fourier-transform of eqn.~(\ref{eq:K-res-coord-II}):
\begin{eqnarray}
   &&  \bigg( \frac{\T{k}^4\alpha_s(\T{q}^2)}{\T{q}^4\alpha_s(\T{k}^2)} \bigg)^{-1} K^{\text{(res.)}}_{II}(\T{k},\T{q}) = \int\limits_{\T{x},\T{y},\T{z},\Tb{z}} \frac{\alpha_s((\T{x}-\Tb{z})^{-2})N_c}{2\pi^2 (2\pi)^2S_\perp} \frac{e^{-i\T{k}(\T{x}-\T{y})} e^{i\T{q}(\T{z}-\T{y})}}{(\T{x}-\Tb{z})^2} \nonumber \\
   &&\times  \bigg[ \delta^{(2)}(\T{z}-\T{x}) - R^{(1)}_{|\T{x}-\Tb{z}|^{-1}}(\Tb{z}-\T{z}) \bigg] + (\T{q}\to -\T{q}) \nonumber \\
%  (\begin{array}{c}
%      \T{x}\to\T{x}+\Tb{z} \\
%      \T{z}\to\T{z}+\Tb{z} 
%  \end{array}) && = \int\limits_{\T{x},\T{y},\T{z},\Tb{z}} \frac{\alpha_s(\T{x}^{-2})N_c}{2\pi^2 (2\pi)^2S_\perp} \frac{e^{i(\T{q}-\T{k})\Tb{z}}e^{-i\T{k}(\T{x}-\T{y})} e^{i\T{q}(\T{z}-\T{y})}}{\T{x}^2} \nonumber \\
%   &&\times  \bigg[ \delta^{(2)}(\T{z}-\T{x}) - R^{(1)}_{|\T{x}|^{-1}}(\T{z}) \bigg] + (\T{q}\to -\T{q}) \nonumber \\
%   && = \delta^{(2)}(\T{q}-\T{k}) \int\limits_{\T{x},\T{z}} \frac{\alpha_s(\T{x}^{-2})N_c}{2\pi^2\T{x}^2} \bigg[ \delta^{(2)}(\T{z}-\T{x}) - e^{-i\T{q}(\T{x}-\T{z})}R^{(1)}_{|\T{x}|^{-1}}(\T{z}) \bigg] + (\T{q}\to -\T{q})\nonumber \\
   && = \delta^{(2)}(\T{q}-\T{k}) \int\limits_{\T{x}} \frac{\alpha_s(\T{x}^{-2})N_c}{2\pi^2\T{x}^2} \bigg[ 1 - e^{-i\T{k}\T{x}}R^{(1)}_{|\T{x}|^{-1}}(\T{q}) \bigg] + (\T{q}\to -\T{q}).
\end{eqnarray}
One can separate-out the divergent part of this integral:
\begin{equation}
    K^{\text{(res.)}}_{II}(\T{k},\T{q})=K^{\text{(res.)}}_{II,\text{div.}}(\T{k},\T{q})+K^{\text{(res.)}}_{II,\text{fin.}}(\T{k},\T{q}),
\end{equation}
where the finite part:
\begin{equation}
    K^{\text{(res.)}}_{II,\text{fin.}}(\T{k},\T{q}) = \delta^{(2)}(\T{q}-\T{k}) \int\limits_{\T{x}} \frac{\alpha_s(\T{x}^{-2})N_c}{2\pi^2\T{x}^2} e^{-i\T{k}\T{x}} \bigg[ 1 - R^{(1)}_{|\T{x}|^{-1}}(\T{q}) \bigg] + (\T{q}\to -\T{q}), \label{eq:K-delta-R1}
\end{equation}
contributes to the term of the Fourier-transformed kernel (\ref{eq:K_BFKL-exact}) proportional to the $\delta^{(2)}(\T{k}-\T{q})$. The integral (\ref{eq:K-delta-R1}) is well-defined, because the integrand is suppressed at large $|\T{x}|$ by $e^{-i\T{k}\T{x}}$ and at small $|\T{x}|$ by the square bracket, which tends to zero because $R^{(1)}_Q(\T{q})\to 1$ for $Q\to\infty$. In turn, the IR-divergent part of $K_{II}^{\text{(res.)}}$ is:
\begin{eqnarray}
     K^{\text{(res.)}}_{II,\text{div.}}(\T{k},\T{q}) &=& \delta^{(2)}(\T{q}-\T{k}) \int\limits_{\T{x}} \frac{\alpha_s(\T{x}^{-2})N_c}{2\pi^2\T{x}^2}  \bigg[ 1 - e^{-i\T{k}\T{x}} \bigg] + (\T{q}\to -\T{q}) \nonumber \\
     &=& \delta^{(2)}(\T{q}-\T{k}) \int\limits_0^{\infty} \frac{dQ_2^2}{Q_2^2} \frac{\alpha_s(Q_2^{2})N_c}{2\pi}  \bigg[ 1 - J_0\bigg(\frac{|\T{k}|}{Q_2} \bigg) \bigg] + (\T{q}\to -\T{q}). \label{eq:KII-div}
\end{eqnarray}

One can see that the integral over $Q_2$ is ill-defined at $Q_2\to 0$ because of the logarithmic divergence.
To eliminate this IR-sensitivity, let us come back to the first term in the square brackets in eqn.~(\ref{eq:Kres-I-FT-00}). In order to isolate the ill-defined IR part of this term, one has to consider the convolution of the kernel $K_I^{\text{(res.)}}(\T{k},\T{q})$ and a test function $f(\T{q})$. Then, this convolution can be split into a finite part, where we promote the denominator $1/(\T{q}-\T{k})^2$ into a ``$(+)$-distribution'' in transverse space (\ref{eq:1/(k-q)^2_+:DEF}), %:
%\begin{equation}
%    \int\limits_{\T{q}}\frac{f(\T{q})}{(\T{k}-\T{q})^2_+} = \int\limits_{\T{q}}\frac{f(\T{q})-f(\T{k})\theta(|\T{k}-\T{q}|<|\T{k}|)}{(\T{k}-\T{q})^2},
%\end{equation}
thus defining the divergence-free part of the kernel $K^{\text{(res.)}}_{I}$ as follows:
\begin{eqnarray}
    && \bigg( \frac{\T{k}^4\alpha_s(\T{q}^2)}{\T{q}^4\alpha_s(\T{k}^2)} \bigg)^{-1} K^{\text{(res.)}}_{I,\text{fin.}}(\T{k},\T{q}) =  -\int\limits_0^{\infty} \frac{dQ_1^2}{Q_1^2} \int\limits_0^{\infty}\frac{dQ_2^2}{Q_2^2} \frac{\alpha_s(Q_2^2)N_c}{\pi^2} \theta(Q_1>Q_2) \label{eq:Kres-I-FT-fin-00}\\
    &&\times\frac{\partial}{\partial \ln Q_2^2}\bigg[ \frac{J_0(|\T{k}-\T{q}|/Q_2)}{(\T{k}-\T{q})_+^2} \frac{\partial}{\partial \ln Q_1^2} J_0(|\T{k}-\T{q}|/Q_1)   \nonumber \\
  && + \frac{J_0(|\T{k}|/Q_2)}{\T{k}^2} R^{(2,1)}(\T{q},Q_1) \frac{\partial}{\partial \ln Q_1^2} J_0(|\T{k}|/Q_1) - R^{(1)}(\T{q},Q_1) \frac{\T{k}\cdot (\T{k}-\T{q})}{\T{k}^2 (\T{k}-\T{q})^2} \nonumber \\
  && \times\frac{\partial}{\partial \ln Q_1^2} \big( J_0(|\T{k}-\T{q}|/Q_1) J_0(|\T{k}|/Q_2) +  J_0(|\T{k}|/Q_1)J_0(|\T{k}-\T{q}|/Q_2)  \big) \bigg] + (\T{q}\to -\T{q}). \nonumber
\end{eqnarray}
While the divergent part of the convolution can be transformed as follows:
\begin{eqnarray}
    && \int\limits_{\T{q}} K^{\text{(res.)}}_{I,\text{div.}}(\T{k},\T{q}) f(\T{q}) = -2f(\T{k}) \int\limits_0^{\infty}\frac{dQ_2^2}{Q_2^2} \frac{\alpha_s(Q_2^2)N_c}{\pi^2}\nonumber \\
    && \times \int\limits_{\T{q}}\frac{\partial J_0(|\T{k}-\T{q}|/Q_2) }{\partial \ln Q_2^2} \bigg(  \int\limits_{Q_2^2}^{\infty} \frac{dQ_1^2}{Q_1^2} \frac{\partial J_0(|\T{k}-\T{q}|/Q_1) }{\partial \ln Q_1^2} \bigg) \frac{\theta(|\T{k}-\T{q}|<|\T{k}|)}{(\T{k}-\T{q})^2}  \nonumber \\
%    && = -2f(\T{k}) \int\limits_0^{\infty}\frac{dQ_2^2}{Q_2^2} \frac{\alpha_s(Q_2^2)N_c}{\pi} \int\limits_0^{\T{k}^2}\frac{d\T{q}^2}{\T{q}^2} \frac{\partial J_0(|\T{q}|/Q_2) }{\partial \ln Q_2^2} \bigg(1-J_0\bigg(\frac{|\T{q}|}{Q_2} \bigg) \bigg)\nonumber \\
%    && = 2f(\T{k}) \int\limits_0^{\infty}\frac{dQ_2^2}{Q_2^2} \frac{\alpha_s(Q_2^2)N_c}{\pi} \int\limits_0^{\T{k}^2}\frac{d\T{q}^2}{\T{q}^2} \frac{\partial J_0(|\T{q}|/Q_2) }{\partial \ln \T{q}^2} \bigg(1-J_0\bigg(\frac{|\T{q}|}{Q_2} \bigg) \bigg)\nonumber \\
    && = -2f(\T{k}) \int\limits_0^{\infty}\frac{dQ_2^2}{Q_2^2} \frac{\alpha_s(Q_2^2)N_c}{2\pi} \bigg[ 1-J_0\bigg(\frac{|\T{k}|}{Q_2} \bigg) \bigg]^2. 
\end{eqnarray}
Thus the functional $K^{\text{(res.)}}_{I,\text{div.}}(\T{k},\T{q})$ is:
\begin{equation}
    K^{\text{(res.)}}_{I,\text{div.}}(\T{k},\T{q}) = -\delta^{(2)}(\T{q}-\T{k}) \int\limits_0^{\infty}\frac{dQ_2^2}{Q_2^2} \frac{\alpha_s(Q_2^2)N_c}{2\pi} \bigg[ 1-J_0\bigg(\frac{|\T{k}|}{Q_2} \bigg) \bigg]^2 + (\T{q}\to -\T{q}). \label{eq:KI-div}
\end{equation}
And the sum of the divergent parts (\ref{eq:KI-div}) and (\ref{eq:KII-div}) is equal to:
\begin{eqnarray}
    && K^{\text{(res.)}}_{I,\text{div.}}(\T{k},\T{q}) + K^{\text{(res.)}}_{II,\text{div.}}(\T{k},\T{q}) \nonumber \\
    &&=  \delta^{(2)}(\T{q}-\T{k}) \int\limits_0^{\infty}\frac{dQ_2^2}{Q_2^2} \frac{\alpha_s(Q_2^2)N_c}{2\pi} J_0\bigg(\frac{|\T{k}|}{Q_2} \bigg) \bigg[ 1-J_0\bigg(\frac{|\T{k}|}{Q_2} \bigg) \bigg] + (\T{q}\to -\T{q}) . \label{eq:KI-div+KII-div}
\end{eqnarray}
The integrand of this equaiton is now suppressed both at small $Q_2$ and large $Q_2$, thus the Fourier-transform of the kernel is well-defined. The eqn.~(\ref{eq:KI-div+KII-div}) together with ~(\ref{eq:K-delta-R1}) contributes to the kernel $K_{\delta}(\T{k},\T{q})$ of eqn.~(\ref{eq:K_BFKL-exact}).   

Now our task is to take eqn.~(\ref{eq:Kres-I-FT-fin-00}), which is IR-safe, and integrate by parts over $Q_1$ and $Q_2$, so that the derivatives act on the  functions $R_{Q_1}^{(j)}(\T{p})$ or $\alpha_s(Q_2)$-factors.  Integrating by parts over $Q_2$ with the assumption that $J_0(k/Q_2)\alpha_s(Q_2)\to 0$ for $Q_2\to 0$ and then integrating-by-parts over $Q_1$ in the boundary terms of integration over $Q_2$, one can put  eqn.~(\ref{eq:Kres-I-FT-fin-00}) into the following form: 
\begin{eqnarray}
    && \bigg( \frac{\T{k}^4\alpha_s(\T{q}^2)}{\T{q}^4\alpha_s(\T{k}^2)} \bigg)^{-1} K^{\text{(res.)}}_{I,\text{fin.}}(\T{k},\T{q}) = \int\limits_0^{\infty} \frac{dQ^2}{Q^2}  \frac{N_c}{2\pi^2} \bigg[ \frac{\partial\alpha_s(Q^2)}{\partial \ln Q^2} \frac{J_0^2(|\T{q}-\T{k}|/Q)}{(\T{k}-\T{q})_+^2} \nonumber \\
    && + \frac{\partial [\alpha_s(Q^2) R^{(2,1)}_Q(\T{q})]}{\partial \ln Q^2}  \frac{J_0^2(|\T{k}|/Q)}{\T{k}^2}  -  \frac{2\T{k}\cdot (\T{k}-\T{q})}{\T{k}^2 (\T{k}-\T{q})^2} \frac{\partial[\alpha_s(Q^2) R^{(1)}_Q(\T{q})]}{\partial \ln Q^2}  J_0(|\T{k}|/Q) J_0(|\T{k}-\T{q}|/Q) \bigg] \nonumber \\
  && +\int\limits_0^{\infty} \frac{dQ_1^2}{Q_1^2}  \int\limits_0^{Q_1^2}\frac{dQ_2^2}{Q_2^2} \frac{N_c}{\pi^2} \frac{\partial \alpha_s(Q_2^2)}{\partial \ln Q_2^2}\bigg[ \frac{J_0(|\T{q}-\T{k}|/Q_2)}{(\T{k}-\T{q})_+^2} \frac{\partial}{\partial \ln Q_1^2} J_0(|\T{q}-\T{k}|/Q_1) \nonumber \\
  && + \frac{J_0(|\T{k}|/Q_2)}{\T{k}^2} R^{(2,1)}_{Q_1}(\T{q}) \frac{\partial}{\partial \ln Q_1^2} J_0(|\T{k}|/Q_1)  - R^{(1)}_{Q_1}(\T{q}) \frac{\T{k}\cdot (\T{k}-\T{q})}{\T{k}^2 (\T{k}-\T{q})^2} \nonumber \\
  &&\times\frac{\partial}{\partial \ln Q_1^2} \big( J_0(|\T{k}-\T{q}|/Q_1) J_0(|\T{k}|/Q_2) +  J_0(|\T{k}|/Q_1)J_0(|\T{k}-\T{q}|/Q_2)  \big) \bigg]  + (\T{q}\to -\T{q}). \label{eq:KI-fin-half-way}
\end{eqnarray}
The first two lines of eqn.~(\ref{eq:KI-fin-half-way}) lead to eqn.~(\ref{eq:DK1-BFKL-res-ex}).
 To derive the remaining part of the result,  integration by parts over $Q_1$ in the last three lines of eqn.~(\ref{eq:KI-fin-half-way}) should be performed. One has to be careful with the $Q_1$-derivative, which can act either on $R_{Q_1}^{(j)}$ or on the upper limit of $Q_2$-integration. Taking all of this into account, we derive (\ref{eq:DK2-BFKL-res-ex}) and (\ref{eq:DK3-BFKL-res-ex}) from eqn.~(\ref{eq:KI-fin-half-way}).

%%%%%%%%%%%%%%%%%%%%%%%%%%%%%%%%%%%%%%%%%%%%%%%%%%%%%%%%%%%%%%%

\bibliographystyle{JHEP}
\bibliography{mybibfile}

\end{document}